\documentstyle[aps,amssymb,epsfig,twocolumn,paper]{revtex}
\def\DEG{{$^\circ$C}}
\def\massalski{{Massalski \cite{BB}}}
\def\str64{BCC$_{AB2}^{[011]}$}
\def\CH40{FCC$_{A2B2}^{[201]}$}
\def\tablefont{\footnotesize}
\def\twophase{$\leftrightarrow$}
\def\picdim{80mm}
\def\ftnsz{\footnotesize}
\def\TcDO19{Trends of Tc alloys are further discussed in Section (\ref{section.Tc.trend}).}
\def\PawSection{as described in Section (\ref{section.firstprinciple})}
\def\us{us-lda}
\def\paw{paw-gga}
\def\tablelineone{{\it Ab initio}}\def\tablelinetwo{result} 

\def\NNalloys{80 }
\def\NNalloysC{57 }
\def\NNalloysS{23 }
\def\NNstructures{176 }
\def\NNprototypes{101 }
\def\NNprototypesPoint{101. }
\def\NNcalc{14080 }
\def\NNTOTcalc{32,402 }
\def\NNsecs{$\sim 9.05\cdot10^8$ }   
\def\NNyears{$\sim 28.7$ }           

\def\NNimmiscibilitytable{23}\def\NNimimscibilitycorrecttable{21}
\def\NNagreementsGGA{89}\def\NNagreementsLDA{84}
\def\NNdisagreementsGGA{9}\def\NNdisagreementsLDA{14}
\def\NNgoodpredictions{21}
\def\NNpossiblepredictions{96}
\def\NNimpossible{48} \def\NNimpossibleunkexp{21} \def\NNimpossibleunvproto{27} 
\def\NNexperiments{284}\def\NNexperimentsaccessible{236}

\def\pro{{}}
\def\unkexp{unknown experimental compound}
\def\unvproto{unavailable prototype}

\def\tableimmisciblescapital{TABLE 4} \def\tableimmisciblesmall{Table 4}
\def\tablesummaryscapital{TABLE 3} \def\tablesummarysmall{Table 3}

\def\TheoryText{Another {\it ab initio} study, relevant for this system, can be found in reference }
\def\TheoryTexts{Other {\it ab initio} studies, relevant for this system, can be found in references }

\def\CitationsAgAu{

\bibitem{ref.AgAu.Wei87} 
S.-H. Wei, A. A. Mbaye, L. G. Ferreira, and A. Zunger, 
Phys. Rev. B {\bf 36}, 4163 (1987).

\bibitem{ref.AgAu.Mohri88a} 
T. Mohri, K. Terakura, T. Oguchi, and K. Watanabe, 
Acta Metall. {\bf 36}, 547 (1988).

\bibitem{ref.AgAu.Johnson90} 
R. A. Johnson, 
Phys. Rev. B {\bf 41}, 9717 (1990).

\bibitem{ref.AgAu.Mohri91} 
T. Mohri, K. Terakura, S. Takizawa, and J. M. Sanchez, 
Acta. Metall. Mater. {\bf 39}, 493 (1991).

\bibitem{ref.AgAu.Mohri93} 
T. Mohri, S. Takizawa, and K. Terakura, 
J. Physics: Cond. Matt. {\bf 5}, 1473 (1993).

\bibitem{ref.AgAu.Lu95a} 
Z. W. Lu, B. M. Klein, and A. Zunger, 
Superlattice Microst. {\bf 18}, 161 (1995).

\bibitem{ref.AgAu.Lu95b} 
Z. W. Lu, B. M. Klein, and A. Zunger, 
J. Phase Equilib. {\bf 16}, 36 (1995).

\bibitem{ref.AgAu.Lu95c}
Z. W. Lu, B. M. Klein, and A. Zunger, 
Model. Simul. Mater Sc. {\bf 3}, 753 (1995).

\bibitem{ref.AgAu.Ozolins98a}
V. Ozolins, C. Wolverton, and A. Zunger, 
Phys. Rev. B {\bf 57}, 4816 (1998).

\bibitem{ref.AgAu.Ozolins98b}
V. Ozolins, C. Wolverton, and A. Zunger, 
Phys. Rev. B {\bf 57}, 6427 (1998).

\bibitem{ref.AgAu.Zunger02}
A. Zunger, L. G. Wang, G. L. W. Hart, and M. Sanati, 
Model. Simul. Mater Sc. {\bf 10}, 685 (2002).
}

\def\ReferenceAgAu{
\TheoryTexts \cite{ref.AgAu.Wei87,ref.AgAu.Mohri88a,ref.AgAu.Johnson90,ref.AgAu.Mohri91,ref.AgAu.Mohri93,ref.AgAu.Lu95a,ref.AgAu.Lu95b,ref.AgAu.Lu95c,ref.AgAu.Ozolins98a,ref.AgAu.Ozolins98b,ref.AgAu.Zunger02}.
}

\def\CitationsAgCd{
}

\def\CitationsAgMg{

\bibitem{ref.AgMg.Liu94}
Y. Liu, R. G. Jordan, and S. L. Qiu, 
Phys. Rev. B {\bf 49}, 4478 (1994).

\bibitem{ref.AgMg.Velikokhatnyi00}
O. I. Velikokhatnyi, S. V. Eremeev, I. I. Naumov, and A. I. Potekaev
J. Physics: Cond. Matt. {\bf 12}, 8825 (2000).

\bibitem{ref.AgMg.Rosengaard94}
N. M. Rosengaard and H. L. Skriver, 
Phys. Rev. B {\bf 49}, 14666 (1994).
}

\def\ReferenceAgMg{
\TheoryTexts \cite{ref.AgMg.Liu94,ref.AgMg.Velikokhatnyi00,ref.AgMg.Rosengaard94}.
}

\def\CitationsAgMo{

\bibitem{ref.AgMo.Dai04}
X. D. Dai, H. R. Gong, and B. X. Liu, 
J. Phys. Soc. Jpn. {\bf 73}, 1222 (2004).
}

\def\CitationsAgNa{
}

\def\CitationsAgNb{
}

\def\CitationsAgPd{

\bibitem{ref.AgPd.Skorodumova95}
N. V. Skorodumova, S. I. Simak, E. A. Smirnova, and Yu. Kh. Vekilov,
Phys. Lett. A {\bf 208}, 157-60 (1995).

\bibitem{ref.AgPd.Bruno94}
E. Bruno, B. Ginatempo, E. S. Guiliano, A. V. Ruban, and Yu. Kh. Vekilov,
Phys. Rep. {\bf 249}, 353 (1994).

\bibitem{ref.AgPd.Bruno95}
E. Bruno, B. Ginatempo, and E. S. Giuliano,
Phys. Rev. B {\bf 52}, 14544 (1995).

\bibitem{ref.AgPd.Bruno98}
E. Bruno, B. Ginatempo, and E. S. Giuliano,
Nuovo Cimento D {\bf 20}D, 1367-76 (1998).

\bibitem{ref.AgPd.Olovsson02}
W. Olovsson, I. A. Abrikosova, and B. Johansson, 
J. Electron Spectrosc. Relat. Phenom. {\bf 127}, 65-9 (2002).

}

\def\ReferenceAgPd{
\TheoryTexts \cite{ref.AgPd.Skorodumova95,ref.AgPd.Bruno94,ref.AgPd.Bruno95,ref.AgPd.Bruno98,ref.AgPd.Olovsson02}.
}

\def\CitationsAgRh{
}

\def\CitationsAgRu{

\bibitem{ref.AgRu.Li04}
J. H. Li, L. T. Kong, and B. X. Liu,
J. Phys. Chem. B {\bf 108}, 16071-16076 (2004).
}

\def\ReferenceAgRu{
\TheoryText \cite{ref.AgRu.Li04}.
}

\def\CitationsAgTc{
}

\def\CitationsAgTi{
}

\def\CitationsAgY{
}

\def\CitationsAgZr{
}

\def\CitationsAlSc{

\bibitem{ref.AlSc.Anton97}
H. Anton and P. C. Schmidt, 
Intermetallics {\bf 5}, 449-65 (1997).

\bibitem{ref.AlSc.Asta98}
M. Asta, S. M. Foiles, and A. A. Quong,
Phys. Rev. B {\bf 57}, 11265 (1998)

\bibitem{ref.AlSc.Asta01}
M. Asta and V. Ozolins,
Phys. Rev. B {\bf 64}, 094104 (2001).

\bibitem{ref.AlSc.Ozolins01}
V. Ozolins and M. Asta,
Phys. Rev. Lett. {\bf 86}, 448 (2001).

\bibitem{ref.AlSc.Mayer03}
B. Mayer, H. Anton, E. Bott, M. Methfessel, J. Sticht, J. Harris, and P. C. Schmidt, P.C. 
Intermetallics {\bf 11}, 23-32 (2003).

\bibitem{ref.AlSc.Clouet04}
E. Clouet, M. Nastar, and C. Sigli,
Phys. Rev. B {\bf 69}, 64109 (2004).
}

\def\ReferenceAgRu{
\TheoryTexts \cite{ref.AlSc.Anton97,ref.AlSc.Asta98,ref.AlSc.Asta01,ref.AlSc.Ozolins01,ref.AlSc.Mayer03,ref.AlSc.Clouet04}.
}

\def\CitationsAuCd{
}

\def\CitationsAuMo{

\bibitem{ref.AuMo.Rodriguez95}
J. A. Rodriguez and M. Kuhn,
Surf. Sci. {\bf 330}, L657-64 (1995).
}

\def\CitationsAuNb{
}

\def\CitationsAuPd{
}

\def\ReferenceAuPd{
\TheoryText \cite{ref.AgAu.Mohri93}.}

\def\CitationsAuPt{
}

\def\CitationsAuRh{
}

\def\CitationsAuRu{

\bibitem{ref.AuRu.Kuhn95}
M. Kuhn, J. A. Rodriguez, J. Hrbek, A. Bzowski, and T. K. Sham,
Surf. Sci. {\bf 341}, L1011-18 (1995).
}

\def\CitationsAuSc{
}

\def\CitationsAuTi{
}

\def\CitationsAuY{
}

\def\CitationsAuZr{
}

\def\CitationsCdMo{
}
			
\def\CitationsCdNb{
}
			
\def\CitationsCdPd{
}
			
\def\CitationsCdPt{
}
			
\def\CitationsCdRh{
}
			
\def\CitationsCdRu{
}

\def\CitationsCdTc{
}

\def\CitationsCdTi{
}

\def\CitationsCdY{

\bibitem{ref.CdY.Ishii2003}
Y. Ishii, K. Nozawa, and T. Fujiwara, 
Mat. Res. Soc. Symp. Proc. {\bf 805}, 129-134 (2003)
}

\def\ReferenceCdY{
\TheoryText \cite{ref.CdY.Ishii2003}.
}

\def\CitationsCdZr{
}

\def\CitationsCrMg{
}

\def\CitationsMoNb{

\bibitem{ref.MoNb.Chao04}
J. Chao, C. Wolverton, J. Sofo, L.-Q. Chen, and Z.-K. Liu,
Phys. Rev. B {\bf 69}, 214202 (2004).

\bibitem{ref.MoNb.Rajput95}
S. S. Rajput, R. Prasad, R. M. Singru, S. Kaprzyk, and A. Bansil,
Mater. Sci. Forum {\bf 175-178}, 925 (1995).

\bibitem{ref.MoNb.MoBecker95}
C. MoBecker and J. Hafner,
J. Physics: Cond. Matt. {\bf 7}, 5843-56 (1995).

}

\def\ReferenceMoNb{
\TheoryTexts \cite{ref.AgPd.Bruno94,ref.MoNb.Chao04,ref.MoNb.Rajput95,ref.MoNb.MoBecker95}.
}

\def\CitationsMoPd{
}

\def\CitationsMoPt{
}
 
\def\CitationsMoRh{
}

\def\CitationsMoRu{

\bibitem{ref.MoRh.Ferreira90}
S. Ferreira, J. Duarte Jr., and S. Frota-Pessoa,
Phys. Rev. B {\bf 41}, 5627 (1990).
}

\def\ReferenceMoRu{
\TheoryText \cite{ref.MoRh.Ferreira90}.
}

\def\CitationsMoTc{

\bibitem{ref.MoTc.Souvatzis04}
P. Souvatzis, M. I. Katsnelson, S. Simak, R. Ahuja, O. Eriksson, and P. Mohn,
Phys. Rev. B {\bf 70}, 12201 (2004).
}

\def\CitationsMoTi{

\bibitem{ref.MoTi.Rubin95}
G. Rubin and A. Finel,
J. Physics: Cond. Matt. {\bf 7}, 3139-52 (1995).
}

\def\ReferenceMoTi{
\TheoryText \cite{ref.MoTi.Rubin95}.
}

\def\CitationsMoY{

\bibitem{ref.MoY.Kong02}
L. T. Kong, J. B. Liu, and B. X. Liu,
J. Mater. Res. {\bf 17}, 528-531 (2002).
}

\def\CitationsMoZr{
}

\def\ReferenceMoZr{
\TheoryText \cite{ref.AlSc.Anton97}.
}

\def\CitationsNbPd{

\bibitem{ref.NbPd.Watson02}
R. E. Watson, M. Weinert, and M. Alatalo,
Phys. Rev. B {\bf 65}, 014103 (2002).
}

\def\ReferenceNbPd{
\TheoryText \cite{ref.NbPd.Watson02}.
}

\def\CitationsNbPt{
}

\def\CitationsNbRh{

\bibitem{ref.NbRh.Rajagopalan04}
M. Rajagopalan and M. Sundareswari,
J. Alloys Compd. {\bf 379}, 8-15 (2004).
}

\def\ReferenceNbRh{
\TheoryText \cite{ref.NbRh.Rajagopalan04}.
}

\def\CitationsNbRu{

\bibitem{ref.NbRu.Sanchez93}
J. M. Sanchez and J. D. Becker, 
Mat. Res. Soc. Symp. Proc. {\bf 291}, 115-27 (1993).

\bibitem{ref.NbRu.Becker93}
J. D. Becker and J. M. Sanchez,
J. Mater. Sci. {\bf A170}, 161-167 (1993).

\bibitem{ref.NbRu.Becker91}
J. D. Becker, J. M. Sanchez, and J. K. Tien,
Mat. Res. Soc. Symp. Proc. {\bf 213}, 113-18 (1991).
}

\def\ReferenceNbRu{
\TheoryTexts \cite{ref.NbRu.Sanchez93,ref.NbRu.Becker93,ref.NbRu.Becker91}.
}

\def\CitationsNbY{

\bibitem{ref.NbY.Kong02}
L. T. Kong, J. B. Liu and B. X. Liu,
J. Phys. Soc. Jpn. {\bf 71}, 141-3 (2002).
}

\def\CitationsNbZr{

\bibitem{ref.NbZr.Leuken90}
H. v. Leuken, A. Lodder, M. T. Czyzyk, F. Springelkamp, and R. A. de Groot,
Phys. Rev. B {\bf 41}, 5613 (1990).

\bibitem{ref.NbZr.Sanchez94}
J. M. Sanchez and J. D. Becker, 
Prog. Theor. Phys. Sul. {\bf 115}, 131-45 (1994).

\bibitem{ref.NbZr.Grad96}
G. B. Grad, A. F. Guillermet, and J. R. Granada, 
Z. Metallkd. {\bf 87}, 726-731 (1996).

\bibitem{ref.NbZr.Grad99}
G. B. Grad, G. M. Benites, G. Aurelio, and A. F. Guillermet,
Mat. Sci. Eeng. A-Struct. {\bf A273-275}, 175-180 (1999).

\bibitem{ref.NbZr.Kudrnovsky89}
J. Kudrnovsky and V. Drchal,
Z. Phys. B Cond. Mat. {\bf 73}, 489-93 (1989).
}

\def\CitationsPdPt{
}

\def\ReferencePdPt{
\TheoryTexts \cite{ref.ZUNGER_PROTOS1,ref.AgAu.Lu95b}.
}

\def\CitationsPdRh{

\bibitem{ref.PdRh.Asato01}
M. Asato, T. Hoshino, and K. Masuda-Jindo, 
J. Magn. Magn. Mater. {\bf 226-230}, 1051-2 (2001).

\bibitem{ref.PdRh.Asato01b}
M. Asato, T. Mizuno, T. Hoshino, and H. Sawada
Mater. Trans. {\bf 42}, 2216-2224 (2001).

\bibitem{ref.PdRh.Wolverton93}
C. Wolverton and D. de Fontaine, 
Mat. Res. Soc. Symp. Proc. {\bf 291}, 431-6 (1993).

\bibitem{ref.PdRh.Marquez03}
F. M. Marquez, C. Cienfuegos, B. K. Pongsai, M. Yu. Lavrentiev, N. L. Allan, J. A. Purton, and G. D. Barrera, 
Model. Simul. Mater Sc. {\bf 11}, 115-26 (2003).

}

\def\CitationsPdRu{

\bibitem{ref.PdRu.Guo02}
H.B. Guo, X.Y. Li, and B. X. Liu,
J. Phys. Soc. Jpn. {\bf 71}, 2933-5 (2002). 
}

\def\CitationsPdTc{
}

\def\CitationsPdTi{

\bibitem{ref.PdTi.Wolverton93}
C. Wolverton, G. Ceder, D. de Fontaine, and H. Dreysse, 
Phys. Rev. B {\bf 48}, 726 (1993).
}

\def\ReferencePdTi{
\TheoryText \cite{ref.PdTi.Wolverton93}.
}

\def\CitationsPdY{
}

\def\CitationsPdZr{
}

\def\CitationsPtRh{
}

\def\ReferencePtRh{
\TheoryText \cite{ref.AgAu.Lu95b}.
}

\def\CitationsPtRu{
}

\def\CitationsPtY{
}

\def\CitationsPtTi{
}

\def\ReferencePtTi{
\TheoryText \cite{ref.PdTi.Wolverton93}.
}

\def\CitationsPtZr{
}

\def\CitationsRhRu{
}

\def\CitationsRhY{
}

\def\CitationsRhTc{
}

\def\CitationsRhTi{
}

\def\ReferenceRhTi{
\TheoryTexts \cite{ref.NbRh.Rajagopalan04,ref.PdTi.Wolverton93}.
}

\def\CitationsRhZr{

\bibitem{ref.RhZr.Szajek89}
A. Szajek and A. Jezierski,
Solid State Commun. {\bf 71}, 917-22 (1989).
}

\def\ReferenceRhZr{
\TheoryTexts \cite{ref.NbRh.Rajagopalan04,ref.RhZr.Szajek89}.
}

\def\CitationsRuTi{
}

\def\CitationsRuTc{
}

\def\CitationsRuY{
}

\def\CitationsRuZr{

\bibitem{ref.RuZr.Mehl91}
M. J. Mehl, J. E. Osburn, D. A. Papaconstantopoulos, and B. M. Klein,
Mat. Res. Soc. Symp. Proc. {\bf 186}, 277-82 (1991).
}

\def\ReferenceRuZr{
\TheoryTexts \cite{ref.NbRu.Sanchez93,ref.NbRu.Becker93,ref.NbRu.Becker91,ref.RuZr.Mehl91}.
}

\def\CitationsTcTi{
}

\def\CitationsTcY{
}

\def\CitationsTcZr{
}

\def\CitationsTiZr{
}

\def\CitationsYZr{
}

\begin{document}
\textheight 230mm

\wideabs
{
\title{\LARGE 
{\sf 
Accuracy of ab initio methods in predicting the \\
crystal structures of metals: review of 80 binary alloys. }}
\author{
Stefano Curtarolo$^{1,3}$, Dane Morgan$^{2}$, Gerbrand Ceder$^{2}$}
\address{
$^1$Department of Mechanical Engineering and Materials Science, Duke University, Durham, NC 27708 \\
$^2$Department of Materials Science and Engineering, MIT, Cambridge, MA 02139 \\
$^3${corresponding author, e-mail: stefano@duke.edu}
}
\date{\today}
\maketitle
\begin{abstract}
Predicting and characterizing the crystal structure of materials is a key
problem in materials research and development. We report the results of 
{\it ab initio} LDA/GGA computations for the following systems: {
AgAu,     AgCd,     AgMg,     AgMo$^*$, AgNa,     AgNb$^*$, AgPd,     AgRh$^*$, AgRu$^*$, AgTc$^*$, 
AgTi,     AgY,      AgZr,     AlSc,     AuCd,     AuMo$^*$, AuNb,     AuPd,     AuPt$^*$, AuRh$^*$, 
AuRu$^*$, AuSc,     AuTc$^*$, AuTi,     AuY,      AuZr,     CdMo$^*$, CdNb$^*$, CdPd,     CdPt,     
CdRh,     CdRu$^*$, CdTc$^*$, CdTi,     CdY,      CdZr,     CrMg$^*$, MoNb,     MoPd,     MoPt,     
MoRh,     MoRu,     MoTc$^*$, MoTi,     MoY$^*$,  MoZr,     NbPd,     NbPt,     NbRh,     NbRu,     
NbTc,     NbY$^*$,  NbZr$^*$, PdPt,     PdRh$^*$, PdRu$^*$, PdTc,     PdTi,     PdY,      PdZr,     
PtRh,     PtRu,     PtY,      PtTc,     PtTi,     PtZr,     RhRu,     RhTc,     RhTi,     RhY,      
RhZr,     RuTi,     RuTc,     RuY,      RuZr,     TcTi,     TcY,      TcZr,     TiZr$^*$, YZr$^*$   
($^*=$ systems in which the {\it ab initio} method predicts that no compounds are stable).
A detailed comparison to experimental data confirms the high accuracy with which 
{\it ab initio} methods can predict ground states. }
\end{abstract}
}


\tcontents

\newpage
\section{Introduction}
\label{section.introduction}

First principles computation, whereby the properties of materials are predicted 
starting from the principles of quantum mechanics, is becoming well integrated 
with more traditional materials research.  
A list of {\it ab initio} studies on binary and ternary alloy phase stability 
up to 1994 can be found in reference \cite{ref.DeFontaine_SSP_1994}.  
Since the earliest, completely {\it ab initio} computation of a binary phase diagram
\cite{ref.DeFontaine_TiRh_1988}, the approaches to compute the total energy of a solid have 
significantly improved, and computing resources have continued to become faster 
and less expensive.  
We believe that a point has been reached where, with a reasonable amount of resources, 
high throughput first principles studies of a large number of alloys can 
be performed \cite{ref.CurtaroloPRL2003,ref.Mg.Curtarolo2005,ref.Mg.Curtarolo2005b,ref.MorganMRS2004,ref.MorganJMST2004,ref.WangCALPHAD2004}.
In this paper we present the results of a first principles study of \NNcalc 
computed total energies on \NNstructures crystal structures in \NNalloys binary alloys.  
To our knowledge this is the largest first principles study of its kind on alloys.  
As we have compared the results in every system to experimental compilations, 
this study also offers a statistical test on the accuracy of some current 
{\it ab initio} approaches in correctly predicting the structure of materials.

For \NNagreementsGGA\, compounds we find unambiguous agreement between experiment 
and the {\it ab initio} computation (Table 5), giving some indication of the predictive power of 
modern {\it ab initio} electronic structure methods.  
For many systems, verification of the {\it ab initio} results is difficult, 
as the systems have been poorly or incompletely characterized, or only high-temperature 
information is available experimentally.  
For most of these system, we make predictions that are consistent with the 
limited available information.  Even though our library of \NNstructures crystal 
structures is, to our knowledge, the largest library of {\it ab initio} energies 
ever produced, there are still \NNimpossibleunvproto \,compounds for which we cannot
verify the experimental structures as they are not in our library.  
We have not included such prototypes because they are extremely rare and complicated (many atoms per unit cell) (Table 8).

Overall, we find remarkably few significant discrepancies between the 
{\it ab initio} predictions and the experimental observations (Table 9).
Based on the experimental data in references \cite{BB,PP}, 
we find only nine compounds for which LDA/GGA seems to 
predict the ground state incorrectly. For four of these nine systems, 
the experimental ground state is within less than 10meV/atom of the 
{\it ab initio} ground state.  For the remaining five systems, there are 
at least two in which further investigation indicates 
that the experimental structure assignment is poorly justified, 
leaving three compounds in which a significant disagreement between 
experiment and {\it ab initio} LDA/GGA is likely.
Such disagreements are addressed in Section (\ref{section.discussion}).

We believe that the low ratio of unambiguous errors (3)
to the number of unambiguous correct predictions (\NNagreementsGGA)
is encouraging, and establishes clearly the potential of predicting 
crystal structure correctly with {\it ab initio} methods.  
	
We also predict the stability of five new crystal structures which, 
{to the best of our knowledge,} 
have not yet been observed in any system. 
An AB$_3$ superstructure of the fcc lattice, stable for CdPt$_3$, PdPt$_3$ and Pd$_3$Pt, 
an AB bcc superstructure for MoTi, 
an AB$_3$ bcc superstructure for MoTi$_3$, Mo$_3$Ti, Nb$_3$Tc, RuTi$_3$ and TcTi$_3$,
an A$_2$B$_2$ hcp superstructure for RhRu, and
an A$_2$B$_4$ hcp superstructure for RhRu$_2$ (Appendix (\ref{section.structures})).
In addition, we find two new crystal structures which are not superstructures 
of fcc, bcc or hcp: MoZr$_3$ and Mo$_5$Ti (for Mo$_5$Ti, MoZr$_5$, and Nb$_5$Ru) 
(Appendix (\ref{section.structures2})).

\section{The library: alloys and structures}
\label{section.library}


{\bf Binary alloys.\, }
Our calculated library contains \NNalloys binary intermetallic alloys.
The alloys include the binaries that can be made from row 5
transition metals, as well as some systems with 
Aluminum, Gold, Magnesium, Platinum, Scandium, Sodium, Titanium, and Technetium.
The alloys are: 
AgAu,     AgCd,     AgMg,     AgMo$^*$, AgNa,     AgNb$^*$, AgPd,     AgRh$^*$, AgRu$^*$, AgTc$^*$, 
AgTi,     AgY,      AgZr,     AuCd,     AuMo$^*$, AuNb,     AuPd,     AuPt$^*$, AuRh$^*$, AuRu$^*$, 
AuSc,     AuTc$^*$, AuTi,     AuY,      AuZr,     AlSc,     CdMo$^*$, CdNb$^*$, CdPd,     CdPt,     
CdRh,     CdRu$^*$, CdTc$^*$, CdTi,     CdY,      CdZr,     CrMg$^*$, MoNb,     MoPd,     MoPt,     
MoRh,     MoRu,     MoTc$^*$, MoTi,     MoY$^*$,  MoZr,     NbPd,     NbPt,     NbRh,     NbRu,     
NbTc,     NbY$^*$,  NbZr$^*$, PdPt,     PdRh$^*$, PdRu$^*$, PdTc,     PdTi,     PdY,      PtRh,     
PtRu,     PtY,      PtTc,     PdZr,     PtTi,     PtZr,     RhRu,     RhTc,     RhTi,     RhY,      
RhZr,     RuTi,     RuTc,     RuY,      RuZr,     TcTi,     TcY,      TcZr,     TiZr$^*$, YZr$^*$,  
where the superscript $^*$ indicates the systems in which the 
{\it high throughput ab initio} 
method predicts that no compounds are stable: 
\NNalloysC alloys form compounds and \NNalloysS are non compound forming.

{\bf Structures and their prototypes.\,} 
The library contains \NNstructures crystal structures.
Many of these have the same structure type but with different compositions of occupancies, 
for example, AB$_3$ and A$_3$B 
(also AB and BA if the point groups of atomic positions of A and B are different), 
so the number of distinct prototypes is \NNprototypesPoint
The various concentrations are listed in the table below.

\begin{center}   
  {\tablefont
    \begin{tabular}{||c|c|c||}  \hline  
      Compounds           & Conc.    & number of  \\ 
      composition         & of B     & prototypes  \\ \hline
         A \&          B  &  0\%     &  4 \\ \hline
    A$_5$B \&     AB$_5$  &  16.66\% &  3 \\ \hline
    A$_4$B \&     AB$_4$  &  20\%    &  2 \\ \hline
    A$_3$B \&     AB$_3$  &  25\%    & 27 \\ \hline 
    A$_2$B \&     AB$_2$  &  33.33\% & 32 \\ \hline
A$_5$B$_3$ \& A$_3$B$_5$  &  37.5\%  &  2 \\ \hline
A$_3$B$_2$ \& A$_2$B$_3$  &  40\%    &  1 \\ \hline
A$_4$B$_3$ \& A$_3$B$_4$  &  42.85\% &  1 \\ \hline
        AB (\& BA$^*$)    &  50\%    & 29 (+3) \\ \hline
    \end{tabular}  
  }
\end{center}
    {TABLE 1.
      Compositions, concentrations and number of prototypes inside the library.
      The library has \NNstructures structures, and
      \NNprototypes distinct prototypes 
      ($^*$ at composition AB, 3 prototypes have 
      different point groups in atomic positions A and B, 
      therefore they represent distinct structure types).
    }
\vspace{3mm}

Of such prototypes, 65 are chosen from the most common 
intermetallic binary structures in the {\small CRYSTMET} database \cite{CC} 
and in the Pauling File \cite{PP}.
Such prototypes can be described by their Strukturbericht 
designation and/or natural prototypes \cite{BB,PP}:
A1, A2, A3, A4, A15, 
B$_{h}$, B1, B2, B3, B4, B8$_{1}$, B8$_{2}$, B10, B11, B19, B27, B32, B33 (B$_{f}$), 
C$_{c}$, C2, C6, C11$_{b}$, C14, C15, C15$_{b}$, C16, C18, C22, C32, C33, C37, C38, C49,
D0$_{a}$, D0$_{3}$, D0$_{9}$, D0$_{11}$, D0$_{19}$, D0$_{22}$, D0$_{23}$, D0$_{24}$, 
D1$_{3}$, D1$_{a}$, D2$_{d}$, D7$_{3}$, D8$_{8}$, 
L1$_{0}$, L1$_{1}$, L1$_{2}$, L6$_{0}$, 
CaIn$_{2}$, CuTe, CuZr$_{2}$, GdSi$_{2}$ (1.4), 
MoPt$_{2}$, NbAs (NbP), NbPd$_{3}$, Ni$_{2}$Si, $\Omega$ (with z=1/4), 
Pu$_{3}$Al (Co$_{3}$V), Ti$_{3}$Cu$_{4}$, W$_{5}$Si$_{3}$, YCd$_{3}$, ZrSi$_{2}$, $\gamma$-Ir.
The rest of the structures (36) are fcc, bcc or hcp superstructures.
Twelve of these superstructures consist of stacking of pure A and B planes along some direction.
Of such prototypes, 12 contain a stacking direction, therefore we can name them  
following the parent lattice and the stacking direction:
\begin{equation}
  {LATTICE}_{stacking}^{[direction]}.
  \nonumber
\end{equation}
For example the designation $FCC_{A2B2}^{[001]}$
indicates a structure prototype with FCC parent lattice and A$2$B$2$ 
stacking along direction $[001]$. The 12 prototypes of the library 
that can be labeled in this manner are:
BCC$_{AB2}^{[011]}$, BCC$_{AB2}^{[211]}$, 
FCC$_{A2B2}^{[001]}$, FCC$_{A2B2}^{[011]}$, 
FCC$_{A2B2}^{[111]}$, FCC$_{A2B2}^{[311]}$, 
FCC$_{AB2}^{[100]}$, FCC$_{AB2}^{[111]}$, 
FCC$_{AB3}^{[001]}$, FCC$_{AB3}^{[011]}$, 
FCC$_{AB3}^{[011]}$, FCC$_{AB3}^{[111]}$.
For the FCC superstructures, a conversion table between 
the lattice-stacking-direction, 
the Sanchez - de Fontaine notation \cite{ref.SANCHEZ_DEFONTAINE},
and the Lu {\it et al.} designations \cite{ref.ZUNGER_PROTOS1,ref.ZUNGER_PROTOS2} 
is included below.

\begin{center}   
  {\tablefont
    \begin{tabular}{||l|c|c|c|c||}  \hline  
      & Space Group & Sanchez - de & Lu {\it et al.} & Here  \\ 
      &   & Fontaine \cite{ref.SANCHEZ_DEFONTAINE}&\cite{ref.ZUNGER_PROTOS1,ref.ZUNGER_PROTOS2}& and \cite{ref.CurtaroloMIT2003}\\ \hline
	AB$_2$     & I4/mmm \#139 & AB$_2$(c) & $\beta1/\beta2$    & $FCC_{AB2}^{[100]}$      \\ \hline 
        AB$_2$     & P$\bar{3}$m1 \#164 &     & $\alpha1/\alpha2$  & $FCC_{AB2}^{[111]}$      \\ \hline
        AB$_3$     & P4/mmm \#123 &     & Z1/Z3           & $FCC_{AB3}^{[001]}$     \\ \hline 
        A$_2$B$_2$ & P4/nmm \#129 & AB(a) & Z2            & $FCC_{A2B2}^{[001]}$      \\ \hline  
        A$_2$B$_2$ & C2/m \#12      &           & W2      & $FCC_{A2B2}^{[311]}$     \\ \hline  
        AB$_3$     & Pmmm \#47      &           & Y1/Y3   & $FCC_{AB3}^{[011]}$      \\ \hline 
        A$_2$B$_2$ & Pmmn \#59 & AB(e)          & Y2      & $FCC_{A2B2}^{[011]}$     \\ \hline  
        A$_2$B$_2$ & I4$_1$/amd \#141 &           & CH or ``40''       & $FCC_{A2B2}^{[201]}$    \\ \hline  
        AB$_3$     & R$\bar{3}$m \#166&           & V1/V3              & $FCC_{AB3}^{[111]}$     \\ \hline 
        A$_2$B$_2$ & R$\bar{3}$m \#166&           & V2                 & $FCC_{A2B2}^{[111]}$    \\ \hline  
    \end{tabular}  
  }
\end{center}

    {TABLE 2.
      Conversion table between the lattice-stacking-direction, 
      the Sanchez - de Fontaine notation \cite{ref.SANCHEZ_DEFONTAINE},
      and the Lu {\it et al.} designations \cite{ref.ZUNGER_PROTOS1,ref.ZUNGER_PROTOS2}.
    }
\vspace{4mm}

The prototypes of stable structures, 
that cannot be described by Strukturbericht designation, 
natural prototypes or lattice-stacking-direction convention,
are described in Appendix \ref{section.structures}.
The complete geometrical description of all the fcc, bcc, 
and hcp superstructures of the library can be found 
in reference \cite{ref.CurtaroloMIT2003}.

\section{High-throughput First Principles calculations}
\label{section.firstprinciple}

{\bf Ultra Soft Pseudopotentials LDA calculations (US-LDA).} 
Most of the energy calculations of the library were performed using
Density Functional Theory in the Local Density Approximation (LDA), with the
Ceperley-Alder form for the correlation energy as parameterized by Perdew-Zunger \cite{ref.PerdewZunger} 
with ultra soft Vanderbilt type pseudopotentials \cite{ref.Vanderbilt}, 
as implemented in {\small VASP} \cite{ref.VASP}.
Calculations are at zero temperature and pressure, and without zero-point
motion. The energy cutoff in an alloy was set to 1.5 times the larger
of the suggested energy cutoffs of the pseudopotentials of the elements 
of the alloy (suggested energy cutoffs are derived by the method described in \cite{ref.VASP}).
Brillouin zone integrations were done using 2000/(number of atoms in unit cell) 
${\bf k}$-points distributed as uniformly as possible on a 
Monkhorst-Pack mesh \cite{ref.MONKHORST_PACK1,ref.MONKHORST_PACK2}.
For a sample set of calculations, we verified that with these energy cutoffs 
and ${\bf k}$-points meshes the absolute energy is converged to better than 10 meV/atom. 
Energy differences between structures are expected to be converged to much smaller tolerances.
Spin polarization was used only for magnetic alloys.
For the non-magnetic ones, spin polarization was used only to verify 
the energies of the ground state structures.
All structures were fully relaxed. 
With such methodology, we consider degenerate structures the 
ones that have energies closer than about 5meV/atom.

We have decided to calculate the entire library 
within the LDA formalism instead of the PAW-GGA method (discussed below)
because of the faster execution speed of LDA.

{\bf PAW-GGA calculations.} 
When several structures are in close competition for the ground state,
we also performed calculations in the Generalized Gradient Approximation (GGA), 
with Projector Augmented-Wave (PAW) pseudopotentials, 
as implemented in {\small VASP} \cite{ref.VASP,ref.PAW_BLOCK,ref.VASP_PAW}.
One would expect the PAW-GGA approach to be somewhat more accurate than the US-LDA.
For the GGA correlation energy, we used the Perdew-Wang parameterization (GGA-PW91) 
\cite{ref.perdew_wang_PW91}. 
Similar to the US-LDA case, all PAW-GGA calculations are performed at 
zero temperature and pressure, and without zero-point motion. 
The energy cutoff in each calculation was set to 1.75 
times the larger of the suggested energy cutoffs of the pseudopotentials of 
the elements of the alloy.
Brillouin zone integrations were done using at least 
$\sim$3500/(number of atoms in unit cell) ${\bf k}$-points distributed as 
uniformly as possible on a 
Monkhorst-Pack mesh \cite{ref.MONKHORST_PACK1,ref.MONKHORST_PACK2}.
Spin polarization was used in all calculations.
We expect such calculations to be able to distinguish energies 
of structures previously degenerate.
This is because the increased density of the ${\bf k}$-points mesh, 
the reduced radial cutoff of the PAW potentials versus the ultrasoft pseudopotentials, 
and the GGA correlation energy implementation. 
With such methodology, we consider degenerate structures the 
ones that have energies closer than $\lesssim$1meV/atom.

For some structures both the US-LDA and PAW-GGA calculations are presented. 
To avoid confusion, we specify the type of calculation in brackets: ``(\us)'' or ``(\paw)''.
{\it In addition, all the results of the PAW-GGA calculations are described in italics}. 

{\bf Symmetries of the pure elements.} 
Both the US-LDA and PAW-GGA calculations reproduce the correct experimental 
crystal structures of the pure elements at room temperature.
The only exception is Sodium. In the two formalisms, Na-hcp is very slightly
favored over Na-bcc and Na-fcc, in agreement with other first principle calculations
\cite{ref.Na2001Achcroft,ref.Na1983McM,ref.Na1998Jaf,ref.Na1986Dac,ref.Na2000Kat}. 
At room temperature Sodium has the bcc structure, and undergoes
a martensitic transformation below 35K, to a closed packed structure 
\cite{ref.Na1948Bar,ref.Na1956Bar,ref.Na1984Over,ref.Na1986Ern,ref.Na1990Smi,ref.Na1990Sch,ref.Na1991Sch,ref.Na1999Blas}.
Therefore our results, Na-hcp stable and the very small energy differences with Na-bcc and Na-fcc,
are consistent with the behavior at low temperature only.

{\bf High-throughput computing.} 
To perform the high number of calculations, 
we have implemented a set of automatic tools to prepare 
initial data, perform calculations and analyze results. 

{\it Preparation of input files.} 
All input files are prepared by starting from the templates based on the prototype chemistry.
The volume of the unit cells is determined as a linear combination
of atomic volumes of pure elements with Vegard's law \cite{ref.Vegard}.
The internal position of the atoms have been taken from geometrical configurations 
(for fcc, bcc, hcp superstructures) or have been extracted 
from the {\small CRYSTMET} and the Pauling File databases \cite{CC,PP}. 

{\it Ab initio Calculations.} 
The calculations are performed by a high-throughput tool, 
called {\small AFLOW} (automatic flow), which searches inside the library 
for potential input files and performs the proper {\it ab initio} calculations.
{\small AFLOW} is able to balance the CPU loads in a multiprocessor and cluster environment, 
maximizing the total throughput of the process \cite{ref.CurtaroloMIT2003}.
{\small AFLOW} extracts the energy and initiates further relaxations
with the atomic positions and unit cell geometries obtained from {\small VASP}. 
Therefore, all the structures are relaxed at least twice, 
and many ($\sim 10\%$ of the total) are relaxed three or more times, 
until proper geometrical and energetic convergence is obtained.
To conclude, for a library of \NNstructures crystal structures in \NNalloys binary alloys, 
with a total of \NNcalc structures, {\small AFLOW} performed \NNTOTcalc {\small VASP} calculations.
The total computing time used for this project was \NNsecs CPU seconds  (\NNyears CPU years), 
which was spread over a large set of computers \cite{ref.REF_CPU}. 
Given the different types of CPUs used, we can only {\it estimate} the 
total amount of computation for the project at roughly $\sim 1.2\cdot10^6$ {\small TERAFLOP}.

{\it Collection and error check of the results.} 
Once each structure has been calculated, 
space group symmetry, bond distances, coordination numbers and unit volumes 
are compared for all the structures of that alloy with the same concentration in the library.
Frequently, several structures, starting from different prototypes, relax to the same
final structure with the same energy. 
Therefore, to identify the correct stable phase, it is mandatory 
to be able to determine the final relaxed structure with the 
highest possible reliability.
Details of the high-throughput error checking tool can be found 
in reference \cite{ref.CurtaroloMIT2003}.
A few stable structures remain unidentified. 
For those we report the unit cell and atomic positions in Appendix \ref{section.structures}.
Such prototypes might be new phases to be checked experimentally. 

{\it Calculation of the formation energies and the convex hull.} 
The formation energy for each structure is determined with respect to the most
stable structure of the pure elements. 
To determine the ground states of a system one needs to find, as a function 
of composition, the ordered compounds that have an energy lower than any 
other structure or any linear combination of structures that gives the 
proper composition. This set of ground state structures forms a 
{\it convex hull}, as all other structures have an energy that falls above 
the set of tie lines that connects the energy of the ground states. 
In thermodynamical terms, the {\it convex hull} represents the Gibbs 
free energy of the alloy at zero temperature.

{\bf Comparison with experiments.} 
To compare the {\it ab initio} results to experimental information,
we relied on the information and references in the Binary Alloy Phase Diagrams
(\massalski) and in the Pauling File \cite{PP}. 
Though in some cases references not included in these were also used, 
no systematic approach to go beyond information in the Pauling File 
or Binary Alloy Phase Diagrams was used.

\section{Discussion and summary of results}
\label{section.discussion}

   In comparing the stable structures predicted by the {\it ab initio} computations 
with available experimental information, we have attempted to classify the 
results in a few distinct categories. Table 5 gives the compounds where 
the {\it ab initio} result and experiments are in unambiguous agreement. 
The fact that there are a large number of compounds (\NNagreementsGGA) in 
Table 5 is a positive statement about the accuracy of LDA/GGA in 
capturing the close energetic competition between the 176 structures in 
our library. For one of the systems in Table 5 (PdTi$_{3}$) we took 
the liberty of modifying the experimental result \cite{BB}
which shows the A15 structure (stoichiometry AB$_{3}$) as a line compound 
at composition PdTi$_{4}$. While off-stoichiometric compounds are obviously 
possible, this usually goes together with significant width of the 
single-phase field. Hence, we concluded (maybe erroneously) that the placement 
of A15 at composition PdTi$_{4}$ in reference \cite{BB} 
is likely a typographical error.

	Table 6 shows compositions which are experimentally known to form 
compounds, but lack a complete identification of the structure type. 
Hence, the {\it ab initio} prediction should be seen as a likely crystal structure 
for the compound. In most cases, the {\it ab initio} structure is consistent with 
the conditions imposed by the limited experimental data 
(e.g. bcc-ordering is seen for Ag-Cd, and we predict B19; or C37 is speculated 
for Au-Sc, in agreement with what we predict). 
Hence, it is likely that with further experimental characterization, 
many of these systems would move to Table 5.

	Table 7a-b contains compositions which are experimentally 
characterized as solid solutions, high-temperature two-phase regions, 
or have not been studied at all in a particular composition range. 
Because of the lack of low-temperature information in these systems, 
an unambiguous statement about the accuracy of the LDA/GGA prediction 
cannot be made. For some systems, the {\it ab initio} result is very 
consistent with the experimentally observed behavior. For example, 
in Ag-Au we find many fcc ordered superstructures with low formation 
energies, indicating a weak ordering interaction. So it is likely 
that the {\it ab initio} predicted structures are stable at low temperature, 
but disorder into a solid solution at elevated temperature, a rather 
common occurrence for noble-metal alloys. For other systems in this 
table the comparison between experiments and {\it ab initio} is more troubling. 
For example, while Mo-Ti is experimentally described as a two-phase system 
with no known compounds down to 400\DEG, we find very strong compound formation. 
While it is possible that these compounds disappear through peritectoid 
reactions below 400\DEG, the large negative formation energies make 
this unlikely. It is more likely that either the experimental 
characterization of this system, or the {\it ab initio} result is significantly 
in error. A system such as Mo-Nb, on the other hand, is not at all 
studied experimentally, and no statement about the accuracy of our 
prediction in this system can be made.

	Table 8 contains the compounds whose crystal structure is not 
present in the library, and hence, it is not possible to make a comparison 
with the {\it ab initio} data. In some cases (e.g. Au$_{10}$Zr$_{7}$), 
this is because the crystal structure is uncommon and has a large unit cell, 
making it less worthwhile to include it in the library. 
In other cases (e.g. Rh$_{5}$Ti) the structure is simply unknown. 
Table 8 also includes structures that are only stable at off-stoichiometric 
compositions (e.g. B19 in Au-Ti), as we did not include disorder 
in the library prototypes.

	Finally, Table 9 contains the compositions for which there seems to 
be a clear disagreement between the experimental data in references \cite{BB,PP} 
and the {\it ab initio} results. Because these systems can point at either 
experimental errors, or shortcomings of the {\it ab initio} method, we discuss 
them here in more detail.
Note that {\it ab initio} errors are most likely due to the LDA or GGA 
approximations or the pseudopotentials being used.
In general, one should consider that our 
calculations only produce the zero-K energy, whereas all the experimental 
results are at non-zero temperature. Given that entropic differences 
between structures can be of the order $0.1\sim1.0k_B$ per atom 
\cite{ref.2002Wal}, very small energy differences 
between the experimentally observed structure and our {\it ab initio} results, 
could be reversed at elevated temperature. 
In particular, 
AuNb$_{3}$-A15, 
NbRh$_{3}$-L1$_{2}$, 
NbRu$_{3}$-L1$_{2}$ 
and NbRu'-L1$_0$
may be in this category. 
This phenomenon is quite common. 
As an example, Wolverton and Ozilins have shown that, in the case of the aluminum-copper system,
the vibrational entropy difference is responsible for stabilizing the 
$\theta$ phase Al$_2$Cu (tetragonal C16) over the competing Al$_2$Cu-$\theta'$ phase
(distortion of $\theta_c$-C1, the cubic fluorite phase with CaF$_2$ prototype),
which has the lowest energy, and, therefore, is the stable structure at zero Kelvin \cite{ref.AlCu.Wolverton2001}.
At temperatures higher than T$\sim 150-200$\DEG, Al$_2$Cu-$\theta$ is more favorable. 

\subsection{Discussion of disagreement between experimental and {ab initio} results}

{\bf AuNb$_{3}$.}
The structure AuNb$_{3}$-A15 has been reported a number of times 
\cite{BB,PP,AuNb.56Woo,AuNb.61Dwi,AuNb.69Flu,AuNb.70Ber,AuNb.70Ehr,AuNb.73Ros,AuNb.73Kha,AuNb.77Kur,AuNb.79Flu,AuNb.81Wir,AuNb.82Sav,AuNb.83Wir,AuNb.86Ram,AuNb.91Di,AuNb.81Bai}.
In particular, R\"oschel {\it et al.} \cite{AuNb.73Ros} observed the A15 structure to grow in quenched samples. 
Ball-milling samples of the material to form disordered bcc, revert upon 
heat treatment in an exothermic reaction to A15, suggesting
that A15 could be the ground state for this composition at low temperature.
The relatively small energy difference between A15 and the AuNb$_{2}$\twophase Nb tie-line 
in our calculation could lead to finite temperature stability of A15. 
Alternatively, the $\sim$7meV/atom energy difference (US-LDA and PAW-GGA) could simply be an error of the {\it ab initio} approach.

{\bf AuY.}		
While we predict the B33 structure to be stable by 26meV/atom below 
the B2 structure, this system is experimentally listed as having a B2 
structure. This experimental classification is based on a paper 
by Chao {\it et al.} \cite{AuY.63Cha,PP}. 
Chao's work mentioned that they only see B2 when 
performing a rapid quench. 
Further work by Kusma \cite{AuY.65Kus} and Dwight \cite{AuSc.67Dwi} 
did not observe the B2 structure. 
The experimental results in this system are therefore somewhat suspect. 
It is possible that B2 is a high temperature structure while B33 is the low temperature form. 
However, it should also be pointed out that the mixing energies in this system are very 
large ( $\sim$1eV), and that the potential error of $\sim$25meV/atom (US-LDA and PAW-GGA) is a small fraction of this.

{\bf Cd$_{3}$Nb.}		
The structure of this compound is one of the more interesting 
discrepancies between {\it ab initio} and experiments to emerge from 
our study. Based on experimental work of Von Holleck \cite{CdNb.66Hol}, a 
single L1$_{2}$ compound is claimed to be stable in this system. 
Computationally, no compounds are stable at all, and we find this 
system to be immiscible. 
The compound with lowest energy is a bcc 
superstructure at composition CdNb, 
and energy $\sim$55meV/atom above the Cd\twophase Nb tie-line 
{\it ($\sim$62meV/atom with PAW-GGA)}.
The Cd$_{3}$Nb-L1$_{2}$ is 70meV/atom above that same tie-line 
{\it ($>$100meV/atom with PAW-GGA)}. 
It is not common for LDA/GGA to make such large and qualitative 
errors in metals, and further investigation of this system is 
suggested, as it will likely lead to a reconsideration of the 
experimental classification, or to some novel understanding of 
the errors that {\it ab initio} methods can produce in metals.

{\bf CdPt$_{3}$.}		
Experimentally this compound is claimed 
to form in the L1$_{2}$ structure, while we predict it to form 
in a new structure for which no prototype yet exists 
(CdPt$_{3}^{proto}$ in Appendix VIII). 
We believe the experimental assignment to be uncertain. 
It is based on a single paper by Nowotny \cite{CdPt.52Now} 
which actually does not observe the L1$_{2}$ structure in this system, 
but assigns it by chemical similarity with the Pt$_{3}$Zn system.

{\bf NbRh$_{3}$.}		
This compound is characterized as an L1$_{2}$ structure, whereas 
our {\it ab initio} results find the Al$_{3}$Pu prototype to be 8meV/atom 
lower in energy {\it (5.3meV/atom with PAW-GGA)}. 
This small energy difference could very well be 
an LDA/GGA error, or could be reversed at elevated temperature by 
entropic effects.

{\bf NbRu$_{3}$.}	
NbRu$_{3}$ is similar to the discrepancy of compound NbRh$_{3}$. 
Experimentally it is found to be an L1$_{2}$ structure, 
whereas we find a D0$_{24}$ structure that is 8meV/atom 
below the L1$_{2}$. With PAW pseudopotentials and the GGA 
exchange correlation corrections the D0$_{24}$-L1$_{2}$ 
energy difference reduces to 2.5meV/atom. 
Because this number is so small, 
extremely large {\bf k}-points sets and high energy cutoff 
were used to converge it. 
The results indicate that L1$_{2}$ 
and D0$_{24}$ are all but degenerate in this system. 
Since D0$_{24}$ and L1$_{2}$ are very similar (D0$_{24}$ is a 
periodically anti-phased version of L1$_{2}$) subtle entropic 
effects can easily reverse the stability between these two.

{\bf NbRu.}		
Experimentally this system is classified to have the L1$_0$ structure, 
whereas the {\it ab initio} results indicate an unusual two-phase region 
between Nb$_{3}$Ru (D0$_{3}$) and NbRu$_{2}$ (C37) covering this 
composition \cite{BB,PP,NbRu.63Ben,NbRu.89Che}. 
The most recent experimental result is based on the assignment 
by Chen \cite{NbRu.89Che} who observes a disordered bcc solution at high 
temperature and a transformation to a tetragonal structure near 
1000\DEG. Between 720$\sim$920\DEG\, they see a further symmetry reduction 
to an orthorhombic phase. Hence, even if the tetragonal structure 
below 1000\DEG\, would be L1$_0$ (which is only speculated by Chen) its 
further transformation to an orthorhombic structure indicates that 
L1$_0$ is not the ground state structure. 
Based on a diffusion couple study, 
Hurley \cite{NbRu.64Hur} believes that a 
two-phase region exists near 50\% Ru, though the results seem to 
be irreproducible. 
Besides the difficulty with interpreting the experimental data, 
significant problems also exists on the {\it ab initio} side with this 
compound. The PAW-GGA result is substantially different from 
the US-LDA result. GGA gives L1$_0$ only 4meV/atom above the tie-line 
(versus 20meV/atom in LDA), 
whereas B19 
seems to be $>$100meV/atom above the tie-line (versus 13meV/atom with LDA).

{\bf PtY.}		
Experimental assignment of the structure of this compound as 
B27 is based on two papers \cite{PdTi.65Dwi,PtY.71Kri}. 
The {\it ab initio} work finds 
B33 to be lowest in energy with B27 higher by 50meV/atom 
{\it (60meV/atom with PAW-GGA)}. 
Given that the experimental results seem sound, this is 
probably a true {\it ab initio} error.

{\bf Pt$_{3}$Zr$_{5}$.}		
The work of Schubert {\it et al.} \cite{PtZr.68Sch} establishes Pt$_{3}$Zr$_{5}$	
as a D8$_8$ structure, whereas our {\it ab initio} (LDA) results put the D8$_8$ structure 
36meV/atom above the tie-line formed by PtZr$_{2}$-C16 and PtZr-B33. 
The GGA result is 26meV/atom above the tie-line. 
In each case the W$_{5}$Si$_{3}$ structure is lower in energy than 
the D8$_8$. Given the large mixing energies in this system 
(of the order of 1 eV/atom), these {\it ab initio} errors seem plausible.

\section{Conclusions}		
\label{section.conclusion}

Overall, the comparison between experimental data and {\it ab initio} results 
is encouraging. A large number of the ground states are predicted 
correctly, even though significant competition exists between various 
structures, indicating that relative energy differences are well 
reproduced by LDA/GGA. In many cases, the direct comparison between 
{\it ab initio} and experiments is difficult, as systems have often not 
been studied completely, or have not been studied to low enough 
temperature to make a reliable statement about the ground state 
structures. For only a small number of cases, there is a clear discrepancy 
between {\it ab initio} and experiments (Table 9). 
On further investigations, we find that some of the experimental 
assignments are poorly justified (e.g. CdPt$_{3}$). 
In a few of the "error systems", the energy difference between the 
observed structure and the {\it ab initio} structure is small, which 
makes it fall within the uncertainty one would expect from
LDA/GGA or from a lack of entropic considerations.

	A few systems display, in our opinion, such a significant 
discrepancy between available experimental data and {\it ab initio} 
results as to make them worthwhile for further study. These 
include Cd$_{3}$Nb (L1$_{2}$ observed, but no compounds from 
{\it ab initio}), Nb-Ru and Pt$_{3}$Zr$_{5}$ (respectively L1$_0$ and 
D8$_8$ observed, but no stable compounds at that composition in 
{\it ab initio}), PtY (significant error between B33 and B27) and 
the complete Mo-Ti system (which from experiments appears to 
be a miscibility gap, but in the {\it ab initio} results display a
number of novel and strongly stabilized compounds).
These systems should be investigated further, either to find 
experimental errors, or to shed new light on {\it ab initio} problems.

{Our findings are summarized in \tablesummarysmall. 
By comparing all the available {\it ab initio} calculations 
with the experimental results \cite{BB,PP}, 
we obtain \NNexperiments \,\,potential structure comparisons.
A subset of these, (\NNimpossible), are not available due by the lack 
of the proper prototype or the unknown experimental compound.
The rest of the structure comparisons, (\NNexperimentsaccessible), 
can be divided in 
\NNagreementsGGA\,\,{\it agreements} in ordered systems ($N_{ac}$),
\NNimimscibilitycorrecttable\,\,{\it agreements} in immiscible systems ($N_{ai}$),
\NNdisagreementsGGA\,\,{\it disagreements} ($N_d$), 
\NNgoodpredictions\,\,{\it good predictions} ($N_{gp}$), and 
\NNpossiblepredictions\,\,{\it possible predictions} ($N_{pp}$).
This division is summarized in the following table. 
}

{\tablefont
  \begin{center}
    \begin{tabular}{||c||} \hline
      {\bf Total Experiments comparisons: \NNexperiments } \\  \hline
        {\bf Available comparisons: \NNexperimentsaccessible} \\ \hline
      \begin{tabular}{c|c|c|c|l|c}
	                 & Experimental       & {\it Ab initio}   & & \#                           & Table \\ \hline
        Agreements       & Compound           & Compound          &$N_{ac}$ & $\NNagreementsGGA$             &   5   \\ \hline
        Agreements       & Immiscibility      & Immiscibility     &$N_{ai}$ & $\NNimimscibilitycorrecttable$ &   3   \\ \hline
	Disagreements    & Comp./ Immisc.     & Incorrect         &$N_d$    & $\NNdisagreementsGGA$          &   9   \\ \hline
	Good             & Comp. unknown,     & Compound          &$N_{gp}$ & $\NNgoodpredictions$           &   6   \\
	predictions      & uncertain, guessed & prediction        &         &                                &       \\ \hline
	Possible         & not-studied,       & Possible          &$N_{pp}$ & $\NNpossiblepredictions$       & 7a-b  \\
	predictions      & solid-solution,    & compound          &         &                                &       \\
	                 & two-phase region   &                   &         &                                &       \\ 
      \end{tabular} \\ \hline 
      {\bf Unavailable comparisons: \NNimpossible } \\ \hline
    \begin{tabular}{l|c|c}
      Reason \ \ \ \ \ \ \ \ \ \ \ \ \ \ \ \ \ \ \ \ \ \ \ \ \ \ \ \ \ \ \ \ \ \ \ \ \ \ \ \ \ \ \ \ \ \ \ \ \ \ \ \ \ \ \ \ \ \ \ \ 
                                            & \#                    & Table \\ \hline  
      unavailable {\it ab initio} prototype & \NNimpossibleunkexp   &   8   \\ \hline
      unknown experimental compound         & \NNimpossibleunvproto &   8   \\ 
    \end{tabular} \\ \hline
    \end{tabular}
  \end{center}
}
    {\tablesummaryscapital.
      {
	Summary of the \NNexperiments\,\,structure comparisons between experimental and {\it ab initio} calculations.
	When available, we use the PAW-GGA calculations, otherwise we use the US-LDA ones.
	If only US-LDA calculations were used, we would obtain $N_{ac}^{LDA}$=\NNagreementsLDA\,\,agreements and
	$N_d^{LDA}$=\NNdisagreementsLDA\,\,disagreements.
      }
    }
\vspace{4mm}

{The reliability of our {\it ab initio} method can be defined as 
the fraction of correct compounds (agreements) over the number of accessible 
compounds (agreements and disagreements):
\begin{equation}
\eta_c\equiv\frac{N_{ac}}{N_{ac}+N_d} \approx 90.8\%.
\end{equation}
Such quantity, $\eta_c$, contains the reliability of the 
experimental results, which can be removed by considering 
only the {\it true unambiguous disagreements} of Table 9 ($N_d^\star=3$).
Hence the reliability of the method becomes:
\begin{equation}
\eta_c^\star\equiv\frac{N_{ac}}{N_{ac}+N_d^\star} \approx 96.7\%.
\end{equation}
Such quantities, $\eta_c$ and $\eta_c^\star$, can be considered as the probabilities 
that the most stable {\it ab initio} compounds reproduce the correct experimental compound. 
The quantity $\eta_c^\star$ should be used if the experimental compound is certain.

If we extend the argument to the agreements with experimental compounds and immiscible systems,
then the reliabilities become:
\begin{eqnarray}
\eta&\equiv&\frac{N_{ac}+N_{ai}}{N_{ac}+N_{ai}+N_d} \approx 92.4\%, \\
\eta^\star&\equiv&\frac{N_{ac}+N_{ai}}{N_{ac}+N_{ai}+N_d^\star} \approx 97.3\%\,
\end{eqnarray}
where the quantity $\eta_c^\star$ should be used if the experimental compound or immiscibility is certain.
Such quantities, $\eta$ and $\eta^\star$, can be considered as the probabilities 
that {\it ab initio} results reproduce the correct experimental compounds or immiscibilities. 

Given the large number of calculations of this project, 
we believe $\eta$, $\eta^\star$, $\eta_c$, and $\eta_c^\star$ to be good estimations of 
state of the art pseudopotential {\it ab initio} accuracies in phase stability prediction.
}

\cleardoublepage
\section{Alloys without ab initio compounds}
\label{section.noncompoundforming}

\tableimmisciblesmall\, gives the alloys for which we find no compounds with negative formation
energy, and the structure with lowest formation energy in the system.
All of these agree with experiments except for the ones described below.

\begin{center}
  {\tablefont
\begin{tabular}{||c|c|r|r||} \hline
\ System \ &\ \ \ \ Structure \ \ \ & E$_f$ (\us) & References   \\ 
   & with lowest E$_f$  & {\ftnsz meV/atom} & \cite{BB,PP,CC} and \\ \hline 
 { Ag-Mo}& AgMo$_3$-BCC                 & 208 & {\ftnsz \cite{AgMo.48Lin,ref.AgMo.Dai04}                                                }\\ 
         & {\ftnsz proto. \#72 in \cite{ref.CurtaroloMIT2003structs}} & &                                                                \\ \hline
 { Ag-Nb}& AgNb$_3$-BCC                 &  62 & {\ftnsz \cite{AgNb.63Kie,AgNb.89Bar}                                                    }\\ 
         & {\ftnsz proto. \#72 in \cite{ref.CurtaroloMIT2003structs}} & &                                                                \\ \hline
 { Ag-Rh}& AgRh$_3$-FCC$_{AB3}^{[111]}$ & 116 & {\ftnsz \cite{ref.CRC,AgRh.59Rud,AgRh.77Bar,AgRh.86Kar}                                 }\\ \hline
 { Ag-Ru}& AgRu$_3$-HEX                 & 199 & {\ftnsz \cite{ref.Hansen,ref.Moffatt,AgRh.86Kar,AgRu.59Rud,ref.AgRu.Li04}                             }\\ 
         & {\ftnsz proto. \#127 in \cite{ref.CurtaroloMIT2003structs}} & &                                                               \\ \hline
 { Ag-Tc}& AgTc-HEX                     & 147 & {\ftnsz \cite{AgTc.68Gul}                                                               }\\ 
         & {\ftnsz proto. \#126 in \cite{ref.CurtaroloMIT2003structs}} & &                                                               \\ \hline
 { Au-Mo}& AuMo$_3$-D0$_{19}$           &  75 & {\ftnsz \cite{ref.Hansen,ref.Molybdenum,Au.86Mas,Au.87MasOkaBre,AuMo.24Dre,AuMo.53Gea,ref.AuMo.Rodriguez95}  }\\ \hline
 { Au-Pt}& Au$_2$Pt-FCC$_{AB2}^{[111]}$ &   9 & {\ftnsz \cite{ref.Hansen,Au.85Oka,Au.87MasOkaBre,AuPt.07Doe,AuPt.30Joh,AuPt.52Dar}      }\\ \hline
 { Au-Rh}& AuRh$_3$-FCC$_{AB3}^{[001]}$ &  91 & {\ftnsz \cite{Au.84Oka,Au.87MasOkaBre,AuRh.59Rud,AuRh.64Rau}                            }\\ \hline
 { Au-Ru}& AuRu$_3$-HEX                 & 165 & {\ftnsz \cite{Au.84Oka,Au.87MasOkaBre,AuRu.59Rud,ref.AuRu.Kuhn95}                                       }\\ 
         & {\ftnsz proto. \#125 in \cite{ref.CurtaroloMIT2003structs}} & &                                                               \\ \hline
 { Au-Tc}& AuTc-HEX                     &  74 & {\ftnsz \cite{Au.84Oka,AuTc.64Ron}                                                      }\\ 
         & {\ftnsz proto. \#126 in \cite{ref.CurtaroloMIT2003structs}} & &                                                               \\ \hline
 { Cd-Mo}& Cd$_5$Mo-HEX                 & 173 & {\ftnsz \cite{BB}                                                                       }\\ 
         & {\ftnsz proto. \#128 in \cite{ref.CurtaroloMIT2003structs}} & &                                                               \\ \hline
 { Cd-Nb}& CdNb-BCC                     &  55 (\us) & {\ftnsz \cite{CdNb.66Hol}                                                         }\\ 
         & {\ftnsz proto. \#71 in \cite{ref.CurtaroloMIT2003structs}} &  {\it 58 (\paw)} & {\ftnsz see note ``Cd-Nb''                   }\\ \hline
 { Cd-Ru}& CdRu-FCC$_{A2B2}^{[001]}$    &  88 & {\ftnsz     see note ``Cd-Ru''                                                          }\\ \hline
 { Cd-Tc}& CdTc-B11                     &  92 & {\ftnsz \cite{CdTc.64Cha}                                                               }\\ \hline
 { Cr-Mg}& CrMg-B11                     & 251 & {\ftnsz \cite{ref.Magnesium,CrMg.38Mon}                                                 }\\ \hline
 { Mo-Tc}& Mo$_5$Tc-A15                 &  53 & {\ftnsz \cite{ref.Shunk,ref.MoTc.Souvatzis04}                                           }\\ 
         &                              &                         & {\ftnsz  see note ``Mo-Tc''                                         }\\ \hline
  { Mo-Y }& Mo$_2$Y-C15                  &  40 & {\ftnsz \cite{MoY.61Gsc,ref.MoY.Kong02}                                                    }\\ \hline
 { Nb-Y }& Nb$_2$Y-C49                  & 132 & {\ftnsz \cite{NbY.Car1,NbY.Car2,NbY.Lun2,NbY.Tay,ref.NbY.Kong02}                        }\\ \hline
 { Nb-Zr}& NbZr$_2$-HEX                 &  22 & {\ftnsz \cite{NbZr.70Rab,NbZr.72Eff,NbZr.72Fle,NbZr.74Fle,NbZr.82Abr,ref.AgPd.Bruno94}                   }\\ 
         & {\ftnsz proto. \#129 in \cite{ref.CurtaroloMIT2003structs}} & & {\ftnsz \cite{ref.NbRu.Sanchez93,ref.NbRu.Becker93,ref.NbRu.Becker91,ref.NbZr.Leuken90,ref.NbZr.Sanchez94,ref.NbZr.Grad96,ref.NbZr.Grad99,ref.NbZr.Kudrnovsky89}} \\ \hline
 { Pd-Rh}& PdRh$_3$-FCC$_{AB3}^{[111]}$ &  35 & {\ftnsz \cite{ref.ZUNGER_PROTOS1,PdRh.59Rau,PdRh.59Rau2,PdRh.88Tur,PdRh.87Shi,ref.PdRh.Asato01,ref.PdRh.Asato01b,ref.PdRh.Wolverton93,ref.PdRh.Marquez03}                             }\\ \hline
 { Pd-Ru}& PdRu-HEX                     &  44 & {\ftnsz \cite{PdRu.59Rud,PdRu.60Dar,PdRu.89Kle,ref.PdRu.Guo02}                                         }\\ 
         & {\ftnsz proto. \#126 in \cite{ref.CurtaroloMIT2003structs}} & &                                                               \\ \hline
 { Ti-Zr}& Ti$_2$Zr-B8$_2$              &  18 & {\ftnsz \cite{ref.Titanium,ref.Zirconium,TiZr.66Far,TiZr.71Cha,TiZr.77Etc,TiZr.82Auf}   }\\ \hline
 { Y-Zr }& YZr$_5$-HEX                  &  35 & {\ftnsz \cite{ref.Zirconium,YZr.59Lun,YZr.61Uy,YZr.62Lun,YZr.72Wan,YZr.74Wan,YZr.76Wan} }\\ 
         & {\ftnsz proto. \#141 in \cite{ref.CurtaroloMIT2003structs}} & &                                                               \\ \hline
\end{tabular}
   }
\end{center}

{{\tableimmisciblescapital.} Systems without intermetallic compounds. 
The second and third columns give, for each system, the structure with the lowest formation 
energy E$_f$ (US-LDA calculations). 
{The table contains \NNimmiscibilitytable\,\,entries. \NNimimscibilitycorrecttable\,\,of these are in agreement with experiments.
References include both experimental and {\it ab initio} studies.}}
\vspace{3mm}

{\bf Cd-Nb.} One compound, Cd$_3$Nb-L1$_2$, has been reported for the system 
Cd-Nb \cite{BB,CdNb.66Hol}. However, we did not find any stable phase
(Cd$_3$Nb-L1$_2$ has formation energy of $\sim$70meV/atom).
{The disagreement of compound Cd$_3$Nb-L1$_2$ is further discussed in the Section (\ref{section.discussion}).} \\

{\bf Cd-Ru.} The authors are not aware of any experimental result for the Cd-Ru system. \\

{\bf Mo-Tc.} Two compounds have been suggested based on thermodynamic models: 
$\beta$Mo$_2$Tc$_3$-A15 and $\sigma$Mo$_3$Tc$_7$-D8$_b$ \cite{BB,ref.Shunk}. 
However, we did not find any stable phase for the system, nor can we check
the suggested compounds, since our library does not have the $\sigma$ phase 
or the off-stoichiometry A15.

\newpage
\section{Alloys with ab initio compounds}
\label{section.compoundforming}


{\bf Ag-Au (Silver - Gold).}
The system Ag-Au has not been studied in great detail at low temperatures,
and no intermetallic compounds have been reported 
\cite{BB,PP,AgAu.54Wag,AgAu.57Whi,AgAu.59Whi,AgAu.61Coo,Au.87MasOkaBre}. 
The solid is reported to be short-range 
ordered fcc, though it is suggested that long-order might exist at low temperature.
At low temperature we find several stable compounds: 
Ag$_4$Au, Ag$_3$Au, Ag$_2$Au, AgAu-L1$_0$, AgAu$_2$ and AgAu$_3$.
In our electronic structure approach, the ground states are degenerate
for Ag$_3$Au, Ag$_2$Au, AgAu$_2$ and AgAu$_3$.
Our best candidates are  
L1$_2$, D0$_{23}$, Al$_3$Pu, NbPd$_3$, D0$_{22}$ and D0$_{24}$ for Ag$_3$Au, 
C37 and MoPt$_2$ for Ag$_2$Au and AgAu$_2$,
and L1$_2$, D0$_{22}$ and D0$_{23}$ for AgAu$_3$, 
as shown in figure (\ref{label.fig.AgAu}).
The considerable degeneracy of the ground states 
indicates that only very small effective interactions exist between Ag and Ag, 
consistent with the experimentally observed complete solid solubility.
For composition Ag$_4$Au, the structure D1$_a$ is degenerate with the tie-line of the two-phase region Ag\twophase Ag$_3$Au.
We conclude that further experimental and theoretical investigations
are necessary to determine the behavior of AgAu.
{\it To address the degenerate structures,
we further investigate Ag$_3$Au, Ag$_2$Au, AgAu$_2$ and AgAu$_3$ with PAW-GGA potentials,
\PawSection.
With PAW, for composition Ag$_3$Au,
L1$_2$ is the most stable compound and D0$_{22}$, D0$_{23}$, Al$_3$Pu, NbPd$_3$, D0$_{24}$ 
are higher by 0.9, 1.6, 2.2, 2.4, and 3.0meV/atom, respectively.
For both composition Ag$_2$Au and AgAu$_2$, 
C37 is the most stable compound and MoPt$_2$ is higher by 1.9meV/atom and 4.3meV/atom, respectively.
For composition AgAu$_3$, L1$_2$ is the most stable compound 
and D0$_{23}$ and D0$_{22}$ are higher by 3.4meV/atom and 3.9meV/atom, respectively.}

\ReferenceAgAu

{\tablefont
  \begin{center}
    \begin{tabular}{||c||} \hline
      {\bf Ag-Au system} \\ \hline
      Low Temperature Phases comparison chart \\ \hline
      \begin{tabular}{c|c|c}
        Composition  & Experimental  & \tablelineone                      \\
        \% Au        & (\massalski)  & \tablelinetwo                     \\ \hline
        20        &short-range order & Ag$_4$Au-D1$_a$/tie-line          \\
        \      & $\gtrsim$ 950\DEG   &                                   \\  \hline
        25        &short-range order & Ag$_3$Au                          \\
        \      & $\gtrsim$ 950\DEG   & L1$_2$/D0$_{23}$/Al$_3$Pu/ (\us)   \\ 
        \            &               & /NbPd$_3$/D0$_{22}$/D0$_{24}$ (\us)\\ 
        \            &               & {\it L1$_2$ stable (\paw)}         \\
        \            &               & {\it D0$_{22}$$\sim$0.9meV/at. }  \\
        \            &               & {\it D0$_{23}$$\sim$1.6meV/at. }  \\
        \            &               & {\it Al$_3$Pu$\sim$2.2meV/at.  }  \\
        \            &               & {\it NbPd$_3$$\sim$2.4meV/at.  }  \\
        \            &               & {\it D0$_{24}$$\sim$3.0meV/at. }  \\
        \            &               & {\it above L1$_2$ (\paw).   }      \\ \hline
        33.3      &short-range order & Ag$_2$Au-C37/MoPt$_2$ (\us)        \\
        \      & $\gtrsim$ 950\DEG   & {\it C37 stable (\paw)  }          \\
        \            &               & {\it MoPt$_2$$\sim$1.9meV/at. }   \\
        \            &               & {\it above C37 (\paw).   }         \\ \hline
        50        &short-range order & AgAu-L1$_0$                       \\
        \      & $\gtrsim$ 950\DEG   &                                   \\ \hline
        66.6      &short-range order & AgAu$_2$-C37/MoPt$_2$ (\us)        \\
        \      & $\gtrsim$ 950\DEG   & {\it C37 stable (\paw)  }          \\
        \            &               & {\it MoPt$_2$$\sim$4.3meV/at. }   \\
        \            &               & {\it above C37 (\paw).   }         \\ \hline
        75.0      &short-range order & AgAu$_3$                          \\
        \      & $\gtrsim$ 950\DEG   & L1$_2$/D0$_{22}$/D0$_{23}$ (\us)   \\ 
        \            &               & {\it L1$_2$ stable (\paw)   }      \\
        \            &               & {\it D0$_{23}$$\sim$3.4meV/at.}   \\
        \            &               & {\it D0$_{22}$$\sim$3.9meV/at.}   \\
        \            &               & {\it above L1$_2$ (\paw).   }      \\ 
      \end{tabular} \\ \hline
    \end{tabular}
  \end{center}
}

\vspace{-3mm}
\begin{center}
  \vspace{-3mm}
  \begin{figure}
    \epsfig{file=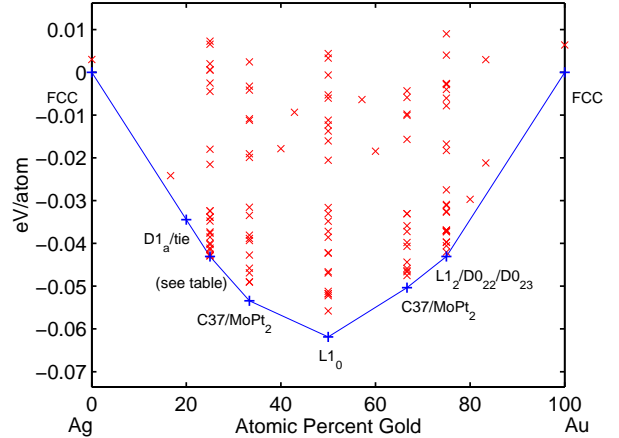,width=\picdim,clip=}
    \caption{AgAu (Silver - Gold) ground state convex hull.}
    \label{label.fig.AgAu}
  \end{figure}
  \vspace{-1mm}
\end{center}

\newpage

{\bf Ag-Cd (Silver - Cadmium).}
At composition AgCd$_3$ we find D0$_{19}$ stable,
while only disordered hcp solid solution has been reported at this composition.
Hence, D0$_{19}$, an ordered hcp superstructure, probably is 
stable at low temperature in this system.
At composition AgCd$_2$  we find a previously unknown compound AgCd$_2$-ZrSi$_2$.
The low temperature $\beta'$, near AgCd stoichiometry, is reported to be ordered bcc in Massalski \cite{BB}.
In particular, the Pauling File reports AgCd-B2 \cite{PP,AgCd.74Ton,AgCd.78Zar,AgCd.78Mat,AgCd.81Mat}.
We find B2, B19 and B27 to have degenerate energies (B19 is quite common for Cd alloys). 
The experimental phase diagram only displays solid solution on the 
Ag-rich side \cite{BB,PP}. 
We find C37 at Ag$_2$Cd and a stable phase with stoichiometry Ag$_3$Cd.
The ground state of Ag$_3$Cd is degenerate:  
three distinct structure (D0$_{19}$, D0$_{22}$, D0$_{24}$)
have similar energy.
{\it To address the degenerate structures, 
we further investigate AgCd and Ag$_3$Cd with PAW-GGA potentials, 
\PawSection. 
With PAW, for compound AgCd,
B19 is the most stable structure, 
and B27 and B2 are higher by 2.8meV/atom and 3.4meV/atom, respectively.
For compound Ag$_3$Cd,
D0$_{22}$ and D0$_{24}$ are still 
degenerate ground states and D0$_{19}$ is 
higher by only 2.2meV/atom. 
It is possible that these compounds have not yet been observed, 
or that they have low ordering temperatures.
}

\vspace{-1mm}
{\tablefont
  \begin{center}
    \begin{tabular}{||c||} \hline
      {\bf Ag-Cd system} \\ \hline
      Low Temperature Phases comparison chart \\ \hline
      \begin{tabular}{c|c|c}
	Composition  & Experimental  & \tablelineone                      \\ 
	\% Cd        & (\massalski)  & \tablelinetwo                     \\ \hline
	25.0         & fcc solid solution    & Ag$_3$Cd                  \\
	\            &               & D0$_{19}$/D0$_{22}$/D0$_{24}$ (\us)\\
	\            &               & {\it D0$_{22}$/D0$_{24}$ (\paw) } \\
	\            &               & {\it D0$_{19}$$\sim$2.2meV/at.  } \\ 
        \            &               & {\it above D0$_{22}$/D0$_{24}$ (\paw) } \\ \hline
	33.3         & fcc solid solution   & Ag$_2$Cd-C37              \\ \hline
	48.5 to 50   & $\beta'$-bcc  & AgCd-B2/B19/B27 (\us).           \\ 
	\            &  B2 \cite{PP} & {\it B19 stable (\paw) }         \\
	\            &               & {\it B27$\sim$2.8meV/at. }       \\ 
	\            &               & {\it B2 $\sim$3.4meV/at. }       \\ 
	\            &               & {\it above B19 (\paw). }         \\ \hline
	66.6         & none          & AgCd$_2$-ZrSi$_2$.               \\ \hline
	64.5 to 81   & hcp solid solution & AgCd$_3$-D0$_{19}$          \\
      \end{tabular} \\ \hline
    \end{tabular}
  \end{center}
}

\vspace{-5mm}
\begin{center}
  \vspace{-3mm}
  \begin{figure}
    \epsfig{file=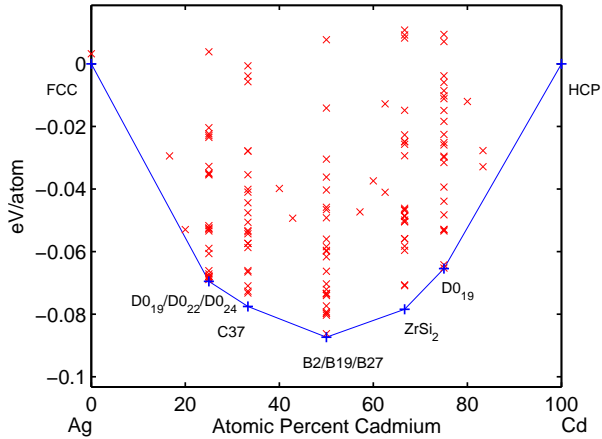,width=\picdim,clip=}
    \caption{AgCd (Silver - Cadmium) ground state convex hull.}
    \label{label.fig.AgCd}
  \end{figure}
  \vspace{-1mm}
\end{center}

\newpage

{\bf Ag-Mg (Silver - Magnesium).}
The phase diagram of the system Ag-Mg is known with reasonable accuracy 
\cite{BB,PP,AgMg.37Hum,AgMg.37Pay,AgMg.85Pro,AgMg.88Kol}.
Our {\it ab initio} method confirms the stability of AgMg-B2. 
In the Ag-rich side of the phase diagram, at stoichiometry Ag$_3$Mg,
we find D0$_{23}$ and D0$_{24}$ to be degenerate.
While Massalski \cite{BB} indicates Ag$_3$Mg to be L1$_2$, it is known from
detailed High Resolution Electronic Microscopy (HREM) 
that Ag$_3$Mg forms Long Period Superstructure (LPS) modulations of
L1$_2$ \cite{AgMg.55Sch,AgMg.58Fuj} 
(LPS D0$_{23}$ is present in our library).
For a detailed discussion of this system see Kulik {\it et-al.} \cite{AgMg.87Kul}.
L1$_2$ is 15.2meV/atom above D0$_{23}$/D0$_{24}$.
On the Mg-rich side, our library does not have prototypes at 80\% Mg concentration, 
so we are not able to find any 
hexagonal AgMg$_4$ (with unknown prototype \cite{AgMg.85Pro,AgMg.88Kol}).
However we find a stable compound AgMg$_3$. 
Two phases, AgMg$_3$-D0$_{a}$ and AgMg$_3$-D0$_{19}$, have degenerate energies.
This is important, since experiments suggest that
AgMg$_4$ was probably identified as hexagonal AgMg$_3$ in early studies.
Unfortunately, we cannot determine the correct {\it ab initio} solution
since we lack the hexagonal AgMg$_4$ prototypes in our library.
{\it To address the degenerate structures,
we further investigate Ag$_3$Mg and AgMg$_3$ with PAW-GGA potentials, 
\PawSection. 
For compound Ag$_3$Mg, with PAW,
D0$_{23}$ is the most stable compound (in agreement with \cite{AgMg.55Sch,AgMg.58Fuj,AgMg.87Kul}),
D0$_{24}$ and L1$_2$ are 1.1meV/atom and 11.3meV/atom
higher than D0$_{23}$, respectively.
In addition, for compound AgMg$_3$, 
D0$_{19}$ is the most stable structure and D0$_{a}$ is $\sim$6.8meV/atom higher than D0$_{19}$.}

\ReferenceAgMg

{\tablefont
  \begin{center}
    \begin{tabular}{||c||} \hline
      {\bf Ag-Mg system} \\ \hline
      Low Temperature Phases comparison chart \\ \hline
      \begin{tabular}{c|c|c}
	Composition  & Experimental  & \tablelineone                      \\ 
	\% Mg        & (\massalski)  & \tablelinetwo                     \\ \hline
	25.0         & L1$_2$        & Ag$_3$Mg-D0$_{23}$/D0$_{24}$ (\us)\\ 
        \            & D0$_{23}$ \cite{AgMg.55Sch,AgMg.58Fuj,AgMg.87Kul}   &  L1$_2$$\sim$15meV/atom above.     \\  
	\	     &               &{\it D0$_{23}$ stable (\paw)}      \\
	\ 	     &               & {\it D0$_{24}$ $\sim$1.1meV/at.}  \\
	\            &               & {\it L1$_2$ $\sim$11.3meV/at. }   \\
  	\            &               & {\it above D0$_{23}$ (\paw) }     \\ \hline
	35.5 to 65.4 & B2            & AgMg-B2                           \\ \hline
        75.8 to 78.2 & cF* (unknown) & AgMg$_3$                          \\
	\            &               & D0$_{19}$ (hP8)/D0$_{a}$ (oP8) (us)\\
	\	     &               & {\it D0$_{19}$ stable (\paw)}     \\
	\            &               & {\it D0$_{a}$ $\sim$6.8meV/at.}   \\
  	\            &               & {\it above D0$_{19}$ (\paw) }     \\ 	 \hline
        80.0         & hP* (unknown) & unavailable (see text)            \\
      \end{tabular} \\ \hline
    \end{tabular}
  \end{center}
}

\begin{center}
  \vspace{-3mm}
  \begin{figure}
    \epsfig{file=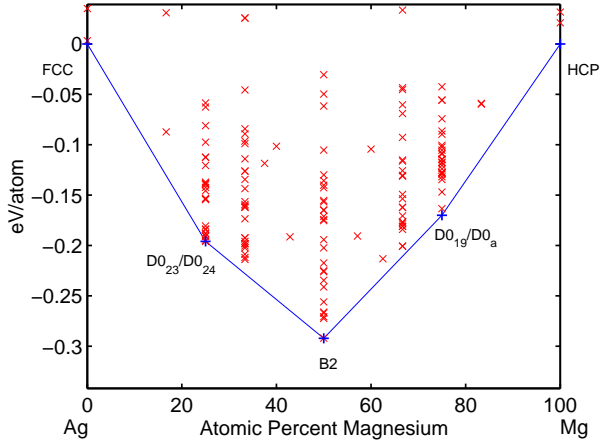,width=\picdim,clip=}
    \vspace{1mm}
    \caption{AgMg (Silver - Magnesium) ground state convex hull.}
    \label{label.fig.AgMg}
  \end{figure}
  \vspace{-1mm}
\end{center}

\newpage

{\bf Ag-Na (Silver - Sodium).}
The phase diagram of the system Ag-Na is known with reasonable accuracy 
and has only one intermetallic compound \cite{BB,PP,AgNa.68Lam,AgNa.69Wee,AgNa.83Cha}.
We confirm the stability of Ag$_2$Na-C15 and find almost all the other compounds to have
positive formation energies energies.
Both with US-LDA and PAW-GGA, Na-hcp is very slighly
favored over Na-bcc and Na-fcc, in agreement with other first principle calculations
\cite{ref.Na2001Achcroft,ref.Na1983McM,ref.Na1998Jaf,ref.Na1986Dac,ref.Na2000Kat}. 
At room temperature Sodium has the bcc structure, and undergoes
a martensitic transformation below 35K, to a closed packed structure 
\cite{ref.Na1948Bar,ref.Na1956Bar,ref.Na1984Over,ref.Na1986Ern,ref.Na1990Smi,ref.Na1990Sch,ref.Na1991Sch,ref.Na1999Blas}.
Therefore our results, Na-hcp stable and the very small energy differences with Na-bcc and Na-fcc,
are consistent with the behaviour at low temperature.

{\tablefont
  \begin{center}
    \begin{tabular}{||c||} \hline
      {\bf Ag-Na system} \\ \hline
      Low Temperature Phases comparison chart \\ \hline
      \begin{tabular}{c|c|c}
	Composition  & Experimental  & \tablelineone                      \\ 
	\% Na        & (\massalski)  & \tablelinetwo                     \\ \hline
	33.0         & C15           & Ag$_2$Na-C15                      \\
      \end{tabular} \\ \hline
    \end{tabular}
  \end{center}
}

\begin{center}
  \vspace{-3mm}
  \begin{figure}
    \epsfig{file=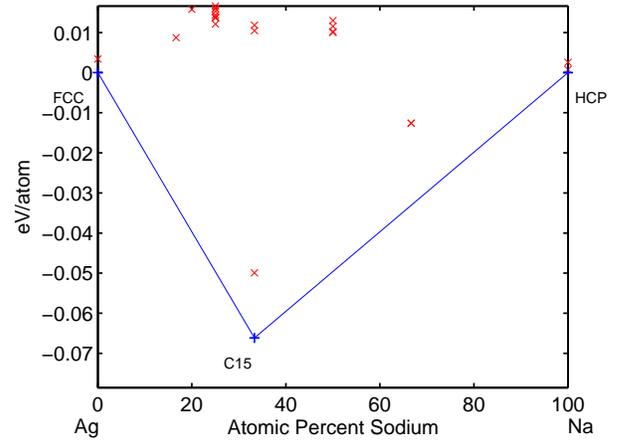,width=\picdim,clip=}
    \vspace{1mm}
    \caption{AgNa (Silver - Sodium) ground state convex hull.}
    \label{label.fig.AgNa}
  \end{figure}
  \vspace{-1mm}
\end{center}

\newpage

{\bf Ag-Pd (Silver - Palladium).}
The system Ag-Pd has not been studied in great detail and no 
intermetallic compounds have been reported \cite{BB,PP,ref.Hansen,AgPd.88Kar}.
The solid is reported to be disordered fcc.
At low temperature we find three stable compounds: AgPd-L1$_1$, Ag$_2$Pd and Ag$_3$Pd. 
In our computations, for Ag$_2$Pd and Ag$_3$Pd, the ground states are degenerate:
our best candidates are C49 or C37 for Ag$_2$Pd, and L1$_2$ or D0$_{22}$ for 
Ag$_3$Pd, as shown in figure (\ref{label.fig.AgPd}). 
{\it To address the degenerate structures,
we further investigate Ag$_2$Pd and Ag$_3$Pd with PAW-GGA potentials, 
\PawSection. 
With PAW, for composition Ag$_2$Pd,
C37 is the most stable compound and 
C49 is higher by 4meV/atom.
For composition Ag$_3$Pd, L1$_2$ and D0$_{22}$ remain degenerate. 
In fact, D0$_{22}$ has an energy 0.4meV/atom greater than L1$_2$, 
too small compared to the numerical accuracy of the {\it ab initio} calculation.}

\ReferenceAgPd

{\tablefont
  \begin{center}
    \begin{tabular}{||c||} \hline
      {\bf Ag-Pd system} \\ \hline
      Low Temperature Phases comparison chart \\ \hline
      \begin{tabular}{c|c|c}
	Composition  & Experimental  & \tablelineone                      \\ 
	\% Ag        & (\massalski)  & \tablelinetwo                     \\ \hline
	50           &solid solution & AgPd-L1$_1$                       \\        
	             & $>$ 900\DEG   &                                   \\ \hline
	66.6         &solid solution & Ag$_2$Pd                          \\
	\            & $>$ 900\DEG   & C37/C49 (\us)                      \\
	\            &               & {\it C37 stable (\paw)   }         \\
	\            &               & {\it C49$\sim$4meV/at.  }         \\
        \            &               & {\it above C37 (\paw)    }         \\ \hline
	75.0         &solid solution & Ag$_3$Pd                          \\
	\            & $>$ 900\DEG   & L1$_2$/D0$_{22}$ (us \& \paw)      \\ 
      \end{tabular} \\ \hline
    \end{tabular}
  \end{center}
}

\begin{center}
  \vspace{-3mm}
  \begin{figure}
    \epsfig{file=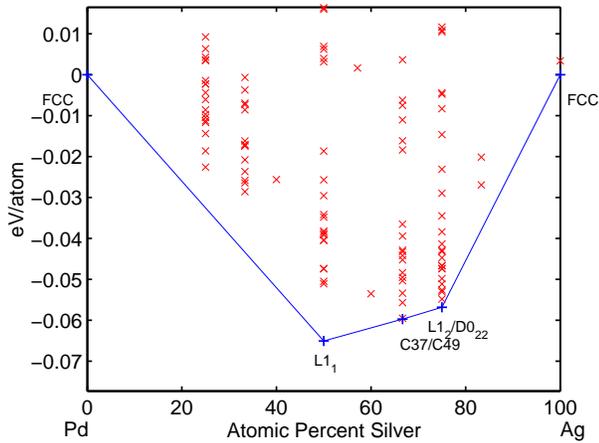,width=\picdim,clip=}
    \vspace{1mm}
    \caption{AgPd (Silver - Palladium) ground state convex hull.}
    \label{label.fig.AgPd}
  \end{figure}
  \vspace{-1mm}
\end{center}

\newpage

{\bf Ag-Ti (Silver - Titanium).}
The stability of AgTi$_2$-C11$_b$ and AgTi-B11 is confirmed
\cite{BB,PP,AgTi.69Ere,AgTi.78PLI}, and no other stable phases 
are found computationally. 
Hence, we conclude that the low temperature part of the phase 
diagram of AgTi is probably accurate.

{\tablefont
  \begin{center}
    \begin{tabular}{||c||} \hline
      {\bf Ag-Ti system} \\ \hline
      Low Temperature Phases comparison chart \\ \hline
      \begin{tabular}{c|c|c}
	Composition  & Experimental  & \tablelineone                      \\ 
	\% Ag        & (\massalski)  & \tablelinetwo                     \\ \hline
	33.3         & C11$_b$       & AgTi$_2$-C11$_b$                  \\ \hline
	50           & B11           & AgTi-B11 ($\gamma$CuTi)           \\
      \end{tabular} \\ \hline
    \end{tabular}
  \end{center}
}

\begin{center}
  \vspace{-3mm}
  \begin{figure}
    \epsfig{file=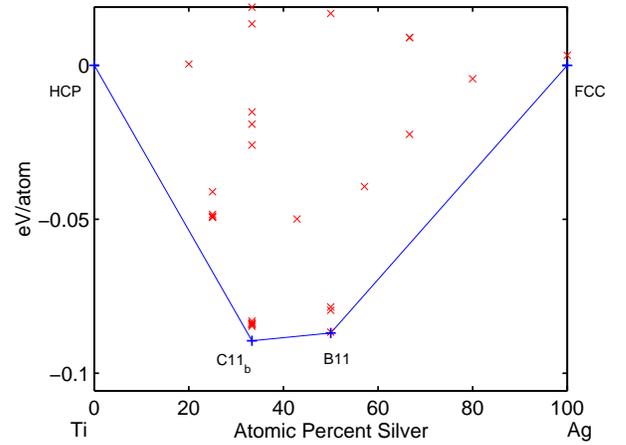,width=\picdim,clip=}
    \vspace{1mm}
    \caption{AgTi (Silver - Titanium) ground state convex hull.}
    \label{label.fig.AgTi}
  \end{figure}
  \vspace{-1mm}
\end{center}

\newpage

{\bf Ag-Y (Silver - Yttrium).}
The stability of AgY-B2 and Ag$_2$Y-C11$_b$ is confirmed \cite{BB,PP,AgY.64GEB}. 
In the Ag-rich region, we do not determine the stability of Ag$_{51}$Y$_{14}$, 
since we do not have the prototype in our library \cite{AgY.70GSC,AgY.70MCM}.
Instead, we find Ag$_3$Y with D0$_a$ structure stable with an 
energy $\sim$22meV/atom below the energy of a two-phase 
region Ag\twophase Ag$_2$Y. 
In the real system, the presence of Ag$_{51}$Y$_{14}$ probably makes structure D0$_a$ metastable.
In the Y-rich region we find AgY$_2$-C37 degenerate 
with the two-phase region Y\twophase AgY (within $\sim$1.7meV/atom)
as shown in figure (\ref{label.fig.AgY}).

{\tablefont
  \begin{center}
    \begin{tabular}{||c||} \hline
      {\bf Ag-Y system} \\ \hline
      Low Temperature Phases comparison chart \\ \hline
      \begin{tabular}{c|c|c}
	Composition  & Experimental  & \tablelineone                      \\ 
	\% Ag        & (\massalski)  & \tablelinetwo                     \\ \hline
	33.3         & two-phase region & AgY$_2$-C37/tie-line           \\
        \            & above 200$^\circ$C &                              \\ \hline
	50           & B2            & AgY-B2                            \\ \hline
	66.6         & C11$_b$       & Ag$_2$Y-C11$_b$                   \\ \hline
        75           & two-phase region & Ag$_3$Y-D0$_a$                 \\
        \            & above 200$^\circ$C & (uncertain)                  \\ \hline
        78.5         & Ag$_{51}$Gd$_{14}$ & unavailable                  \\ 
      \end{tabular} \\ \hline
    \end{tabular}
  \end{center}
}

\begin{center}
  \vspace{-3mm}
  \begin{figure}
    \epsfig{file=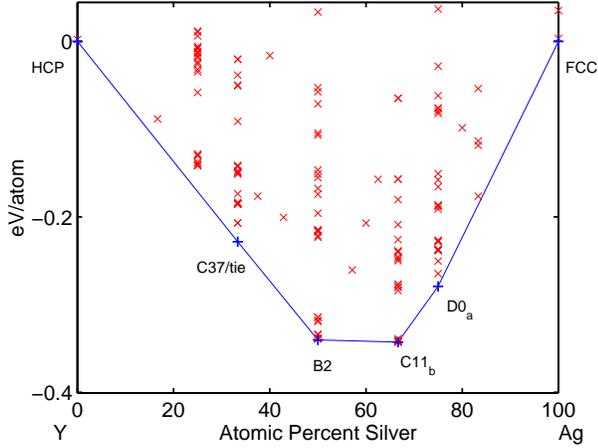,width=\picdim,clip=}
    \vspace{1mm}
    \caption{AgY (Silver - Yttrium) ground state convex hull.}
    \label{label.fig.AgY}
  \end{figure}
  \vspace{-1mm}
\end{center}

\newpage

{\bf Ag-Zr (Silver - Zirconium).}
The phase diagram of Ag-Zr is known partially, and it has been 
estimated from thermodynamic
properties \cite{BB,PP,AgZr.77Bar,AgZr.87Kar}. The stability of
the two known phases AgZr-B11 ($\gamma$-CuTi prototype) 
and AgZr$_2$-C11$_b$ is confirmed by our calculation.
In addition to the known intermetallic compounds, 
we find a stable phase Ag$_2$Zr, which is not present in Massalski \cite{BB}:
the C6 and C32 structures are degenerate. 
To conclude, we find AgZr$_3$-FCC$_{AB3}^{[001]}$
degenerate with the two-phase region 
Zr\twophase AgZr$_2$ (within $\sim$1.3meV/atom).
{\it To address the degenerate structures,
we further investigate Ag$_2$Zr with PAW-GGA potentials, 
\PawSection. 
With PAW,
C32 is the most stable compound and 
C6 is 2meV/atom above C32.} 

{\tablefont
  \begin{center}
    \begin{tabular}{||c||} \hline
      {\bf Ag-Zr system} \\ \hline
      Low Temperature Phases comparison chart \\ \hline
      \begin{tabular}{c|c|c}
	Composition  & Experimental  & \tablelineone                      \\ 
	\% Ag        & (\massalski)  & \tablelinetwo                     \\ \hline
	25.0      & two-phase region & AgZr$_3$                          \\ 
        \            & estimated     & FCC$_{AB3}^{[001]}$/tie-line      \\
	\            & above 700\DEG &                                   \\ \hline
	33.3         & C11$_b$       & AgZr$_2$-C11$_b$                  \\ \hline
	50           & B11           & AgZr-B11 ($\gamma$CuTi)           \\ \hline
	66.6         & two-phase region & Ag$_2$Zr                       \\
        \            & estimated     & C6/C32 (\us)                       \\
	\            & above 700\DEG & {\it C32 stable (\paw)    }        \\
	\            &               & {\it C6$\sim$2meV/at.    }        \\
	\            &               & {\it above C32 (\paw)     }        \\
     \end{tabular} \\ \hline 
    \end{tabular}
  \end{center}
}

\begin{center}
  \vspace{-3mm}
  \begin{figure}
    \epsfig{file=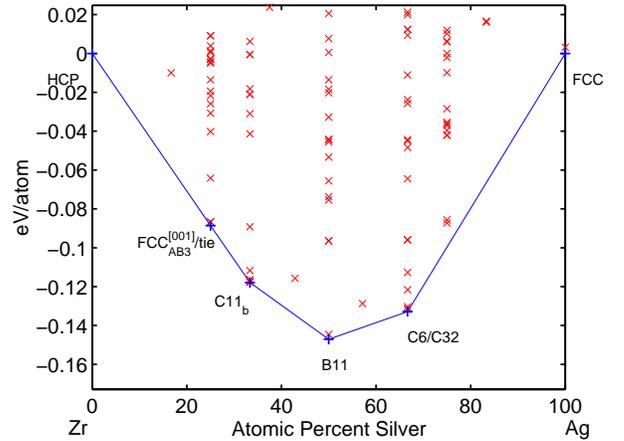,width=\picdim,clip=}
    \vspace{1mm}
    \caption{AgZr (Silver - Zirconium) ground state convex hull.}
    \label{label.fig.AgZr}
  \end{figure}
  \vspace{-1mm}
\end{center}
\newpage

{\bf Al-Sc (Aluminum - Scandium).}
The experimental phase diagram has compounds at compositions:
Al$_3$Sc, Al$_2$Sc, AlSc, and AlSc$_2$ \cite{BB,PP,AlSc.65Nau,AlSc.73Dri,AlSc.79Fuj}.
The {\it ab initio} technique confirms the stability of Al$_3$Sc-L1$_2$, Al$_2$Sc-C15,
AlSc-B2, and AlSc$_2$-B8$_2$. In the Sc-rich region of the phase diagram, 
we find a new hexagonal stable phase AlSc$_3$-D0$_{19}$, 
which is not present in Massalski \cite{BB}. 
Only very limited experimental data is available for this side of the phase diagram.

\ReferenceAgRu

{\tablefont
  \begin{center}
    \begin{tabular}{||c||} \hline
      {\bf Al-Sc system} \\ \hline
      Low Temperature Phases comparison chart \\ \hline
      \begin{tabular}{c|c|c}
	Composition  & Experimental  & \tablelineone                      \\ 
	\% Al        & (\massalski)  & \tablelinetwo                     \\ \hline
	25.0       & two-phase region& AlSc$_3$-D0$_{19}$                \\
        \            & above 0$^\circ$C   &                              \\ \hline
	33.3         & B8$_2$        & AlSc$_2$-B8$_2$                   \\ \hline
	50           & B2            & AlSc-B2                           \\ \hline
	66.6         & C15           & Al$_2$Sc-C15                      \\ \hline
        75           & L1$_2$        & Al$_3$Sc-L1$_2$                   \\ 
      \end{tabular} \\ \hline
    \end{tabular}
  \end{center}
}

\begin{center}
  \vspace{-3mm}
  \begin{figure}
    \epsfig{file=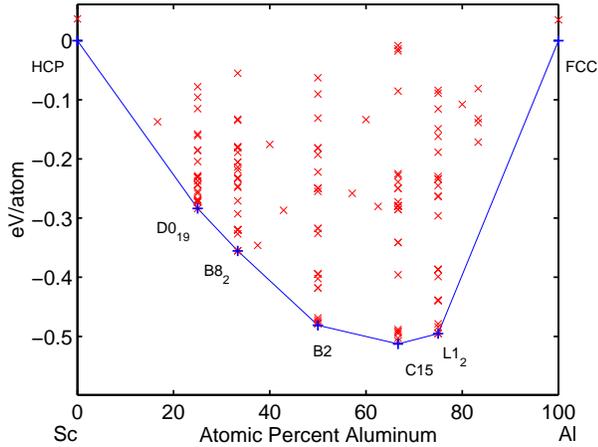,width=\picdim,clip=}
    \vspace{1mm}
    \caption{AlSc (Aluminum - Scandium) ground state convex hull.}
    \label{label.fig.AlSc}
  \end{figure}
  \vspace{-1mm}
\end{center}

\newpage

{\bf Au-Cd (Gold - Cadmium).}
Several compounds have been reported for the system Au-Cd,
and the phase diagram is known with reasonable accuracy
\cite{BB,PP,ref.Hansen,Au.85Oka,Au.87MasOkaBre,AuCd.61Hir,AuCd.64Fil,AuCd.69Hir,AuCd.73Kom}.
We confirm $\beta'$-AuCd-B19 and the 
high-temperature phase $\beta$-AuCd-B2 (with energy $\sim$6meV/atom higher than B19).
At 25\% Cd composition, we do not find the compound 
$\alpha_1$Au$_3$Cd-Ag$_3$Mg because we do not have the
proper prototype in our library. 
Instead, we find structures D0$_{24}$, D0$_{19}$, and Al$_3$Pu (Co$_3$V) 
with degenerate energies. 
Such structures have hexagonal lattices and they are 
candidates for the phase field $\alpha_2$ which has 
been reported to be a long-period superstructure with hP lattice \cite{BB}.
At 33\% Cd composition, we find three phases 
Au$_2$Cd-C37, Au$_2$Cd-MoPt$_2$, and Au$_2$Cd-C49
with energies 4meV/atom, 14meV/atom and 16meV/atom above 
the tie-line Au$_3$Cd \twophase AuCd, respectively.
At 50\% Cd composition, the prototype of the phase 
$\beta''$-AuCd is not known, but it is reported to
have a hexagonal structure. 
For $\beta''$, we suggest two candidates: AuCd-L1$_0$ and AuCd-\CH40 (CH ``40'' in reference \cite{PdPt.91Zun}),
which have energies higher by 5.4meV/atom and 5.5meV/atom with respect to B19. 
None of these structures has hexagonal lattice type, and
all our hexagonal prototypes have much higher energies.

{\tablefont
  \begin{center}
    \begin{tabular}{||c||} \hline
      {\bf Au-Cd system} \\ \hline
      Low Temperature Phases comparison chart \\ \hline
      \begin{tabular}{c|c|c}
	Composition  & Experimental  & \tablelineone                      \\ 
	\% Cd        & (\massalski)  & \tablelinetwo                     \\ \hline
	$\sim$25 &$\alpha_1$-Ag$_3$Mg & unavailable                      \\ \hline
        $\sim$25.5 to 35.5 & $\alpha_2$-hP? (unknown)& Au$_3$Cd          \\
        \            &               & D0$_{24}$/D0$_{19}$/Al$_3$Pu      \\
        \            &               & hP candidates for $\alpha_2$      \\ \hline
        33.3       &two-phase region& Au$_3$Cd \twophase AuCd            \\
        \            &               & Au$_2$Cd-C37, MoPt$_2$, C49       \\
        \            &               & above tie-line                    \\ \hline
	$\sim$47.5 & $\beta'$-B19    & AuCd-B19                          \\
	$\sim$50 & $\beta''$ unknown & L1$_0$/\CH40 $\sim$5meV/at.       \\
	43 to 57 & $\beta$-B2 (high T)& B2 $\sim$6meV/at.                \\
	\	     &	 	     & above B19                         \\ \hline
        61.6 to 62.1 & Au$_3$Cd$_5$-D8$_m$ & unavailable                 \\ \hline
	72 to 76 &$\epsilon'$-unknown& AuCd$_3$-L6$_0$                   \\
	\    & tentative phase field &                                   \\ \hline
 	$\sim$82 &$\eta'$-unknown    & nothing stable                    \\
	\    & tentative phase field & at 83.3\%                         \\ 
      \end{tabular} \\ \hline
    \end{tabular}
  \end{center}
}

At concentration $\sim$75\% Cd, a stable phase $\epsilon'$, with unknown structure, is 
believed to exists \cite{BB}. 
We confirm the existence of a stable compound and our best prediction is AuCd$_3$-L6$_0$.
At concentration $\sim$82\% Cd another stable phase $\eta'$ 
is believed to exist. We are not able to confirm $\eta'$. 
In fact at 83.3\% Cd, the phase with lowest energy is $\sim$60meV/atom 
higher than the two tie-line AuCd$_3$\twophase Cd.
{\it To address the degenerate structures,
we further investigate Au$_3$Cd with PAW-GGA potentials, 
\PawSection. 
Also with PAW,
D0$_{24}$, Al$_3$Pu and D0$_{19}$ remain degenerate.}

\vspace{-3mm}
\begin{center}
  \vspace{-3mm}
  \begin{figure}
    \epsfig{file=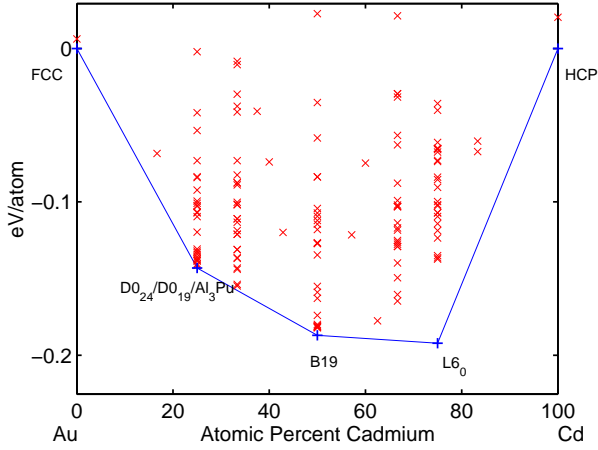,width=\picdim,clip=}
    \caption{AuCd (Gold - Cadmium) ground state convex hull.}
    \label{label.fig.AuCd}
  \end{figure}
  \vspace{-1mm}
\end{center}

\newpage

{\bf Au-Nb (Gold - Niobium).}
The phase diagram of the system Au-Nb is based on three known compounds 
(Au$_2$Nb, Au$_2$Nb$_3$, AuNb$_3$) 
\cite{BB,PP,ref.Niobium,Au.85Oka,Au.87MasOkaBre,AuNb.56Woo,AuNb.61Dwi,AuNb.69Flu,AuNb.70Ber,AuNb.70Ehr,AuNb.73Ros,AuNb.73Kha,AuNb.77Kur,AuNb.79Flu,AuNb.81Wir,AuNb.82Sav,AuNb.83Wir,AuNb.86Ram,AuNb.91Di,AuNb.81Bai}.
We do not confirm the stability of phase Au$_2$Nb-C32 \cite{BB,PP,ref.Niobium}. 
Instead of C32, we find a stable compound Au$_2$Nb-C6 
and C32 with energy higher by $\sim$12meV/atom with respect to C6.
We can not conclude anything about stability of Au$_2$Nb$_3$ since 
our library does not contain any A$_2$B$_3$ prototype.
The Nb-rich side of the experimental phase diagram is drawn with the assumption 
that AuNb$_3$-A15 is stable at low temperature ($\sim$500\DEG) \cite{BB}.
Our calculations suggest that AuNb$_3$-A15 might not be stable at low-temperature 
as it is $\sim$7meV/atom above the two phase field AuNb$_2$\twophase Nb.
In addition, we find a stable phase AuNb$_2$ at 66.6\% Nb composition, 
with bcc parent lattice, and space group Fmmm \#69.
The compound, labeled as \str64, has AB2 stacking along [011] direction.
If the prototype \str64 were not included in the calculation,
the phase AuNb$_3$-A15 would be stable, as shown in figure (\ref{label.fig.AuNb}). 
To conclude, we think it might be worthwhile to reconsider 
the thermodynamic modeling of the Nb-rich side of the phase diagram 
with the inclusion of AuNb$_2$-\str64, 
to obtain the proper temperature and concentration ranges of phase A15.
{\it To address the disagreements with the experimental results, 
we further investigate Au$_2$Nb and AuNb$_3$ with PAW-GGA potentials, 
\PawSection. 
For the compound Au$_2$Nb
C32 is the most stable compound and 
C6 is higher by 3.2meV/atom.
For composition AuNb$_3$,
A15 is still unstable, being $\sim$6.5meV/atom above the tie-line AuNb$_2$\twophase Nb.   
}
{The disagreement at composition AuNb$_3$ is further discussed in Section (\ref{section.discussion}).}

{\tablefont
  \begin{center}
    \begin{tabular}{||c||} \hline
      {\bf Au-Nb system} \\ \hline
      Low Temperature Phases comparison chart \\ \hline
      \begin{tabular}{c|c|c}
	Composition  & Experimental  & \tablelineone                      \\ 
	\% Nb        & (\massalski)  & \tablelinetwo                     \\ \hline
	33.3         & C32           & Au$_2$Nb-C6 (\us).                \\ 
        \            &               & C32 $\sim$12meV/at.               \\
	\	     &	             & above C6 (\us).                   \\ 
        \            &               & {\it C32 stable (\paw)  }         \\ 
        \            &               & {\it C6$\sim$3.2meV/at. }         \\
        \            &               & {\it above C32 (\paw).  }         \\ \hline
	60   &Au$_2$Nb$_3$ (unknown) & unavailable                       \\ 
        \            & tI10 I4/mmm   &                                   \\ \hline
	66.6        &two-phase region& AuNb$_2$-\str64                   \\ 
        \           & (calculated)   &                                   \\ \hline
        73 to 83     & A15           &two-phase region                   \\
       	\	     &               & AuNb$_2$\twophase Nb.             \\
        \	     &               & AuNb$_3$-A15 is $\sim$7meV/at.    \\
	\            &               & above tie-line (\us).             \\
	\            &               & {\it A15 is $\sim$6.5meV/at.}     \\
        \            &               & {\it above tie-line (\paw). }     \\
	\            &               & See Section (\ref{section.discussion}). \\ 
      \end{tabular} \\ \hline
    \end{tabular}
  \end{center}
}

\begin{center}
  \vspace{-3mm}
  \begin{figure}
    \epsfig{file=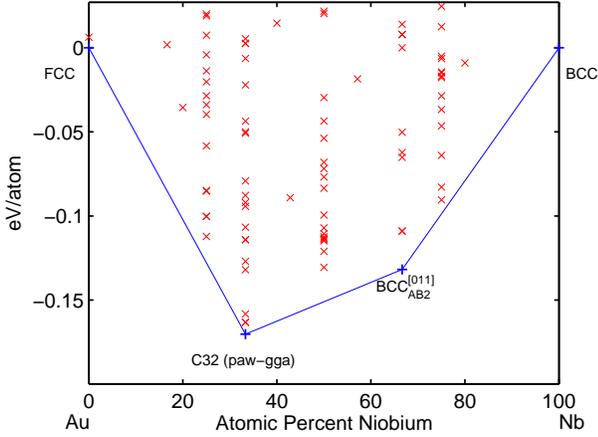,width=\picdim,clip=}
    \vspace{1mm}
    \caption{AuPd (Gold - Niobium) ground state convex hull.}
    \label{label.fig.AuNb}
  \end{figure}
  \vspace{-1mm}
\end{center}

\newpage

{\bf Au-Pd (Gold - Palladium).}
Two ordered compounds have been suggested for the system Au-Pd 
in the fcc solid solution (Au,Pd): 
Au$_3$Pd and AuPd$_3$ \cite{AuPd.06Rue,AuPd.65Nag,AuPd.66Mat,AuPd.71Kaw}.
The low-temperature part of the phase diagram has been constructed
with the addition of a phase field AuPd that is believed to 
exist below $\sim$100\DEG 
\cite{BB,PP,Au.85Oka,Au.87MasOkaBre}.
In the Au-rich part of the phase diagram, we 
confirm the stability of a compound Au$_3$Pd,
but we find three structures, D0$_{23}$, D0$_{22}$ and L1$_2$,
with degenerate energy.
At composition Au$_4$Pd, we find the structure D1$_a$ to be degenerate with the tie-line Au\twophase Au$_3$Pd. 
At composition 33.3\% Pd, we find a stable compound
Au$_2$Pd, but again, two structures C49 and C37 are degenerate. 
At composition 50\%, we confirm the existence of
a stable phase AuPd. The prototype is not known.
Our best guess is AuPd-\CH40 (CH ``40'' in reference \cite{PdPt.91Zun}), 
and L1$_0$ with an energy higher by 8meV/atom with respect to \CH40.
Finally, in the Pd-rich part of the phase diagram,
we find a compound AuPd$_3$ to be degenerate with the tie-line of the two-phase region
AuPd\twophase Pd. For this compound, three structures, D0$_{23}$, D0$_{22}$ and L1$_2$, have degenerate energies.
The considerable degeneracy of the ground states is indicative of week effective interactions and consistent with 
the significant miscibility of the two elements inside each other. 
The near degeneracy of structures such as D0$_{23}$, D0$_{22}$ and L1$_2$, 
which are related to each other by periodic antiphase boundaries, indicates that more complicated 
Long Period Superstructures (LPS) might be present.
{\it To address the small energy differences between structures,
we further investigate Au$_3$Pd, Au$_2$Pd, AuPd, and AuPd$_3$ with PAW-GGA potentials, 
\PawSection. 
For the compound Au$_3$Pd,
D0$_{23}$ is the most stable compound and 
D0$_{22}$ and L1$_2$ are higher by 9.4meV/atom and 12.7meV/atom, respectively.
For the compound AuPd$_3$,
D0$_{23}$ is the most stable compound and 
L1$_2$ and D0$_{22}$ are higher by 1.8meV/atom and 4.8meV/atom, respectively.
Also with PAW, for the compound Au$_2$Pd, 
C37 and C49 remain degenerate.
For the compound AuPd, 
\CH40 is the most stable compound and L1$_0$ is higher by 10.8meV/atom.}

\ReferenceAuPd

{\tablefont
  \begin{center}
    \begin{tabular}{||c||} \hline
      {\bf Au-Pd system} \\ \hline
      Low Temperature Phases comparison chart \\ \hline
      \begin{tabular}{c|c|c}
	Composition  & Experimental  & \tablelineone                      \\ 
	\% Pd        & (\massalski)  & \tablelinetwo                     \\ \hline
	12 to 32     &Au$_3$Pd-L1$_2$& Au$_3$Pd                          \\
	\            & (guessed)     & D0$_{23}$/D0$_{22}$/L1$_2$ (\us)  \\ 
	\            &               & {\it D0$_{23}$ stable (\paw)  }   \\ 
	\            &               & {\it D0$_{22}$$\sim$9.4meV/at. }  \\ 
	\            &               & {\it L1$_2$$\sim$12.7meV/at. }    \\
	\	     &		     & {\it above D0$_{23}$ (\paw)   }   \\ 
	20           &               & Au$_4$Pd-D1$_a$/tie-line (\us)    \\ \hline
	33.3        &solid solution& Au$_2$Pd                            \\
	\            &               & C49/C37 (\us/{\it \paw})          \\  \hline
	50           & AuPd (guessed)& AuPd-\CH40 (\us)                  \\
	\            & unknown       & L1$_0$ $\sim$8meV/at.             \\
	\	     & prototype     & above \CH40 (\us)                 \\ 
	\	     &		     & {\it \CH40 stable (\paw)      }   \\ 
	\            &               & {\it L1$_0$ $\sim$10.8meV/at.}    \\
	\	     &               & {\it above \CH40 (\paw).}         \\ \hline
        68 to 90     & L1$_2$        & AuPd$_3$-D0$_{23}$/D0$_{22}$/     \\
	\            & (guessed)     &             /L1$_2$/tie-line (\us)\\
	\            &               & {\it D0$_{23}$ stable (\paw)  }   \\ 
	\            &               & {\it L1$_2$$\sim$1.8meV/at. }     \\
	\            &               & {\it D0$_{22}$$\sim$4.8meV/at. }  \\ 
	\	     &		     & {\it above D0$_{23}$ (\paw)   }   \\ 
      \end{tabular} \\ \hline
    \end{tabular}
  \end{center}
}

\begin{center}
  \vspace{-3mm}
  \begin{figure}
    \epsfig{file=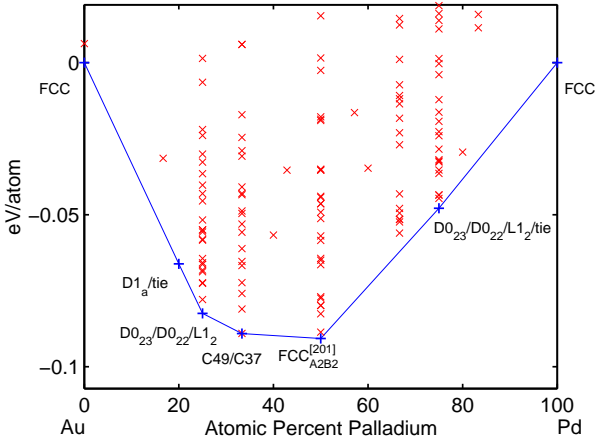,width=\picdim,clip=}
    \vspace{1mm}
    \caption{AuPd (Gold - Palladium) ground state convex hull.}
    \label{label.fig.AuPd}
  \end{figure}
  \vspace{-1mm}
\end{center}

\newpage

{\bf Au-Sc (Gold - Scandium).}
The phase diagram of the system Au-Sc is known only in the 
Au-rich region, and a total of three intermetallic compounds have been reported: Au$_4$Sc, Au$_3$Sc, and AuSc 
 \cite{BB,PP,Au.87MasOkaBre,AuSc.62Ald,AuSc.71Reu,AuSc.65Rid,AuSc.76Mie,AuSc.76Mat,AuSc.67Dwi}.
At 20\% Sc, we confirm Au$_4$Sc-D1$_a$. 
At 25\% Sc, we find a two-phase region, in agreement with experiments.
In fact, Au$_3$Sc-D0$_a$, the structure with lowest energy, 
is higher by $\sim$7meV/atom with respect to the tie-line  
Au$_4$Sc-D1$_a$\twophase Au$_2$Sc-C11$_b$.
At 33\% and at 50\% concentrations of Sc, we find 
two stable phases Au$_2$Sc and AuSc, but in both cases 
two structures are degenerate C11$_b$ and MoPt$_2$ for Au$_2$Sc, and 
B2 and B19 for AuSc. 
The experimental results indicate Au$_2$Sc-C11$_b$ and AuSc-B2.
The AuSc$_2$-C37 (Co$_2$Si prototype) compound has been speculated 
to exist by similarity with other Au-(Rare Earth)$_2$ systems \cite{AuSc.62Ald}.
Our {\it ab initio} method confirms AuSc$_2$-C37.
{\it To address the degenerate structures,
we further investigate Au$_2$Sc and AuSc with PAW-GGA potentials, 
\PawSection. 
For compound Au$_2$Sc, with PAW,
Au$_2$Sc-C11$_b$ is the most stable compound and 
Au$_2$Sc-MoPt$_2$ is higher by 1.0meV/atom. 
In addition, for compound AuSc,
Au$_2$Sc-B2 is the most stable compound and 
Au$_2$Sc-B19 is higher by 2.2meV/atom. 
While these energy differences are very small, 
both calculations are in agreement with experiments.}

{\tablefont
  \begin{center}
    \begin{tabular}{||c||} \hline
      {\bf Au-Sc system} \\ \hline
      Low Temperature Phases comparison chart \\ \hline
      \begin{tabular}{c|c|c}
	Composition  & Experimental  & \tablelineone                      \\ 
	\% Sc        & (\massalski)  & \tablelinetwo                     \\ \hline
	20           & D1$_a$   & Au$_4$Sc-D1$_a$                        \\ \hline
	25           &two-phase region& two-phase region		 \\
	\	     &		     & Au$_3$Sc-D0$_a$ $\sim$7meV/at.    \\
	\            &               & above Au$_4$Sc\twophase Au$_2$Sc. \\ \hline
	33.3         &C11$_b$        & Au$_2$Sc-C11$_b$/MoPt$_2$ (\us)   \\
	\            &               & {\it C11$_b$ stable (\paw)}       \\
	\            &               & {\it MoPt$_2$ $\sim$1.0meV/at. }  \\ 
	\            &               & {\it above C11$_b$ (\paw). }      \\  \hline
	50           & B2            & AuSc-B2/B19 (\us)                 \\
	\            &               & {\it B2 stable (\paw) }           \\
	\            &               & {\it B19 $\sim$2.2meV/at. }       \\ 
	\            &               & {\it above B2 (\paw). }           \\  \hline
	66.6         &C37 (Co$_2$Si) & AuSc$_2$-C37                      \\
        \            & (guessed)     &                                   \\ 
      \end{tabular} \\ \hline
    \end{tabular}
  \end{center}
}

\begin{center}
  \vspace{-3mm}
  \begin{figure}
    \epsfig{file=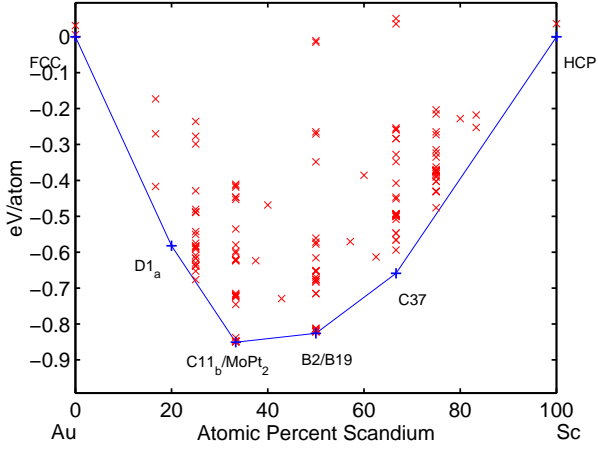,width=\picdim,clip=}
    \vspace{1mm}
    \caption{AuSc (Gold - Scandium) ground state convex hull.}
    \label{label.fig.AuSc}
  \end{figure}
  \vspace{-1mm}
\end{center}

\newpage

{\bf Au-Ti (Gold - Titanium).}
Five ordered compounds have been reported for the system Au-Ti at low temperature:
Au$_4$Ti, Au$_2$Ti, $\alpha$AuTi, $\beta$AuTi, and AuT$_3$
\cite{BB,PP,ref.Titanium,AgTi.78PLI,AuTi.52Rau,AuTi.54Mcq,AuTi.56Pie,AuTi.62Sch,AuTi.62Pie,AuTi.64Von,AuTi.67Vie,AuTi.67Vet,AuTi.68Pie,AuTi.89Sha,PtTi.69Sin,PdTi.70Don,PtZr.43Wal,PdZr.62Sto}.
We confirm the stability of $\alpha$AuTi-B11, Au$_4$Ti-D1$_a$ and AuTi$_3$-A15.
At 25\% Ti, we find a two-phase region, in agreement with experiments.
In fact, Au$_3$Ti-D0$_a$, the structure with lowest energy, 
is higher by $\sim$11meV/atom with respect to the tie-line Au$_4$Ti\twophase Au$_2$Ti.
For compound Au$_2$Ti, the experimental phase is C11$_b$. 
At such composition we find two structures with degenerate energies:
Au$_2$Ti-C11$_b$ and Au$_2$Ti-MoPt$_2$.
We do not confirm an off-stoichiometry stable phase $\beta$AuTi-B19, 
because we do not have such prototype in our library. The stoichiometric
B19 has an energy $\sim$50meV/atom above $\alpha$AuTi-B11.
At 42.8\% concentration of Ti, we find a previously unknown compound Au$_4$Ti$_3$-Cu$_4$Ti$_3$. 
This compound is degenerate with the tie-line of the two-phase field Au$_2$Ti\twophase AuTi, 
therefore its existence is uncertain.
{\it To address the degenerate structures Au$_2$Ti-C11$_b$/MoPt$_2$ 
and the energy difference between $\alpha$AuTi-B11 and $\beta$AuTi-B19 phases,
we further investigate Au$_2$Ti and AuTi with PAW-GGA potentials, 
\PawSection. 
With PAW,
Au$_2$Ti-C11$_b$ is the most stable compound and 
Au$_2$Ti-MoPt$_2$ is higher by 3.7meV/atom.
For phase AuTi, 
AuTi-B11 is the most stable compound (as in the ultrasoft pseudopotential case), and
AuTi-B19 is higher by 53.6meV/atom. } 

{\tablefont
  \begin{center}
    \begin{tabular}{||c||} \hline
      {\bf Au-Ti system} \\ \hline
      Low Temperature Phases comparison chart \\ \hline
      \begin{tabular}{c|c|c}
	Composition  & Experimental  & \tablelineone                      \\ 
	\% Ti        & (\massalski)  & \tablelinetwo                     \\ \hline
	18-21	     & D1$_a$        & Au$_4$Ti-D1$_a$                   \\ \hline
	25           &two-phase region& two-phase region		 \\
	\	     &above 500\DEG  & Au$_3$Ti-D0$_a$ $\sim$11meV/at.   \\
	\            &               & above Au$_4$Ti\twophase Au$_2$Ti. \\ \hline
	33.3         & C11$_b$       & Au$_2$Ti                          \\
	\            &               & C11$_b$/MoPt$_2$ (\us)             \\
	\            &               & {\it C11$_b$ stable (\paw)}        \\
	\            &               & {\it MoPt$_2$$\sim$3.7meV/at.}    \\
        \            &               & {\it above C11$_b$ (\paw).}        \\  \hline 
	42.8         & two-phase region & Au$_4$Ti$_3$                   \\
	\            & above 500\DEG &  Cu$_4$Ti$_3$/tie                 \\ \hline 
	50           & $\alpha$AuTi-B11 & AuTi-B11                       \\
	50-51        & $\beta$AuTi-B19 & B19$\sim$50meV/at.              \\ 
	\            &               & {\it B19$\sim$53.6meV/at. (\us)}   \\
        \            &               & {\it above B11 (\paw).}            \\  \hline 
	38-52        & $\gamma$AuTi-B2 & B2$\sim$135meV/at.              \\
	\            & high-temperature & above B11 (\us).                \\  \hline 
        75           & A15           & AuTi$_3$-A15                      \\ 
      \end{tabular} \\ \hline
    \end{tabular}
  \end{center}
}

\begin{center}
  \vspace{-3mm}
  \begin{figure}
    \epsfig{file=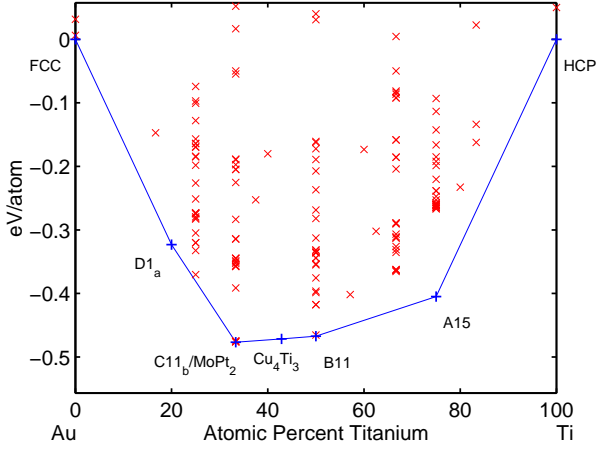,width=\picdim,clip=}
    \vspace{1mm}
    \caption{AuTi (Gold - Titanium) ground state convex hull.}
    \label{label.fig.AuTi}
  \end{figure}
  \vspace{-1mm}
\end{center}

\newpage

{\bf Au-Y (Gold - Yttrium).}
Four intermetallic compounds have been reported for the 
system Au-Y: Au$_3$Y, Au$_2$Y, AuY, and AuY$_3$ \cite{BB,PP,Au.87MasOkaBre,AuSc.65Rid,AuSc.76Mie,AuSc.76Mat,AuSc.67Dwi,AuY.63Cha,AuY.65Kir,AuY.65Kus,AuY.68Sad}.
With our {\it ab initio} method, we confirm stability of Au$_3$Y-D0$_a$ and Au$_2$Y-C11$_b$.
At equal composition, AuY, we do not confirm B2: 
we find a stable B33 and phase B2 higher by $\sim$26meV/atom.
No experimental compound is reported for composition AuY$_2$, 
though we obtain AuY$_2$-C37.
At 75\% composition of Y, we do not find a stable AuY$_3$ 
compound reported experimentally \cite{PP,AuSc.76Mie}: 
the structure with lowest energy 
is AuY$_3$-D0$_{11}$ with an energy higher by $\sim$30meV/atom
with respect to the tie-line.
{\it We further investigate AuY with PAW-GGA potentials, 
\PawSection. 
With PAW,
B33 is the most stable compound and B2 is $\sim$25meV/atom higher than B33.}
{The disagreement at composition AuY is further discussed in Section (\ref{section.discussion}).}

{\tablefont
  \begin{center}
    \begin{tabular}{||c||} \hline
      {\bf Au-Y system} \\ \hline
      Low Temperature Phases comparison chart \\ \hline
      \begin{tabular}{c|c|c}
	Composition  & Experimental  & \tablelineone                      \\ 
	\% Y         & (\massalski)  & \tablelinetwo                     \\ \hline
	25           & D0$_a$   & Au$_3$Y-D0$_a$               \\ \hline
	33.3         & C11$_b$       & Au$_2$Y-C11$_b$                   \\ \hline
	50           & B2            & AuY-B33 (\us)                      \\
	\            &               & B2$\sim$26meV/at.                 \\
	\	     &		     & above B33 (\us).                   \\
	\            &               & {\it B33 stable (\paw)  }          \\
	\            &               & {\it B2$\sim$25meV/at. }          \\
	\	     &		     & {\it above B33 (\paw).  }         \\
	\            &               & See Section (\ref{section.discussion}).  \\ \hline
        66.6         & not studied   & AuY$_2$-C37                       \\  \hline
        75           & hyp           & nothing stable                    \\
	\     & (uncertain prototype)& D0$_{11}$$\sim$30meV/at.          \\
	\            &               & above tie-line                    \\ 
      \end{tabular} \\ \hline
    \end{tabular}
  \end{center}
}

\begin{center}
  \vspace{-3mm}
  \begin{figure}
    \epsfig{file=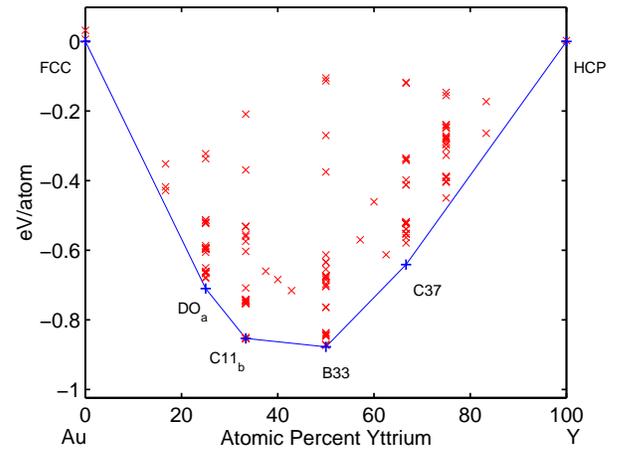,width=\picdim,clip=}
    \vspace{1mm}
    \caption{AuY (Gold - Yttrium) ground state convex hull.}
    \label{label.fig.AuY}
  \end{figure}
  \vspace{-1mm}
\end{center}

\newpage

{\bf Au-Zr (Gold - Zirconium).}
Several ordered compounds have been reported for the system Au-Zr: 
Au$_4$Zr, Au$_3$Zr, Au$_2$Zr, Au$_{10}$Zr$_7$, Au$_4$Zr$_5$, AuZr$_2$, and AuZr$_3$
\cite{BB,PP,AuTi.62Sch,AuZr.48Rau,AuZr.62Sch2,AuZr.62Nev,AuZr.62Stol}.
Our {\it ab initio} method confirms the stability of Au$_3$Zr-D0$_a$, Au$_2$Zr-C11$_b$  
and AuZr$_3$-A15. 
{At composition AuZr$_2$, Massalski and the Pauling file 
report C11$_b$ and CuZr$_2$ type compounds, respectively \cite{BB,PP}.
C11$_b$ (MoSi$_2$ prototype) and CuZr$_2$ are very similar structures, 
both tetragonal (tI6) and with the same space group \#139. 
C11$_b$ belongs to the bcc superstructure family while
CuZr$_2$, a slightly distorted version of C11$_b$, is a closed packed 
structure \cite{PP}, therefore, within the formalism of the atomic environments, 
they represents two different prototypes \cite{PP}.
Our library contains both prototypes. 
However, in our calculations,
AuZr$_2$-C11$_b$ and AuZr$_2$-CuZr$_2$ are both unstable
with degenerate energies higher by $\sim$20meV/atom 
with respect to the tie-line AuZr\twophase AuZr$_3$.} 
We can not say anything about Au$_4$Zr, Au$_{10}$Zr$_7$ and Au$_4$Zr$_5$,
because our library lacks prototypes at the proper concentrations.
At AuZr composition we find two structures with degenerate energies:
B11 and FCC$_{A2B2}^{[001]}$. 
This prediction is unreliable because such compounds appear in the
two-phase region Au$_{10}$Zr$_7$\twophase Au$_4$Zr$_5$, that we are not 
able to describe properly. 
At 42.8\% concentration of Zr, we find Au$_4$Zr$_3$-Cu$_4$Ti$_3$. 
As before, this prediction is unreliable because it appears in the
two-phase region Au$_2$Zr\twophase Au$_{10}$Zr$_7$, that we are not 
able to describe.
To conclude, the inclusion of prototypes of 
Au$_{10}$Zr$_7$ and Au$_4$Zr$_5$ in our calculations,
might change the stability of the predicted AuZr and Au$_4$Zr$_3$ phases.
{\it {To address the degenerate structures,
we further investigate AuZr$_2$ with PAW-GGA potentials,
\PawSection.
With PAW,
AuZr$_2$-CuZr$_2$ and AuZr$_2$-C11$_b$ are still degenerate,
but become stable with an energy 6.6meV/atom lower than the
tie-line AuZr\twophase AuZr$_3$.}
}

{\tablefont
  \begin{center}
    \begin{tabular}{||c||} \hline
      {\bf Au-Zr system} \\ \hline
      Low Temperature Phases comparison chart \\ \hline
      \begin{tabular}{c|c|c}
	Composition  & Experimental  & \tablelineone                      \\ 
	\% Zr        & (\massalski)  & \tablelinetwo                     \\ \hline
	20           & Au$_4$Zr      & unavailable                       \\
	\            & oP20 - Pnma   &                                   \\ \hline
	25           & D0$_a$        & Au$_3$Zr-D0$_a$                   \\ \hline 
	33.3         & C11$_b$       & Au$_2$Zr-C11$_b$                  \\ \hline 
	42.8       &two-phase region & Au$_4$Zr$_3$-Cu$_4$Ti$_3$         \\
	\     &Au$_2$Zr\twophase Au$_{10}$Zr$_7$& unreliable, see text.  \\ \hline        
	45           &Au$_{10}$Zr$_7$& unavailable                       \\
	\            & tI34          &                                   \\ \hline	
	50           &two-phase region & AuZr-B11/FCC$_{A2B2}^{[001]}$   \\
	\ &Au$_{10}$Zr$_7$\twophase Au$_4$Zr$_5$& unreliable, see text.  \\ \hline
	55.5	     & Au$_4$Zr$_5$  & unavailable                       \\
	\	     & unknown	     &                                   \\
	\	     & prototype     &                                   \\ \hline
	66.6      & C11$_b$ \cite{BB}& two-phase region (\us)            \\
	\        & CuZr$_2$ \cite{PP}& AuZr$_2$-CuZr$_2$/C11$_b$         \\
	\	     &		     & $\sim$20meV/at. above tie-line    \\
	\            &               & AuZr\twophase AuZr$_3$ (\us).     \\  
	\	     &	             & {\it CuZr$_2$/C11$_b$ stable }    \\
	\	     &	             & {\it $\sim6.6meV/at.$ below }     \\
	\	     &		     & {\it AuZr\twophase AuZr$_3$ (\paw).}      \\ \hline
	75	     & A15	     & AuZr$_3$-A15                      \\ 
      \end{tabular} \\ \hline
    \end{tabular}
  \end{center}
}

\begin{center}
  \vspace{-3mm}
  \begin{figure}
    \epsfig{file=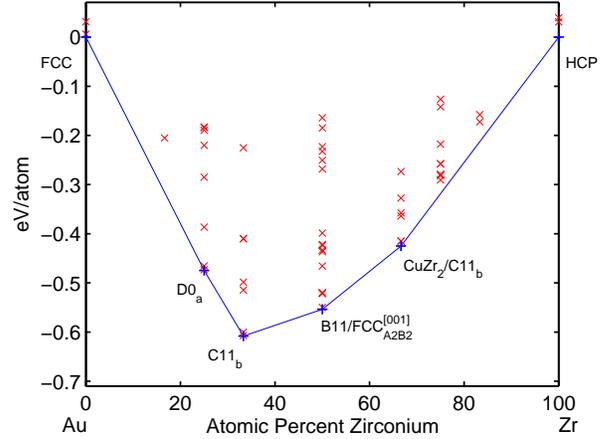,width=\picdim,clip=}
    \vspace{1mm}
    \caption{AuZr (Gold - Zirconium) ground state convex hull.}
    \label{label.fig.AuZr}
  \end{figure}
  \vspace{-1mm}
\end{center}

\newpage

{\bf Cd-Nb (Cadmium - Niobium).}
The phase diagram for the system Cd-Nb is not known \cite{BB}. 
It has been reported that Cd$_3$Nb has L1$_2$ prototype \cite{CdNb.66Hol}.
However, we are not able to find any stable compound, so we 
propose that the system is not compound-forming, but will display
low-temperature immiscibility. In our calculations, Cd$_3$Nb-L1$_2$ is
$\sim$70meV/atom higher in energy than the phase separation of Cd\twophase Nb.
In addition, the structure with the lowest formation energy is a 
bcc superstructure at composition CdNb and energy $\sim$55meV/atom 
higher then Cd\twophase Nb. 
{\it {To address the disagreement with the experimental result,
we further investigate the relevant compounds with PAW-GGA potentials, 
\PawSection. 
With PAW, Cd$_3$Nb-L1$_2$ is
$>$100meV/atom higher in energy than the tie-line Cd\twophase Nb.
The structure with the lowest formation energy is a 
bcc superstructure at composition CdNb$_3$, and energy $\sim$58meV/atom above Cd\twophase Nb.}}
It is unlikely that the {\it ab initio} calculations have such a large errors.
{The disagreement at composition Cd$_3$Nb is further discussed in Section (\ref{section.discussion}).}

{\tablefont
  \begin{center}
    \begin{tabular}{||c||} \hline
      {\bf Cd-Nb system} \\ \hline
      Low Temperature Phases comparison chart \\ \hline
      \begin{tabular}{c|c|c}
	Composition  & Experimental  & \tablelineone                      \\ 
	\% Nb        & (\massalski)  & \tablelinetwo                     \\ \hline
	25           & Cd$_3$Nb-L1$_2$  & immiscible                     \\
	\            &               & See Section (\ref{section.discussion}).  \\
      \end{tabular} \\ \hline
    \end{tabular}
  \end{center}
}

{\bf Cd-Pd (Cadmium - Palladium).}
Four Cd-Pd compounds have been identified experimentally at low-temperature
($\gamma$, $\gamma_1$, $\gamma'$ and $\beta_1$ \cite{CdPd.80Arn,CdPd.82Neu,BB}). 
However, the phase boundaries are unknown.
We confirm $\beta_1$-CdPd-L1$_0$.
In the Cd-rich part of the phase diagram, we find Cd$_3$Pd 
(near $\gamma'$ at 74\%Cd). We are not able to determine the exact 
structure of Cd$_3$Pd but our best candidates are 
D0$_{19}$, D0$_{24}$, NbPd$_3$-type, and Al$_3$Pu-type (Co$_3$V).
In the Pd-rich part of the phase diagram, we find a stable phase 
CdPd$_3$. As before, we cannot determine its structure, precisely.
Our guesses are D0$_{22}$ and NbPd$_3$-type, which have degenerate energies.
At concentration 33.3\% Cd, we find a new compound CdPd$_2$-C37 
which is not present in Massalski \cite{BB}.
{\it To address the degenerate structures,
we further investigate CdPd$_3$ and Cd$_3$Pd with PAW-GGA potentials, 
\PawSection. 
Compounds CdPd$_3$-D0$_{22}$ and CdPd$_3$-NbPd$_3$ remain degenerate with PAW
(the energy difference is 1.1meV/atom).
In addition, for Cd$_3$Pd composition,
NbPd$_3$/Al$_3$Pu remain degenerate compounds, and
D0$_{22}$, D0$_{19}$, D0$_{24}$  are higher by 4.7meV/atom, 5.6meV/atom, 6.7meV/atom, respectively.}

\vspace{-1mm}
{\tablefont
  \begin{center}
    \begin{tabular}{||c||} \hline
      {\bf Cd-Pd system} \\ \hline
      Low Temperature Phases comparison chart \\ \hline
      \begin{tabular}{c|c|c}
	Composition  & Experimental  & \tablelineone                      \\ 
	\% Cd        & (\massalski)  & \tablelinetwo                     \\ \hline
	25           & Pd phase      & CdPd$_3$                          \\
	\            & above 100$^\circ$C & D0$_{22}$/NbPd$_3$ (\us)      \\
	\            &               & D0$_{22}$/NbPd$_3$ (\paw)          \\ \hline
	33.3         & two-phase region & CdPd$_2$-C37                   \\ 
	\            & above 100$^\circ$C,&                              \\
	\            & between $\beta_1$-Pd&                             \\ \hline
	37-55     & $\beta_1$-L1$_0$ & CdPd-L1$_0$                       \\ \hline
	74     & $\gamma'$-(unknown) & Cd$_3$Pd-D0$_{19}$ or             \\
	\            &               & D0$_{24}$/NbPd$_3$/Al$_3$Pu (\us)   \\
	\            &               & {\it NbPd$_3$/Al$_3$Pu stable (\paw)  }  \\
	\	     &		     & {\it D0$_{22}$ $\sim$ 4.7meV/at.,      }  \\
	\	     &		     & {\it D0$_{19}$ $\sim$ 5.6meV/at.,      }  \\
	\	     &		     & {\it D0$_{24}$ $\sim$ 6.7meV/at.       }  \\
	\	     &		     & {\it above NbPd$_3$/Al$_3$Pu  (\paw)    } \\ \hline
	77-80 & $\gamma_1$-(unknown) & unavailable                       \\ \hline
	81-83      & $\gamma$-D8$_3$ & unavailable                       \\ 
      \end{tabular} \\ \hline
    \end{tabular}
  \end{center}
}
\begin{center}
  \vspace{-4mm}
  \begin{figure}
    \epsfig{file=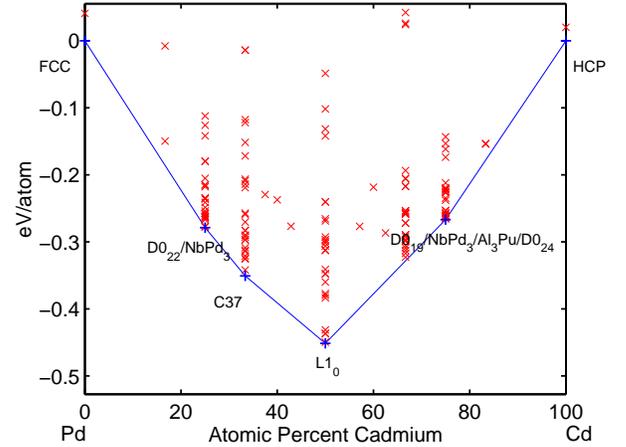,width=\picdim,clip=}
    \caption{CdPd (Cadmium - Palladium) ground state convex hull.}
    \label{label.fig.CdPd}
  \end{figure}
  \vspace{-1mm}
\end{center}

\newpage

{\bf Cd-Pt (Cadmium - Platinum).}
The phase diagram of the system Cd-Pt is partially known \cite{CdPt.52Now,BB},
with several compounds of unknown structure.
We confirm the stability of phase $\alpha_1'$-CdPt-L1$_0$. 
In the Cd-rich part of the phase diagram, the prototypes of Cd$_2$Pt 
and $\gamma_1$ are unknown.
For Cd$_2$Pt, our best guess are two degenerate 
structures, Cd$_2$Pt-C37 and Cd$_2$Pt-C16, which are also degenerate with the hull.
For $\gamma_1$, at stoichiometry Cd$_3$Pt,
we find two degenerate structures D0$_{11}$, D0$_{a}$ and D0$_{22}$.
In the Pt-rich part of the phase diagram, we do not find
a stable compound $\alpha'$-CdPt$_3$-L1$_2$. 
Instead, we find a stable orthorhombic CdPt$_3$ with fcc superstructure 
and Cmmm \#65 space group. The prototype, labeled 
as CdPt$_3^{proto}$, is described in Appendix 
(\ref{proto.AB3.CdPt11relaxed}). 
In addition, CdPt$_3$-L1$_2$ is found with an energy $\sim$25meV/atom above CdPt$_3^{proto}$.
{\it {To address the degenerate structures and the disagreement with experimental results for compound CdPt$_3$,
we further investigate CdPt$_3$, Cd$_2$Pt and Cd$_3$Pt with PAW-GGA potentials, 
\PawSection. 
With PAW, 
CdPt$_3$-CdPt$_3^{proto}$ is the most stable structure and
CdPt$_3$-L1$_2$ higher by $\sim$10.4meV/atom.
For compound Cd$_2$Pt,
C16 is the most stable compound and 
C37 is $\sim$33meV/atom higher than C16.
For compound Cd$_3$Pt,
D0$_{11}$ is the most stable compound and 
D0$_{a}$ and D0$_{22}$ are higher by $\sim$52meV/atom and $\sim$60meV/atom, respectively.}}
{The disagreement at composition CdPt$_3$ is further discussed in Section (\ref{section.discussion}).}

{\tablefont
  \begin{center}
    \begin{tabular}{||c||} \hline
      {\bf Cd-Pt system} \\ \hline
      Low Temperature Phases comparison chart \\ \hline
      \begin{tabular}{c|c|c}
	Composition  & Experimental  & \tablelineone                      \\ 
	\% Pt        & (\massalski)  & \tablelinetwo                     \\ \hline
        $\sim$13 to 17&$\gamma$-(unknown)& two-phase region              \\ \hline
        24 to 26&$\gamma_1$-(unknown)&  Cd$_3$Pt                         \\
	\            &               & D0$_{11}$/D0$_{a}$/D0$_{22}$ (\us)\\ 
	\            &               & {\it D0$_{11}$ stable (\paw)    } \\ 
	\            &               & {\it D0$_{a}$$\sim$52meV/at.  }  \\
	\            &               & {\it D0$_{22}$$\sim$60meV/at.  }  \\
        \            &               & {\it above D0$_{11}$ (\paw)     } \\ \hline
        $\sim$26 to 28&$\gamma_2$-(unknown)&unavailable                  \\ \hline
        $\sim$31 to 38&Cd$_2$Pt-(unknown)& Cd$_2$Pt                      \\
        \            &               & C37/C16/tie-line (\us)            \\
        \            &               & {\it C16 stable (\paw)          } \\
        \            &               & {\it C37$\sim$33meV/at.        }  \\
	\            &               & {\it above C16 (\paw)           } \\ \hline
        $\sim$49 to 51&$\alpha'_1$-L1$_0$& CdPt-L1$_0$                   \\ \hline
        $\sim$75 &$\alpha'$-L1$_2$   & CdPt$_3^{proto}$                  \\
	\            &       &  Appendix (\ref{proto.AB3.CdPt11relaxed}) \\
	\            &               &  L1$_2$$\sim$25meV/at.            \\
	\            &               & above CdPt$_3^{proto}$.           \\ 
	\            &               & {\it CdPt$_3^{proto}$ stable (\paw) } \\
	\            &               & {\it L1$_2$$\sim$10.4meV/at. above }     \\
	\            &               & {\it CdPt$_3^{proto}$ (\paw).} \\ 
	\            &               & See Section (\ref{section.discussion}).\\ 
      \end{tabular} \\ \hline
    \end{tabular}
  \end{center}
}

\begin{center}
  \vspace{-5mm}
  \begin{figure}
    \epsfig{file=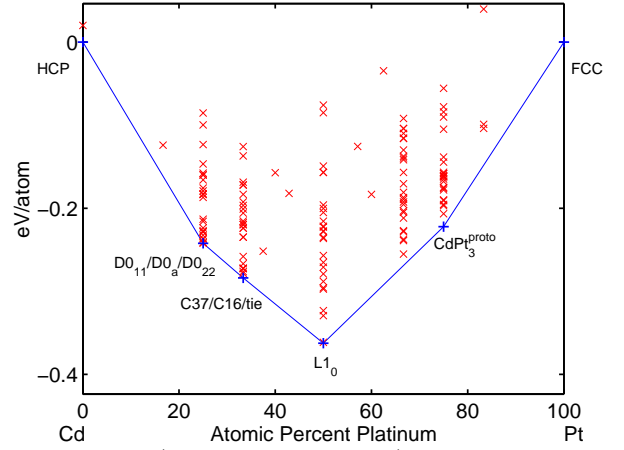,width=\picdim,clip=}
    \caption{CdPt (Cadmium - Platinum) ground state convex hull.}
    \label{label.fig.CdPt}
  \end{figure}
  \vspace{-1mm}
\end{center}

\newpage

{\bf Cd-Rh (Cadmium - Rhodium).}
No experimental phase diagram is available \cite{BB}.
With our {\it ab initio} technique we find two stable compounds, Cd$_2$Rh-C37, 
and Cd$_3$Rh. Our best guesses for Cd$_3$Rh are Cd$_3$Rh-D0$_{24}$ and Cd$_3$Rh-Al$_3$Pu-type (Co$_3$V),
which have degenerate energies (2meV/atom difference).
Interesting metastable phases are Cd$_3$Rh-D0$_{19}$, 
(11 meV/atom above Cd$_3$Rh), and Cd$_2$Rh-C49  
(8meV/atom above Cd$_2$Rh-C37).
{\it To address the degenerate structures,
we further investigate Cd$_3$Rh with PAW-GGA potentials, 
\PawSection. 
With PAW,
Cd$_3$Rh-Al$_3$Pu is the ground state, and
Cd$_3$Rh-D0$_{24}$ has an energy 5meV/atom higher.}

{\tablefont
  \begin{center}
    \begin{tabular}{||c||} \hline
      {\bf Cd-Rh system} \\ \hline
      Low Temperature Phases comparison chart \\ \hline
      \begin{tabular}{c|c|c}
	Composition  & Experimental  & \tablelineone                      \\ 
	\% Cd        & (\massalski)  & \tablelinetwo                     \\ \hline
	66.6         & not studied   & Cd$_2$Rh-C37                      \\ 
	\            &               & C49 metastable                    \\ \hline
      75           & not studied   & Cd$_3$Rh                            \\
	\            &               & D0$_{24}$/Al$_3$Pu (\us)            \\ 
 	\            &               & D0$_{19}$ $\sim$ 11meV/at.         \\ 
        \            &               & above D0$_{24}$/Al$_3$Pu  (\us)     \\
	\            &               & {\it Al$_3$Pu stable (\paw)              } \\
	\            &               & {\it D0$_{24}$ $\sim$ 5meV/at.         } \\
        \            &               & {\it above Al$_3$Pu (\paw)               } \\
      \end{tabular} \\ \hline
    \end{tabular}
  \end{center}
}

\begin{center}
  \vspace{-3mm}
  \begin{figure}
    \epsfig{file=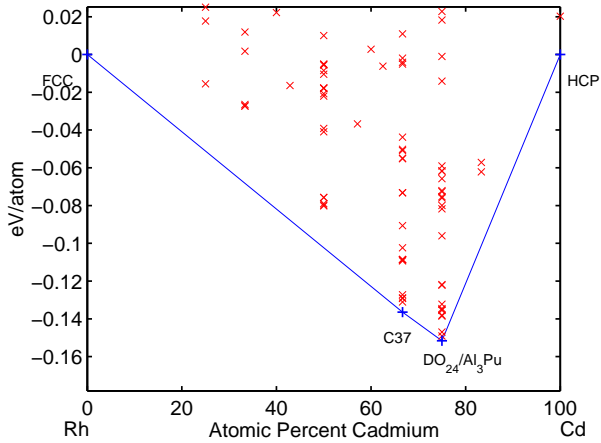,width=\picdim,clip=}
    \vspace{1mm}
    \caption{CdRh (Cadmium - Rhodium) ground state convex hull.}
    \label{label.fig.CdRh}
  \end{figure}
  \vspace{-1mm}
\end{center}

\newpage

{\bf Cd-Ti (Cadmium - Titanium).}
Little is known for the system Cd-Ti at high-temperature \cite{BB,PP,CdTi.62Cha,CdTi.72Rob}.
At low temperature only two stable intermetallic compounds have 
been reported \cite{CdTi.72Rob}.
We confirm the stability of Ti$_2$Cd-C11$_b$.
At composition CdTi, Massalski and the Pauling file, 
report B11 and CdTi compounds, respectively \cite{BB,PP}.
B11 ($\gamma$CuTi prototype) and CdTi are very similar structures, 
both tetragonal (tP4) and with the same space group \#129. 
B11 belongs to the bcc superstructure family while
CdTi, a slightly distorted version of B11, is a closed packed 
structure \cite{PP}. 
Therefore, within the formalism of the atomic environments, 
they represents two different prototypes \cite{PP}.
The {\it ab initio} calculation is able to relax one structure into the other, easily.
For CdTi we confirm the experimental results.
In addition, at concentration 42.8\%Cd, we find a compound Cd$_3$Ti$_4$-Cu$_4$Ti$_3$, 
degenerate with the tie-line of the two-phase region CdTi$_2$\twophase CdTi.
We have not found any other stable or metastable compound: 
we conclude that the low temperature part of the phase diagram is complete.
Note that this system is very similar to Ag-Ti.

{\tablefont
  \begin{center}
    \begin{tabular}{||c||} \hline
      {\bf Cd-Ti system} \\ \hline
      Low Temperature Phases comparison chart \\ \hline
      \begin{tabular}{c|c|c}
	Composition  & Experimental  & \tablelineone                      \\ 
	\% Cd        & (\massalski)  & \tablelinetwo                     \\ \hline
	33.3         & C11$_b$       & CdTi$_2$-C11$_b$                  \\ \hline
	42.8         &two-phase region& Cd$_3$Ti$_4$                     \\
	\            &above 200\DEG  & Cu$_4$Ti$_3$/tie-line             \\ \hline
	50           & B11           & CdTi-B11 (CdTi)                   \\
	\            & CdTi \cite{PP}&                                   \\ 
      \end{tabular} \\ \hline
    \end{tabular}
  \end{center}
}

\begin{center}
  \vspace{-3mm}
  \begin{figure}
    \epsfig{file=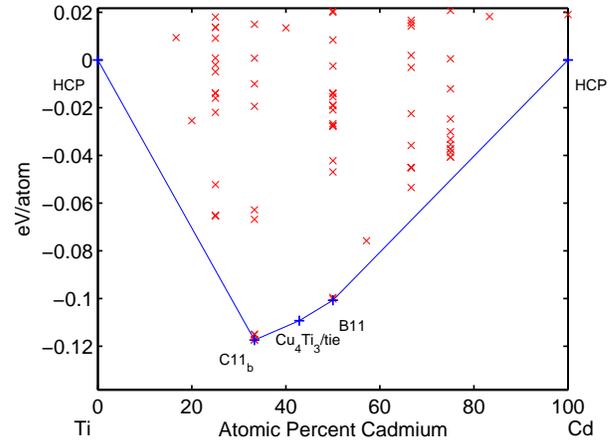,width=\picdim,clip=}
    \vspace{1mm}
    \caption{CdTi (Cadmium - Titanium) ground state convex hull.}
    \label{label.fig.CdTi}
  \end{figure}
  \vspace{-1mm}
\end{center}

\newpage

{\bf Cd-Y (Cadmium - Yttrium).}
The stability of compounds YCd-B2, YCd$_2$-C6 and YCd$_3$ 
(Cd$_3$Er prototype) \cite{BB,PP,CdY.68Elm,CdY.69Ryb,CdY.73Bru} is confirmed.
The Yttrium-rich side of the phase diagram is poorly known. 
In that region we find two degenerate phases Y$_2$Cd-C49 and Y$_2$Cd-C37, 
both of which are close to the energy of the two 
phase region Y\twophase CdY (figure (\ref{label.fig.CdY})). 
We do not find any other stable or metastable phase.
{\it To address the degenerate structures,
we further investigate Y$_2$Cd with PAW-GGA potentials, 
\PawSection. 
With PAW,
C37 is the most stable compound and 
C49 is higher by 7.7meV/atom.} 

\ReferenceCdY

{\tablefont
  \begin{center}
    \begin{tabular}{||c||} \hline
      {\bf Cd-Y system} \\ \hline
      Low Temperature Phases comparison chart \\ \hline
      \begin{tabular}{c|c|c}
	Composition  & Experimental  & \tablelineone                      \\ 
	\% Cd        & (\massalski)  & \tablelinetwo                     \\ \hline
	33.3         & not studied/  & CdY$_2$                           \\ 
	\          & two-phase region& C37/C49 (\us)                      \\
	\            &               & {\it C37 stable (\paw)                 } \\
	\            &               & {\it C49 $\sim$ 7.7meV/at.            } \\
        \            &               & {\it above C37 (\paw)                  } \\ \hline
	50           & B2            & CdY-B2                            \\ \hline
	66.6         & C6            & Cd$_2$Y-C6                        \\ \hline
        75           & Cd$_3$Er      & Cd$_3$Y-Cd$_3$Er                  \\ \hline
        80.4 &Cd$_{45}$Y$_{11}$-Cd$_{45}$Sm$_{11}$&unavailable           \\  \hline
        81.7 &Cd$_{58}$Y$_{13}$-Pu$_{13}$Zn$_{58}$&unavailable           \\  \hline
        85.7         & Cd$_{6}$Y     &unavailable                        \\ 
      \end{tabular} \\ \hline
    \end{tabular}
  \end{center}
}

\begin{center}
  \vspace{-3mm}
  \begin{figure}
    \epsfig{file=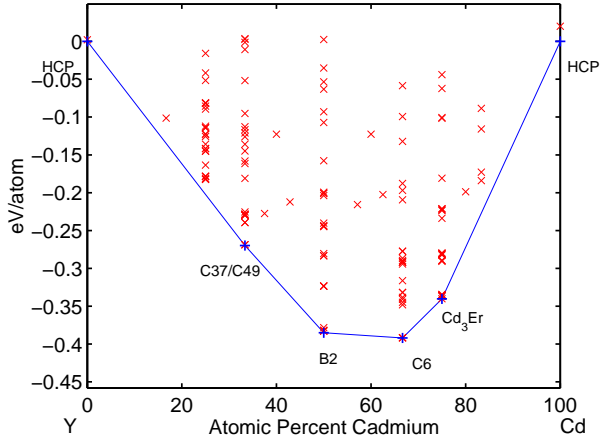,width=\picdim,clip=}
    \vspace{1mm}
    \caption{CdY (Cadmium - Yttrium) ground state convex hull.}
    \label{label.fig.CdY}
  \end{figure}
  \vspace{-1mm}
\end{center}

\newpage

{\bf Cd-Zr (Cadmium - Zirconium).}
The Cd-Zr system is poorly characterized \cite{ref.Zirconium,BB,CdZr.65Ros}, and
compounds have been identified at four compositions 
Cd$_3$Zr, Cd$_2$Zr, CdZr, and CdZr$_2$. 
Massalski \cite{BB} reports Cd$_3$Zr-L1$_0$-AuCu, which we consider a misprint for Cd$_3$Zr-L1$_2$-AuCu$_3$,
while {\small CRYSTMET} and Pauling File databases \cite{PP,CC}
report Cd$_3$Zr-D0$_a$-$\beta$Cu$_3$Ti from reference \cite{CdZr.69Aru}.
We confirm the stability of Cd$_3$Zr-L1$_2$ and CdZr$_2$-C11$_b$.
At composition CdZr, Massalski and the Pauling file, 
report B11 and CdTi compounds, respectively \cite{BB,PP}.
B11 ($\gamma$CuTi prototype) and CdTi are very similar structures, 
both tetragonal (tP4) and with the same space group \#129. 
B11 belongs to the bcc superstructure family while
CdTi, a slightly distorted version of B11, is a closed packed 
structure \cite{PP}.
Therefore, within the formalism of the atomic environments, 
they represents two different prototypes \cite{PP}.
The {\it ab initio} calculation is able to relax one structure into the other, easily.
For CdZz, we find CdZr-L1$_0$ instead of CdZr-B11, 
which has an energy $\sim$18meV/atom higher than L1$_0$.
The prototype of Cd$_2$Zr is not known: our best guess is 
Cd$_2$Zr-C11$_b$.
In addition, we find a stable compound CdZr$_3$-A15 
not present in Massalski \cite{BB}, and a metastable 
L1$_2$ which is 8.5meV/atom higher than A15.
{\it {To address the disagreement with the experimental results for 
compound CdZr, we further investigate L1$_0$ and B11 with PAW-GGA potentials, 
\PawSection. 
With PAW,
CdZr-B11 is the most stable compound and 
CdZr-L1$_0$ is higher by 13meV/atom.}} 

{\tablefont
  \begin{center}
    \begin{tabular}{||c||} \hline
      {\bf Cd-Zr system} \\ \hline
      Low Temperature Phases comparison chart \\ \hline
      \begin{tabular}{c|c|c}
	Composition  & Experimental  & \tablelineone                      \\ 
	\% Cd        & (\massalski)  & \tablelinetwo                     \\ \hline
	25           &two-phase region& CdZr$_3$-A15                     \\
	\            &not studied    & L1$_2$ $\sim$ 8.5meV/at.          \\
        \            &               & above A15                         \\ \hline
	33.3         & C11$_b$       & CdZr$_2$-C11$_b$                  \\ \hline
	50           & B11           & CdZr-L1$_0$ (\us)                 \\ 
	\            &               & B11 $\sim$ 18meV/at.              \\
        \            &               & above L1$_0$ (\us).               \\
        \            &               & {\it B11 stable (\paw)}        \\
	\            &               & {\it L1$_0$ $\sim$13meV/at. }       \\
        \            &               & {\it above L1$_0$ (\paw). }       \\ \hline
	66.6 & cubic (unknown)       & Cd$_2$Zr-C11$_b$                  \\ \hline
        75 & L1$_2$ \cite{BB} & Cd$_3$Zr-L1$_2$                           \\ 
        \  & D0$_a$ \cite{CdZr.69Aru,PP,CC} & 
      \end{tabular} \\ \hline
    \end{tabular}
  \end{center}
}

\begin{center}
  \vspace{-3mm}
  \begin{figure}
    \epsfig{file=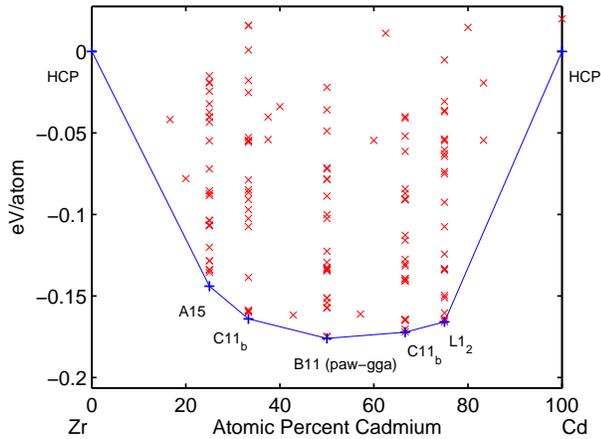,width=\picdim,clip=}
    \vspace{1mm}
    \caption{CdZr (Cadmium - Zirconium) ground state convex hull.}
    \label{label.fig.CdZr}
  \end{figure}
  \vspace{-1mm}
\end{center}

\newpage

{\bf Mo-Nb (Molybdenum - Niobium).}
The system MoNb has not been studied in great detail and no experimental 
intermetallic compounds have been found \cite{BB,PP,ref.Niobium,MoNb.69Rud}.
A bcc solid solution is reported from 2400\DEG$\,$ up to the melt.
We predict four stable compounds at low temperature: 
MoNb$_2$-C11$_b$, MoNb-B2, Mo$_2$Nb-C11$_b$, and Mo$_3$Nb-D0$_3$, 
as shown in figure (\ref{label.fig.MoNb}).

\ReferenceMoNb

{\tablefont
  \begin{center}
    \begin{tabular}{||c||} \hline
      {\bf Mo-Nb system} \\ \hline
      Low Temperature Phases comparison chart \\ \hline
      \begin{tabular}{c|c|c}
	Composition  & Experimental  & \tablelineone                      \\ 
	\% Mo        & (\massalski)  & \tablelinetwo                     \\ \hline
	33.3         &not studied & MoNb$_2$-C11$_b$                     \\ \hline
	50           &not studied & MoNb-B2                              \\ \hline
	66.6         &not studied & Mo$_2$Nb-C11$_b$                     \\ \hline
        75           &not studied & Mo$_3$Nb -D0$_3$                     \\ 
      \end{tabular} \\ \hline
    \end{tabular}
  \end{center}
}

\begin{center}
  \vspace{-3mm}
  \begin{figure}
    \epsfig{file=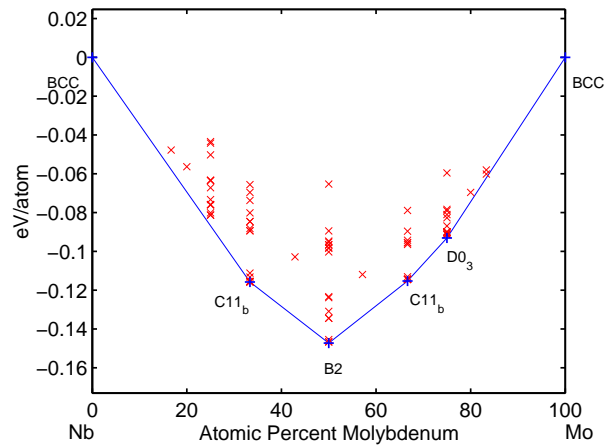,width=\picdim,clip=}
    \vspace{1mm}
    \caption{MoNb (Molybdenum - Niobium) ground state convex hull.}
    \label{label.fig.MoNb}
  \end{figure}
  \vspace{-1mm}
\end{center}

\newpage

{\bf Mo-Pd (Molybdenum - Palladium).}
The phase diagram of this alloy is known from experimental investigations and thermodynamic
calculations \cite{BB,PP,ref.Molybdenum,Calphad.90Kau,MoPd.64Mal}.
The only compound is at composition MoPd$_2$ and is listed in Massalski
as having approximatively the MoPt$_2$ structure \cite{BB,PP,MoPd.64Mal}.
Our {\it ab initio} method finds that the MoPt$_2$ structure is 
8meV/atom higher than MoPd$_2$-ZrSi$_2$.
In addition, we find a stable phase MoPd$_4$-D1$_a$ (MoNi$_4$ prototype), previously unknown.

{\tablefont
  \begin{center}
    \begin{tabular}{||c||} \hline
      {\bf Mo-Pd system} \\ \hline
      Low Temperature Phases comparison chart \\ \hline
      \begin{tabular}{c|c|c}
	Composition  & Experimental  & \tablelineone                      \\ 
	\% Pd        & (\massalski)  & \tablelinetwo                     \\ \hline
	66 to 67     &$\sim$ MoPt$_2$& MoPd$_2$-ZrSi$_2$                 \\ 
	\            &               & MoPt$_2$ $\sim$ 8meV/at.          \\
	\            &               & ZrSi$_2$                          \\ \hline
        80           &disorder fcc   & MoPd$_4$-D1$_a$                   \\ 
	\            &   Pd-A1 &                                         \\ 
      \end{tabular} \\ \hline
    \end{tabular}
  \end{center}
}

\begin{center}
  \vspace{-3mm}
  \begin{figure}
    \epsfig{file=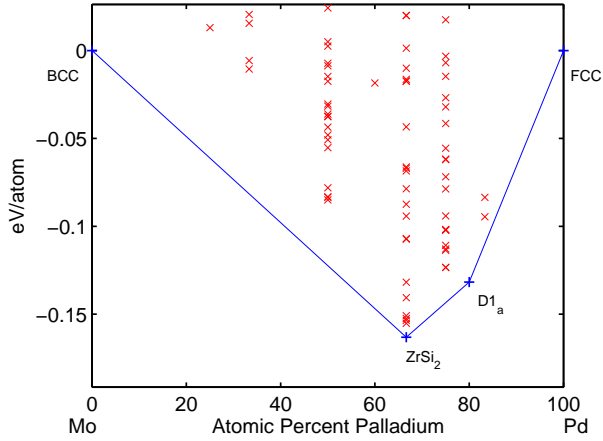,width=\picdim,clip=}
    \vspace{1mm}
    \caption{MoPd (Molybdenum - Palladium) ground state convex hull.}
    \label{label.fig.MoPd}
  \end{figure}
  \vspace{-1mm}
\end{center}

\newpage

{\bf Mo-Pt (Molybdenum - Platinum).}
The phase diagram of this alloy is known from experimental investigations and thermodynamic modeling \
\cite{BB,PP,MoPd.64Mal,MoPt.54Rau,MoPt.56Sch,MoPt.59Kap,MoPt.63Sch,MoPt.64Roo,MoPt.68Ock,MoPt.72Flu,MoPt.73Flu,MoPt.74Flu,PdTi.65Dwi}.
We confirm the stability of MoPt-B19 and MoPt$_2$ (with MoPt$_2$ prototype).
In the Pt-rich part of the phase diagram, we find a compound 
MoPt$_4$-D1$_a$ (MoNi$_a$ prototype), previously unknown.
The observed high-temperature phase A15, 
reported as Mo$_6$Pt in Massalski \cite{BB} and as Mo$_{3.2}$Pt$_{0.8}$ in the Pauling File \cite{PP}, 
is off-stoichiometry.
Our {\it ab initio} method finds a stoichiometric Mo$_3$Pt-A15 to be
$\sim$45meV/atom above the two-phase region Mo\twophase MoPt-B19.
For a detailed experimental study of such A15 phase, see reference \cite{MoPt.74Flu}.

{\tablefont
  \begin{center}
    \begin{tabular}{||c||} \hline
      {\bf Mo-Pt system} \\ \hline
      Low Temperature Phases comparison chart \\ \hline
      \begin{tabular}{c|c|c}
	Composition  & Experimental  & \tablelineone                      \\ 
	\% Pt        & (\massalski)  & \tablelinetwo                     \\ \hline
	18.5   & Mo$_6$Pt-A15 \cite{BB} & Mo$_3$Pt-A15                   \\
        20     & Mo$_{3.2}$Pt$_{0.8}$-A15 \cite{PP} &$\sim$45meV/at. above   \\
               & above $\sim$1200\DEG & Mo\twophase MoPt-B19             \\ \hline
	31.5 to 45 &$\epsilon'$-D0$_{19}$ & unavailable                   \\ 
        \            &above 1000$^\circ$C & at such composition          \\ \hline
	50           & B19           & MoPt-B19                          \\ \hline
	66.6         & MoPt$_2$      & MoPt$_2$                          \\ \hline
        80           &two-phase region& MoPt$_4$-D1$_a$                  \\
        \        &MoPt$_2$\twophase Pt&                                 \\ 
      \end{tabular} \\ \hline
    \end{tabular}
  \end{center}
}

\begin{center}
  \vspace{-3mm}
  \begin{figure}
    \epsfig{file=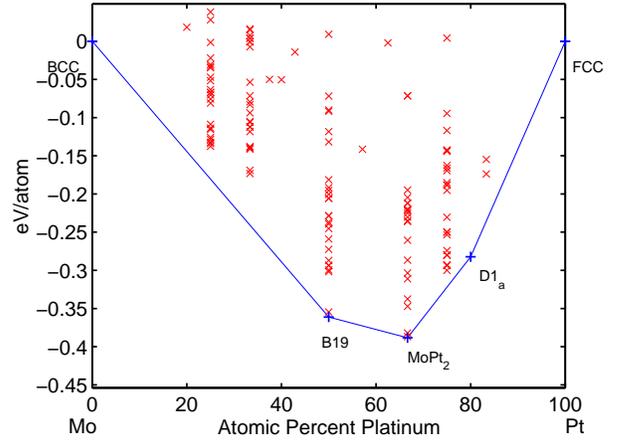,width=\picdim,clip=}
    \vspace{1mm}
    \caption{MoPt (Molybdenum - Platinum) ground state convex hull.}
    \label{label.fig.MoPt}
  \end{figure}
  \vspace{-1mm}
\end{center}

\newpage

{\bf Mo-Rh (Molybdenum - Rhodium).}
The phase diagram of the system Mo-Rh is known above $\sim$900\DEG
and it is based on thermodynamic calculations and experimental results
\cite{BB,PP,Calphad.90Kau,MoRh.58Haw,MoRh.60And,MoRh.66Gie,MoRh.91Gur}.
We confirm the stability of the two known compounds MoRh-B19 \cite{BB} 
and MoRh$_3$-CdMg$_3$ \cite{PP,MoRh.66Gie}.
At concentration 66.6\% Rh, we find the stable phase MoRh$_2$-C37 
(prototype Co$_2$Si), previously not identified \cite{BB}.

{\tablefont
  \begin{center}
    \begin{tabular}{||c||} \hline
      {\bf Mo-Rh system} \\ \hline
      Low Temperature Phases comparison chart \\ \hline
      \begin{tabular}{c|c|c}
	Composition  & Experimental  & \tablelineone                      \\ 
	\% Rh        & (\massalski)  & \tablelinetwo                     \\ \hline
	$\sim$50     & B19           & MoRh-B19                          \\ \hline
	66.6      & two-phase region & Mo$_2$Rh-C37                      \\ 
        \            & above 900$^\circ$C &                              \\ \hline
	$\sim$75     & CdMg$_3$ \cite{PP,MoRh.66Gie} & MoRh$_3$-CdMg$_3$ \\
      \end{tabular} \\ \hline
    \end{tabular}
  \end{center}
}

\begin{center}
  \vspace{-3mm}
  \begin{figure}
    \epsfig{file=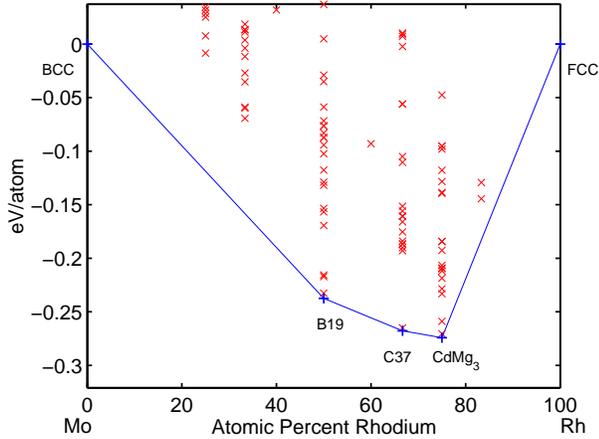,width=\picdim,clip=}
    \vspace{1mm}
    \caption{MoRh (Molybdenum - Rhodium) ground state convex hull.}
    \label{label.fig.MoRh}
  \end{figure}
  \vspace{-1mm}
\end{center}

\newpage

{\bf Mo-Ru (Molybdenum - Ruthenium).}
The phase diagram of this system is known at medium and high-temperature
\cite{BB,PP,MoRu.88Kle}. There is a $\sigma$ phase around 38\% Ru, which
is quite common when bcc (Mo) and hcp (Ru) elements are mixed. We are not able to
confirm the $\sigma$ phase since we do not have the proper prototype in our library.
At concentration 75\% Ru, 
we find the stable phase MoRu$_3$-D0$_{19}$ (formation energy $\sim$ 60meV/atom), 
not present in Massalski \cite{BB}, indicating that it might be a low-temperature ordered structure.

\ReferenceMoRu

{\tablefont
  \begin{center}
    \begin{tabular}{||c||} \hline
      {\bf Mo-Ru system} \\ \hline
      Low Temperature Phases comparison chart \\ \hline
      \begin{tabular}{c|c|c}
	Composition  & Experimental  & \tablelineone                      \\ 
	\% Ru        & (\massalski)  & \tablelinetwo                     \\ \hline
        75           & disorder hcp  & MoRu$_3$-D0$_{19}$                \\
        \            & Ru-A3 above 800$^\circ$C &                        \\
      \end{tabular} \\ \hline
    \end{tabular}
  \end{center}
}

\begin{center}
  \vspace{-3mm}
  \begin{figure}
    \epsfig{file=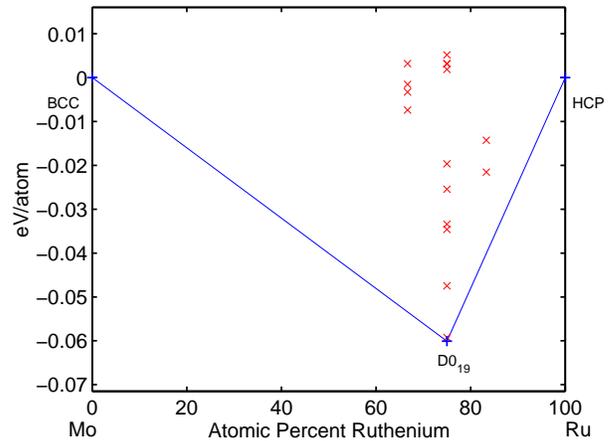,width=\picdim,clip=}
    \vspace{1mm}
    \caption{MoRu (Molybdenum - Ruthenium) ground state convex hull.}
    \label{label.fig.MoRu}
  \end{figure}
  \vspace{-1mm}
\end{center}

\newpage

{\bf Mo-Ti (Molybdenum - Titanium).}
To our knowledge, no intermetallic compounds have been 
reported for the system Mo-Ti \cite{BB,PP,ref.Titanium,MoTi.72Ron,MoTi.75Bag,MoTi.88Mof},
and it is considered a non-compound forming system \cite{PP,CC}.
In contrast to the experimental assessment, we find seven stable compounds.
Four of these have bcc superstructures:
orthorhombic MoTi$_3$, degenerate with the two-phase region Ti\twophase MoTi$_2$,
with space group Immm \#71 and prototype MoTi$_3^{proto}$ (Appendix (\ref{proto.AB3.MoTi81}));
orthorhombic Mo$_3$Ti with space group Immm \#71 and prototype Mo$_3$Ti$^{proto}$
(Appendix (\ref{proto.AB3.MoTi81}));
trigonal MoTi$_2$-BCC$_{AB2}^{[211]}$, and
orthorhombic MoTi with space group Imma \#74 and prototype 
MoTi$^{proto}$ (Appendix (\ref{proto.AB.MoTi80})).
In addition, we find a stable compound Mo$_2$Ti-C11$_b$,
a compound Mo$_4$Ti-D1$_a$ degenerate with the tie-line, 
and finally a monoclinic Mo$_5$Ti which has space group C2/m \#12 
and prototype Mo$_5$Ti$^{proto}$ 
(Appendix (\ref{proto.AB5.MoTi128relaxed})).
Given the relatively large value of the formation energy, 
the experimental miscibility gap is surprising. 
This system warrants further study to sort out the apparent 
discrepancy  between experiments and {\it ab initio} computation.

\ReferenceMoTi

{\tablefont
  \begin{center}
    \begin{tabular}{||c||} \hline
      {\bf Mo-Ti system} \\ \hline
      Low Temperature Phases comparison chart \\ \hline
      \begin{tabular}{c|c|c}
	Composition  & Experimental  & \tablelineone                      \\ 
	\% Mo        & (\massalski)  & \tablelinetwo                     \\ \hline
	25           &two-phase region, & MoTi$_3^{proto}$/tie-line      \\
	\ &$(\beta$Ti,Mo)-A2$\leftrightarrow(\alpha$Ti)-A3,& Appendix (\ref{proto.AB3.MoTi81})\\
	\            &above 400$^\circ$C  &                              \\ \hline
	33.3         &same as above         & MoTi$_2$-BCC$_{AB2}^{[211]}$ \\  \hline
 	50           &same as above               & MoTi$^{proto}$       \\ 
	\            &                 & Appendix (\ref{proto.AB.MoTi80})\\ \hline
	66.6         &same as above      & Mo$_2$Ti-C11$_b$              \\ \hline
        75           &same as above      & Mo$_3$Ti$^{proto}$            \\
	\            &                 & Appendix (\ref{proto.AB3.MoTi81})\\ \hline
	80           &same as above     & Mo$_4$Ti-D1$_a$/tie-line       \\ \hline
        83.3         &same as above     & Mo$_5$Ti$^{proto}$             \\
	\            &      &  Appendix (\ref{proto.AB5.MoTi128relaxed}) \\ 
      \end{tabular} \\ \hline
    \end{tabular}
  \end{center}
}

\begin{center}
  \vspace{-3mm}
  \begin{figure}
    \epsfig{file=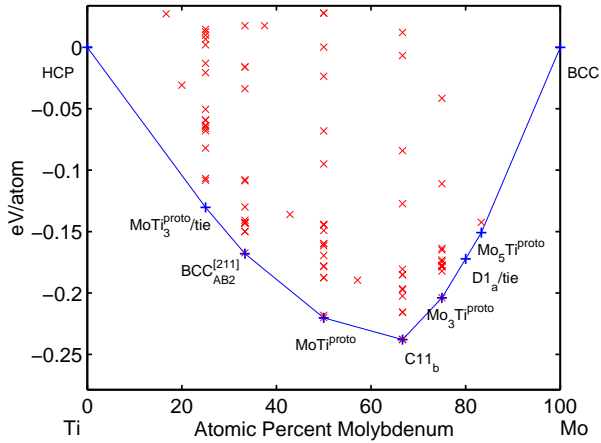,width=\picdim,clip=}
    \vspace{1mm}
    \caption{MoTi (Molybdenum - Titanium) ground state convex hull.}
    \label{label.fig.MoTi}
  \end{figure}
  \vspace{-1mm}
\end{center}

\newpage

{\bf Mo-Zr (Molybdenum - Zirconium).}
Not much is known of the system Mo-Zr to produce a reliable 
phase diagram \cite{BB,PP,ref.Zirconium}.
We confirm the stability of the only known compound Mo$_2$Zr-C15.
Experimentally, the Zr-rich side of the phase diagram has not been 
explored in detail at low temperature, and it is reported to have a 
two-phase region Mo$_2$Zr-C15\twophase ($\alpha$Zr)-A3 above 400$^\circ$C.
However, we find two possible new compounds:
the orthorhombic phase MoZr$_3$ with space group Imma \#74, prototype MoZr$_3^{proto}$ (Appendix (\ref{proto.AB3.MoZr124relaxed})), 
and the monoclinic compound MoZr$_5$-Mo$_5$Ti$^{proto}$ with space group C2/m \#12, and prototype Mo$_5$Ti$^{proto}$
(Appendix (\ref{proto.AB5.MoZr132relaxed})).
Both compounds MoZr$_3^{proto}$ and MoZr$_5$-Mo$_5$Ti$^{proto}$ are quasi-degenerate 
with respect to the two-phase region Mo$_2$Zr\twophase ($\alpha$Zr).
Hence, the existence of MoZr$_3^{proto}$ and MoZr$_5$-Mo$_5$Ti$^{proto}$ is uncertain.
It is useful mentioning that we find a metastable orthorhombic phase MoZr 
with space group Imma \#74 and prototype similar to MoTi$^{proto}$
(Appendix (\ref{proto.AB.MoTi80})) only $\sim$2meV/atom above 
the tie-line Mo$_2$Zr\twophase ($\alpha$Zr).

\ReferenceMoZr

{\tablefont
  \begin{center}
    \begin{tabular}{||c||} \hline
      {\bf Mo-Zr system} \\ \hline
      Low Temperature Phases comparison chart \\ \hline
      \begin{tabular}{c|c|c}
	Composition  & Experimental  & \tablelineone                      \\ 
	\% Mo        & (\massalski)  & \tablelinetwo                     \\ \hline
	16.6         & two-phase region & MoZr$_5$-Mo$_5$Ti$^{proto}$/tie-line      \\
	\            & above 400$^\circ$C & Appendix (\ref{proto.AB5.MoZr132relaxed}) \\ \hline
	25           &same as above  & MoZr$_3^{proto}$/tie-line         \\
	\            &               & Appendix (\ref{proto.AB3.MoZr124relaxed}) \\ \hline
	50           &same as above  & $\sim$MoTi$^{proto}$ $\sim$2meV/at. \\
        \            &          & above Mo$_2$Zr\twophase ($\alpha$Zr) \\ \hline
	60-67        & C15           & Mo$_2$Zr-C15                      \\ 
      \end{tabular} \\ \hline
    \end{tabular}
  \end{center}
}

\begin{center}
  \vspace{-3mm}
  \begin{figure}
    \epsfig{file=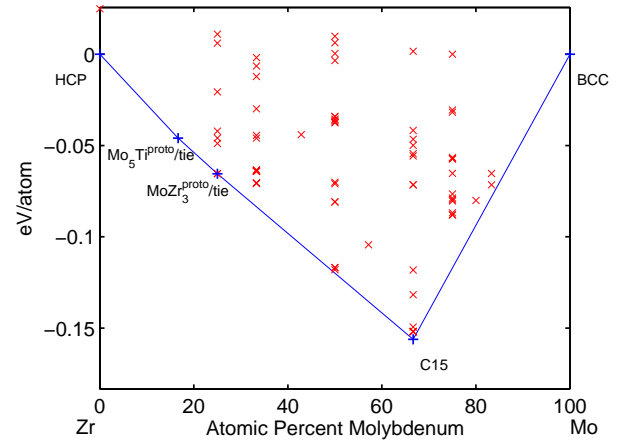,width=\picdim,clip=}
    \vspace{1mm}
    \caption{MoZr (Molybdenum - Zirconium) ground state convex hull.}
    \label{label.fig.MoZr}
  \end{figure}
  \vspace{-1mm}
\end{center}


\newpage

{\bf Nb-Pd (Niobium - Palladium).}
The Pd-Nb phase diagram is known with reasonable accuracy in the Pd-rich 
region \cite{BB,PP,ref.Niobium,PdNb.80GIE}. 
The stability of the experimental phase NbPd$_2$-MoPt$_2$ and $\alpha$NbPd$_3$-D0$_{22}$ is confirmed.
In the Nb-rich region, at concentration 33\% Nb, we find
an orthorhombic compound Nb$_2$Pd-BCC$_{AB2}^{[011]}$, 
not present in Massalski \cite{BB}.
The energy of this phase is $\sim$11meV/atom below the tie-line of the two-phase region Nb\twophase NbPd$_2$.

\ReferenceNbPd

{\tablefont
  \begin{center}
    \begin{tabular}{||c||} \hline
      {\bf Nb-Pd system} \\ \hline
      Low Temperature Phases comparison chart \\ \hline
      \begin{tabular}{c|c|c}
	Composition  & Experimental  & \tablelineone                      \\ 
	\% Pd        & (\massalski)  & \tablelinetwo                     \\ \hline
	33.3         &two-phase region& Nb$_2$Pd-BCC$_{AB2}^{[011]}$     \\
	\            &above 700$^\circ$C  & $\sim$11meV/at. below        \\ 
	\            &               & Nb\twophase NbPd$_2$             \\ \hline
	66.6         & MoPt$_2$      & NbPd$_2$-MoPt$_2$                 \\ \hline
        75      & $\alpha$-D0$_{22}$ & NbPd$_3$-D0$_{22}$                \\ 
      \end{tabular} \\ \hline
    \end{tabular}
  \end{center}
}

\begin{center}
  \vspace{-3mm}
  \begin{figure}
    \epsfig{file=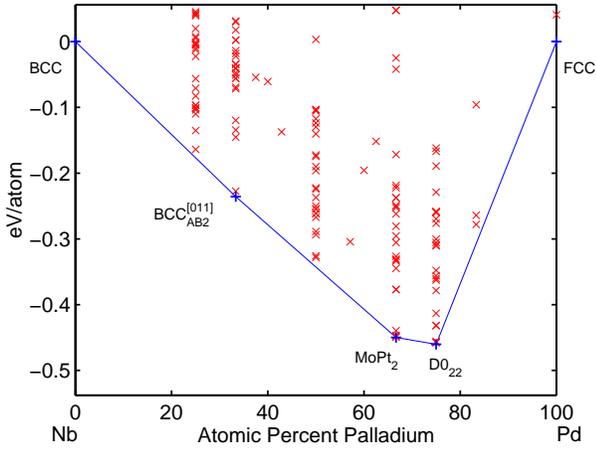,width=\picdim,clip=}
    \vspace{1mm}
    \caption{NbPd (Niobium - Palladium) ground state convex hull.}
    \label{label.fig.NbPd}
  \end{figure}
  \vspace{-1mm}
\end{center}

\newpage

{\bf Nb-Pt (Niobium - Platinum).}
Several intermetallic compounds have been reported for the system Nb-Pt
\cite{BB,PP,ref.Niobium,NbPt.61Mat,NbPt.64Gie,NbPt.64Mal,NbPt.65Gie,NbPt.85Wat}.
We confirm the experimental phases Nb$_3$Pt-A15 and $\alpha$NbPt$_3$-D0$_a$
($\beta$Cu$_3$Ti prototype).
The prototype of NbPt$_2$ is not reported in Massalski \cite{BB} 
(orthorhombic oI6, with space group Immm).
We find NbPt$_2$-MoPt$_2$, in agreement with references \cite{PP,NbPt.64Mal,NbPt.65Gie}.
At 50\% concentration we do not confirm the stability of NbPt-B19. 
Instead of B19 we find NbPt-L1$_0$, and B19 
to be higher by $\sim$11meV/atom above L1$_0$.
It is possible that L1$_0$ is a ground state and B19 a high-temperature state.
We can not say anything about the $\sigma$-phase D8$_b$ since we do 
not have such prototype in our library.
{\it We further investigated NbPt-L1$_0$ and NbPt-B19 with PAW-GGA potentials, 
\PawSection. 
With PAW,
L1$_0$ is the most stable compound and 
B19 is $\sim$18.5meV/atom higher than L1$_0$.}

{\tablefont
  \begin{center}
    \begin{tabular}{||c||} \hline
      {\bf Nb-Pt system} \\ \hline
      Low Temperature Phases comparison chart \\ \hline
      \begin{tabular}{c|c|c}
	Composition  & Experimental  & \tablelineone                   \\ 
	\% Pt        & (\massalski)  & \tablelinetwo                  \\ \hline
	19 to 28     &A15            & Nb$_3$Pt-A15                   \\ \hline
	31 to 38     &D8$_b$         & unavailable                    \\
        \            &               & $\sigma$-phase                 \\ \hline
	$\sim$50 &Nb$_{1-x}$Pt$_{1+x}$-B19 & NbPt-L1$_0$              \\
        \            &               & B19 $\sim$ 11meV/at.           \\
        \            &               & above L1$_0$ (\us)              \\ 
	\	     &		     & {\it L1$_0$ stable (\paw)      }\\ 
	\	     &		     & {\it B19 $\sim$ 18.5meV/at.   }\\ 
        \            &               & {\it above L1$_0$ (\paw)       }\\ \hline
	$\sim$67 & MoPt$_2$ \cite{PP,NbPt.64Mal,NbPt.65Gie}& NbPt$_2$-MoPt$_2$  \\ \hline
        $\sim$76     &D0$_a$    & NbPt$_3$-D0$_a$           \\ 
      \end{tabular} \\ \hline
    \end{tabular}
  \end{center}
}

\begin{center}
  \vspace{-3mm}
  \begin{figure}
    \epsfig{file=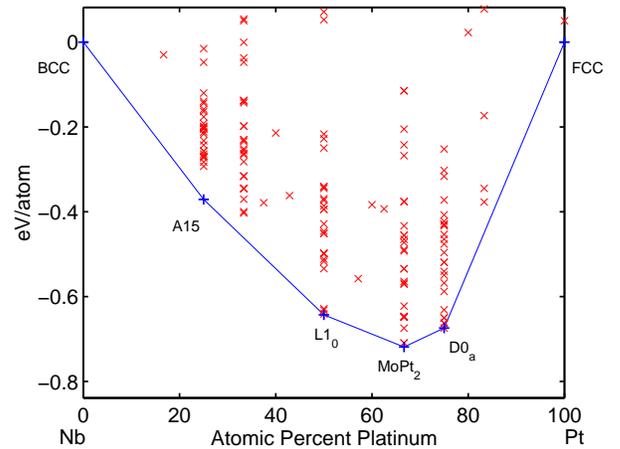,width=\picdim,clip=}
    \vspace{1mm}
    \caption{NbPt (Niobium - Platinum) ground state convex hull.}
    \label{label.fig.NbPt}
  \end{figure}
  \vspace{-1mm}
\end{center}

\newpage

{\bf Nb-Rh (Niobium - Rhodium).}
The system Nb-Rh is poorly characterized in the range of concentration 
50$\sim$80\% Rh \cite{BB,PP,ref.Niobium,NbPt.64Gie,NbRh.64Rit,NbRh.64Rit2,Calphad.90Kau}.
We confirm the stability of Nb$_3$Rh-A15, NbRh-L1$_0$.
NbRh-B19, which is observed as a high-temperature state, is $\sim$27meV/atom above L1$_0$,
and it is possibly stabilized by entropic effects.
Note: in Massalski and in the Pauling File \cite{BB,PP} 
there is no phase at 50\% composition,
and Nb$_{0.96}$Rh$_{1.04}$-L1$_0$ appears off-stoichiometry.
At concentration 75\% Rh, we find the stable phase $\eta$-Al$_3$Pu (Co$_3$V) and
$\kappa$NbRh$_3$-L1$_2$ to be higher by 8meV/atom.
Hence, we think that $\eta$ prevails at low-temperature over 
$\kappa$, in contrast with the sketched phase diagram of reference \cite{BB}.
We can not say anything about D8$_b$ and $\xi$(Nb$_{2}$Rh$_{3}$)
since we do not have the $\sigma$-phase D8$_b$ and any A$_{2}$B$_{3}$ 
prototypes in our library.
{\it To address the structures with similar energy,
we further investigate NbRh$_3$ with PAW-GGA potentials, 
\PawSection. 
With PAW,
NbRh$_3$-Al$_3$Pu is the most stable compound and 
NbRh$_3$-L1$_2$ is higher by 5.3meV/atom.} 
{The disagreement at composition NbRh$_{3}$ is further discussed in Section (\ref{section.discussion}).}

\ReferenceNbRh

{\tablefont
  \begin{center}
    \begin{tabular}{||c||} \hline
      {\bf Nb-Rh system} \\ \hline
      Low Temperature Phases comparison chart \\ \hline
      \begin{tabular}{c|c|c}
	Composition  & Experimental  & \tablelineone                   \\ 
	\% Rh        & (\massalski)  & \tablelinetwo                  \\ \hline
	25 & $\alpha'$(Nb$_3$Rh)-A15                   &Nb$_3$Rh-A15  \\ \hline
	28.5 to 39.5 & $\sigma$(Nb$_{13}$Rh$_{7}$)-D8$_b$ &unavailable\\ \hline
	$\sim$51.5 to 52& Nb$_{0.96}$Rh$_{1.04}$-L1$_0$ &NbRh-L1$_0$  \\
	\            & ($\lesssim 1337$\DEG)          &                \\ \hline
	$\sim$56 to 62& $\epsilon$(Nb$_{2}$Rh$_{3}$)-B19   &NbRh-B19  \\ 
	\            & (high-temperature             & $\sim$27meV/at.          \\
        \            & $\gtrsim 1330$\DEG)         &above L1$_0$ (\us)    \\ \hline
	59 to $\sim$64&$\xi$(Nb$_{2}$Rh$_{3}$)-Nb$_{2}$Rh$_{3}$ &unavailable\\ \hline
	$\sim$67 to 70 & $\eta$(Nb$_{13}$Rh$_{27}$)-Al$_3$Pu &see below \\
	\            & ($\equiv$ Co$_3$V in \cite{BB})   & for Al$_3$Pu \\ \hline
	$\sim$71 to 79&$\kappa$(NbRh$_{3}$)-L1$_2$ &NbRh$_{3}$-Al$_3$Pu (\us)\\
        \            &                          & L1$_2$$\sim$8meV/at. \\
        \            &                          & above Al$_3$Pu (\us)  \\ 
        \            &                          & {\it Al$_3$Pu stable (\paw) } \\ 
        \            &                          & {\it L1$_2$$\sim$5.3meV/at.} \\
        \            &                          & {\it above Al$_3$Pu (\paw)  } \\
	\            &               & See Section (\ref{section.discussion}).\\ 
      \end{tabular} \\ \hline
    \end{tabular}
  \end{center}
}

\begin{center}
  \vspace{-3mm}
  \begin{figure}
    \epsfig{file=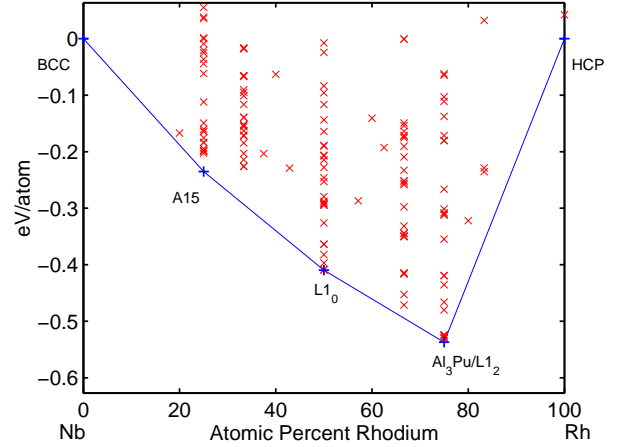,width=\picdim,clip=}
    \vspace{1mm}
    \caption{NbRh (Niobium - Rhodium) ground state convex hull.}
    \label{label.fig.NbRh}
  \end{figure}
  \vspace{-1mm}
\end{center}

\newpage

{\bf Nb-Ru (Niobium - Ruthenium).}
Very little is known for the alloy Nb-Ru, especially at low-temperature 
\cite{BB,PP,ref.Niobium,NbRu.63Ben,NbRu.63Rau,NbRu.64Hur,NbRu.89Tsu,NbRu.89Che}. 
We do not confirm the stability of any low-temperature compound
at composition NbRu, but instead find a two-phase field between Nb$_3$Ru and NbRu$_2$,
even though a high and low-temperature phase of NbRu has been observed \cite{NbRu.63Ben,NbRu.89Che}.
In the Ru-rich side of the phase diagram, we find NbRu$_3$-D0$_{24}$ to
be 8meV/atom lower than L1$_2$ which is suggested experimentally.
At 66\% Ru we find  NbRu$_2$-C37 (with oP12-Co$_2$Si prototype).
With respect to the two-phase field Nb$_3$Ru\twophase NbRu$_2$, 
the structures closest to the tie-line at NbRu composition are
B19 ($\sim$13meV/atom), L1$_0$ ($\sim$20meV/atom), 
B27 ($\sim$23meV/atom), B33 ($\sim$39meV/atom), and B2 ($\sim$45meV/atom).
Experiments have not found any other stable compound. 
However, in the Nb-rich side of the phase diagram, we find Nb$_3$Ru-D0$_3$ 
and a monoclinic Nb$_5$Ru with C2/m \#12 space group and prototype Mo$_5$Ti$^{proto}$.
The structure Nb$_5$Ru-Mo$_5$Ti$^{proto}$ is described in Appendix 
(\ref{proto.AB5.NbRu132relaxed}).
{\it {To address the disagreements at compositions NbRu and NbRu$_3$
we further investigate the relevant compounds with PAW-GGA potentials, 
\PawSection. 
For composition NbRu, the PAW-GGA result is substantially different from 
the US-LDA result. GGA gives L1$_0$ only 4meV/atom above the tie-line 
(versus 20meV/atom in LDA), 
whereas B19 was difficult to converge numerically, and 
seems to be $>$100meV/atom above the tie-line.
For composition NbRu$_3$, also with PAW, 
D0$_{24}$ is the most stable compound and 
L1$_2$ is higher by 2.5meV/atom. 
Because this number is so small, 
extremely large {\bf k}-points sets and high energy cutoff 
were used to converge it.}} 
{The disagreements at compositions NbRu and NbRu$_3$ are further discussed in Section (\ref{section.discussion}).}

\ReferenceNbRu

{\tablefont
  \begin{center}
    \begin{tabular}{||c||} \hline
      {\bf Nb-Ru system} \\ \hline
      Low Temperature Phases comparison chart \\ \hline
      \begin{tabular}{c|c|c}
	Composition  & Experimental  & \tablelineone                     \\ 
	\% Ru        & (\massalski)  & \tablelinetwo                     \\ \hline
	16.6         & disorder Nb-A2& Nb$_5$Ru-Mo$_5$Ti$^{proto}$       \\
        \            &               & Appendix (\ref{proto.AB5.NbRu132relaxed})  \\ \hline
	25           & disorder Nb-A2& Nb$_3$Ru-D0$_3$                   \\ \hline
	$\sim$50     & NbRu$'$-L1$_0$& two-phase region                  \\
        \            & (room-temperature)& Nb$_3$Ru\twophase NbRu$_2$.   \\ 
	\            &               & {\it L1$_0\sim$4meV/at. (\paw) }         \\
	\            &               & {\it B19 $>$100meV/at.  }            \\
	\	     &		     & {\it above tie-line (\paw).}  \\ \hline
        $\sim$50     & NbRu-B2       & B19$\sim$13meV/at. (\us)          \\
        \            & (high-temperature)& L1$_0$$\sim$20meV/at.         \\
        \            &               & B2 $\sim$45meV/at.                \\
	\	     &               & above the tie-line                \\
	\            &               & See Section (\ref{section.discussion}). \\ \hline
	66.6         & two-phase region  & NbRu$_2$-C37                  \\
	\ &above $\sim$700$^\circ$C  &                                   \\ \hline
 	75           & L1$_2$        & NbRu$_3$-D0$_{24}$ (\us)          \\
	\            &               & L1$_2$$\sim$8meV/at.              \\
	\	     &		     & above D0$_{24}$ (\us)             \\
	\            &               & {\it D0$_{24}$ stable (\paw) }    \\
	\            &               & {\it L1$_2$$\sim$2.5meV/at.    }    \\
	\	     &		     & {\it above D0$_{24}$ (\paw)  }    \\	
	\            &               & See Section (\ref{section.discussion}). \\ 
      \end{tabular} \\ \hline
    \end{tabular}
  \end{center}
}

\begin{center}
  \vspace{-3mm}
  \begin{figure}
    \epsfig{file=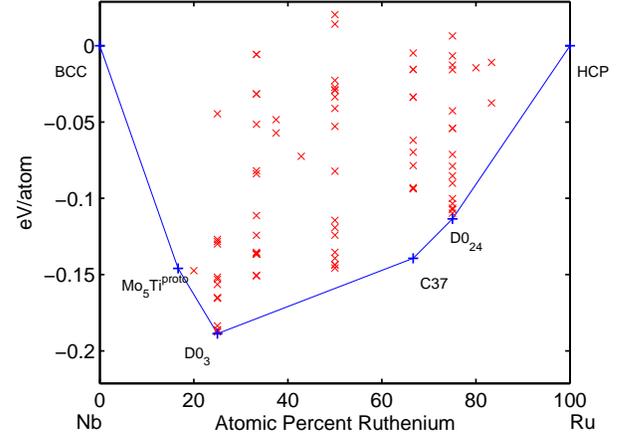,width=\picdim,clip=}
    \vspace{1mm}
    \caption{NbRu (Niobium - Ruthenium) ground state convex hull.}
    \label{label.fig.NbRu}
  \end{figure}
  \vspace{-1mm}
\end{center}

\newpage

{\bf Nb-Tc (Niobium - Technetium).}
The phase diagram of the system Nb-Tc is not known and only one 
intermetallic compound, NbTc$_3$ (metallic and superconductor)
has been reported with $\alpha$Mn structure 
\cite{BB,PP,ref.Niobium,NbTc.61Com,NbTc.63Van,NbTc.70Gio,Tc.62Dar}.
The compound NbTc$_3$ is classified as phase Nb$_{0.15}$Tc$_{0.85}$ in the Pauling File \cite{PP}.
We do not have $\alpha$Mn in our library of prototypes, and, 
at such stoichiometry, we do not find any stable compound.
In the Tc-rich side of the phase diagram, we find a two-phase 
field NbTc\twophase Tc, as shown in figure (\ref{label.fig.NbTc}). 
At 50\% concentration, we obtain NbTc-B2, and in the Nb-rich side of the diagram, 
we find Nb$_2$Tc-C11$_b$ and an orthorhombic phase Nb$_3$Tc 
with space group Immm \#71, bcc superstructure, and prototype 
Nb$_3$Tc$^{proto}$ described in Appendix (\ref{proto.AB3.NbTc81}).
\TcDO19

{\tablefont
  \begin{center}
    \begin{tabular}{||c||} \hline
      {\bf Nb-Tc system} \\ \hline
      Low Temperature Phases comparison chart \\ \hline
      \begin{tabular}{c|c|c}
	Composition  & Experimental  & \tablelineone                      \\ 
	\% Tc        & (\massalski)  & \tablelinetwo                     \\ \hline
	25           & not studied,  & Nb$_3$Tc$^{proto}$                \\
        \            & unknown       & Appendix (\ref{proto.AB3.NbTc81}) \\ \hline
	33.3         & same as above & Nb$_2$Tc-C11$_b$                  \\ \hline
	50           & same as above & NbTc-B2                           \\ \hline
        75     & NbTc$_3$-$\alpha$Mn \cite{BB} & nothing stable          \\
	85     & Nb$_{0.15}$Tc$_{0.85}$-$\alpha$Mn \cite{PP} &           \\
      \end{tabular} \\ \hline
    \end{tabular}
  \end{center}
}

\begin{center}
  \vspace{-3mm}
  \begin{figure}
    \epsfig{file=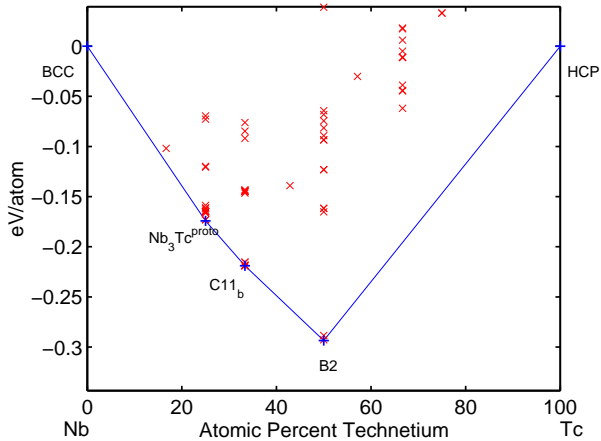,width=\picdim,clip=}
    \vspace{1mm}
    \caption{NbTc (Niobium - Technetium) ground state convex hull.}
    \label{label.fig.NbTc}
  \end{figure}
  \vspace{-1mm}
\end{center}

\newpage

{\bf Pd-Pt (Palladium - Platinum).}
The low temperature part of the phase diagram is believed to have
a miscibility gap at a temperature of about 770\DEG.
This miscibility gap is predicted on the basis of the difference of melting points between 
Pd an Pt \cite{BB,PP,ref.Moffatt,PdPt.59Rau,PdPt.73Haw}. 
Instead of such gap, we find three unknown stable compounds 
with fcc superstructures (Pd and Pt are both fcc). 
We find PdPt-L1$_1$, and two orthorhombic phases, Pd$_3$Pt and PdPt$_3$, with 
space group Cmmm \#65 and prototypes described in Appendix (\ref{proto.AB3.PdPt11relaxed}).
The prototypes are labeled as Pd$_3$Pt$^{proto}$ and PdPt$_3^{proto}$.
The compound Pd$_3$Pt is degenerate with respect the two-phase fields Pd-A1\twophase PdPt-L1$_1$,
therefore its existence is uncertain.
As shown in figure (\ref{label.fig.PdPt}), all the stable phases have 
small formation energy ($<$ 50meV/atom) making them difficult to determine experimentally.
However, we believe the experimental phase diagram to be in error.
{\it We further investigate PdPt with PAW-GGA potentials, 
\PawSection. 
With PAW,
PdPt-L1$_1$ is the most stable compound and 
PdPt-L1$_0$ is higher by 5.5meV/atom.} 
Our results are in disagreement with previous FLAPW-LDA calculations,
where L1$_0$ is found to be the most stable compound \cite{PdPt.91Zun}.

\ReferencePdPt

{\tablefont
  \begin{center}
    \begin{tabular}{||c||} \hline
      {\bf Pd-Pt system} \\ \hline
      Low Temperature Phases comparison chart \\ \hline
      \begin{tabular}{c|c|c}
	Composition  & Experimental  & \tablelineone                      \\ 
	\% Pt        & (\massalski)  & \tablelinetwo                     \\ \hline
	25           & predicted     & Pd$_3$Pt$^{proto}$/tie-line       \\ 
	\           &two-phase region& (uncertain)                       \\ 
        \            &Pd-A1\twophase Pt-A1  & Appendix (\ref{proto.AB3.PdPt11relaxed})  \\ \hline
	50           & same as above & PdPt-L1$_1$                       \\
        \            &               & L1$_0$$\sim$11meV/at.             \\
        \            &               & higher than L1$_1$ (\us)           \\
        \            &               & {\it L1$_1$ stable            }   \\
        \            &               & {\it L1$_0$$\sim$5.5meV/at.    }   \\
        \            &               & {\it higher than L1$_1$ (\paw) }   \\ \hline
	75           & same as above & PdPt$_3^{proto}$                  \\
	\            &               & Appendix (\ref{proto.AB3.PdPt11relaxed})  \\ 
      \end{tabular} \\ \hline
    \end{tabular}
  \end{center}
}

\begin{center}
  \vspace{-3mm}
  \begin{figure}
    \epsfig{file=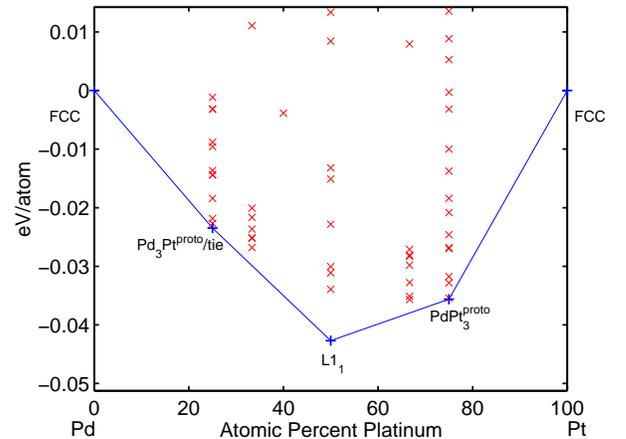,width=\picdim,clip=}
    \vspace{1mm}
    \caption{PdPt (Palladium - Platinum) ground state convex hull.}
    \label{label.fig.PdPt}
  \end{figure}
  \vspace{-1mm}
\end{center}

\newpage

{\bf Pd-Tc (Palladium - Technetium).}
The phase diagram for the system Pd-Tc is sketched based on the 
solid solubility data \cite{BB,PP,ref.Moffatt,Tc.62Dar,Tc.63Nie,PdTc.80Hai,RuTc.75Ale}.
Experimental results report two solid solutions 
(fcc Pd-rich) and (hcp Tc-rich) with a two-phase region in between.
No intermetallic compounds have been reported \cite{BB,PP}. 
However, we find one stable phases PdTc$_3$-D0$_{19}$.
In addition, at 50\% concentration, we find a hcp superstructure
(trigonal lattice, hP4, space group P$\bar{3}$m1) which has an energy
$\sim$3meV/atom higher than the tie-line of the 
two-phase region PdTc$_3$\twophase Pd.
\TcDO19

{\tablefont
  \begin{center}
    \begin{tabular}{||c||} \hline
      {\bf Pd-Tc system} \\ \hline
      Low Temperature Phases comparison chart \\ \hline
      \begin{tabular}{c|c|c}
	Composition  & Experimental  & \tablelineone                      \\ 
	\% Pd        & (\massalski)  & \tablelinetwo                     \\ \hline
	25          &two-phase region& PdTc$_3$-D0$_{19}$                \\
        \           & above $\sim$1000\DEG&                              \\
        \           & Pd-A1\twophase Tc-A3&                              \\ \hline
	50          &two-phase region& two-phase region                  \\
        \           & above $\sim$1000\DEG& PdTc$_3$\twophase Pd         \\
        \           & Pd-A1\twophase Tc-A3& hcp superstr. $\sim$3meV/at.           \\ 
	\	    &                & above tie-line     \\ 
      \end{tabular} \\ \hline
    \end{tabular}
  \end{center}
}

\begin{center}
  \vspace{-3mm}
  \begin{figure}
    \epsfig{file=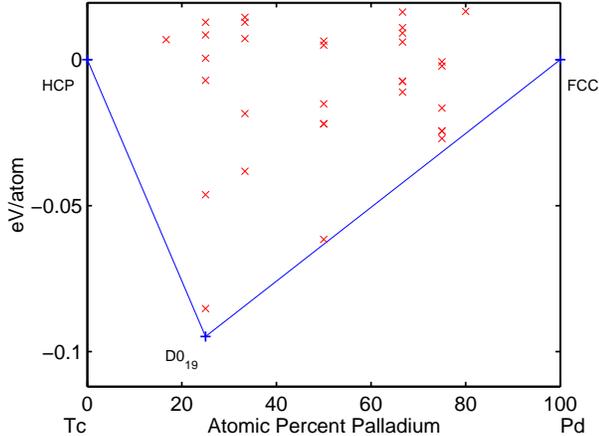,width=\picdim,clip=}
    \vspace{1mm}
    \caption{PdTc (Palladium - Technetium) ground state convex hull.}
    \label{label.fig.PdTc}
  \end{figure}
  \vspace{-1mm}
\end{center}

\newpage

{\bf Pd-Ti (Palladium - Titanium).}
This system has been subject of conflicting results for 
several years \cite{BB,PP,AuTi.89Sha,PdTi.60Rud,PdTi.65Dwi,PdTi.65Ros,PdTi.68Rau,PdTi.68Kra,PdTi.68Sch,PdTi.70Har,PdTi.70Don,PdTi.72Ere,PdTi.76Mie,PdTi.83Siv,PdTi.87Sha,PdTi.89Sel,PdTi.91Pic,PdTi.93Mik,PdTi.95Zha}. 
Our {\it ab initio} method confirms the stability 
of the compounds PdTi$_2$-C11$_b$, PdTi$_3$-A15, and Pd$_3$Ti-D0$_{24}$.
Near 80\% Pd, Long Period Superstructures (LPS) modulations
of L1$_2$ are observed \cite{PdTi.68Rau}. While we do not have
such off-stoichiometric LPS, we find L1$_2$ to be only
6meV/atom above D0$_{24}$ at Pd$_3$Ti composition.
The low energy difference between L1$_2$ and D0$_{24}$ indicates
that it may be easy to form antiphase boundaries and LPS near
this composition.
We find Pd$_2$Ti-MoPt$_2$ which is an orthorhombic distortion of C11$_b$
(MoPt$_2$ is orthorhombic, while C11$_b$ is tetragonal).
We find Pd$_2$Ti-C49 and Pd$_2$Ti-C11$_b$ to be higher by 
3meV/atom and 14meV/atom above MoPt$_2$, respectively.
At low temperature, at concentration 50\%, we find a stable compound $\alpha$TiPd, 
but two prototypes are degenerate: L1$_0$ and B19 
(which is reported experimentally).
We can not find the reported phase PdTi$_4$-A15 
(off-stoichiometry) \cite{BB}, which we think should appear 
inside the two-phase region of Ti $\sim$ PdTi$_3$, or at composition PdTi$_3$.
While off-stoichiometric compounds are obviously 
possible, this usually goes together with significant width of the 
single-phase field.  
Hence, we concluded (maybe erroneously) that the placement 
of A15 at composition  PdTi$_{4}$ in reference \cite{BB} 
is likely a typographical error.
{\it To address the degenerate structures,
we further investigate $\alpha$TiPd with PAW-GGA potentials, 
\PawSection. 
With PAW,
PdTi-B19 is the most stable compound and 
PdTi-L1$_0$ is higher by 10meV/atom.} 

\ReferencePdTi

{\tablefont
  \begin{center}
    \begin{tabular}{||c||} \hline
      {\bf Pd-Ti system} \\ \hline
      Low Temperature Phases comparison chart \\ \hline
      \begin{tabular}{c|c|c}
	Composition  & Experimental  & \tablelineone                      \\ 
	\% Pd        & (\massalski)  & \tablelinetwo                     \\ \hline
	20 and 25    &A15 at 20\%& PdTi$_3$-A15 at 25\%            \\ \hline
	33.3         & C11$_b$       & PdTi$_2$-C11$_b$                  \\ \hline
	47 to 53     &$\alpha$(TiPd)-B19& PdTi-B19/L1$_0$ (\us)           \\ 
        \            &               & {\it B19 stable (\paw)      }            \\ 
        \            &               & {\it L1$_0$$\sim$10meV/at. }            \\
        \            &               & {\it above B19 (\paw)       }            \\ \hline 
	60     &Pd$_3$Ti$_2$ $\sim$ Au$_2$V& unavailable                 \\ \hline
	62.5   &Pd$_5$Ti$_3$ $\sim$ C11$_b$ & unavailable                \\ \hline
	66.6         & orthorhombic & Pd$_2$Ti-MoPt$_2$                  \\ 
        \            & distortion   & (distortion of C11$_b$)        \\
        \            & of C11$_b$    & C49$\sim$3meV/at.                 \\
        \            &               & C11$_b$$\sim$14meV/at.            \\
        \            &               & above MoPt$_2$                    \\ \hline
        75           & D0$_{24}$     & Pd$_3$Ti-D0$_{24}$                \\ 
        \            &               & L1$_2$$\sim$6meV/at.              \\
        \            &               & above D0$_{24}$                   \\ \hline
        80           & L1$_2$        & L1$_2$ metastable                 \\ 
	\            &               & at 75\% (see text)                \\ 
      \end{tabular} \\ \hline
    \end{tabular}
  \end{center}
}

\begin{center}
  \vspace{-3mm}
  \begin{figure}
    \epsfig{file=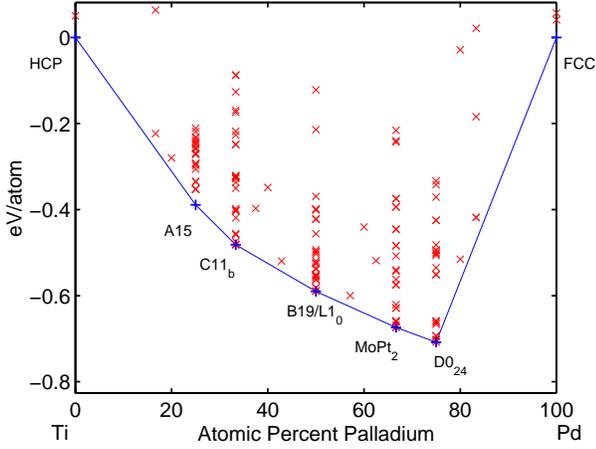,width=\picdim,clip=}
    \vspace{1mm}
    \caption{PdTi (Palladium - Titanium) ground state convex hull.}
    \label{label.fig.PdTi}
  \end{figure}
  \vspace{-1mm}
\end{center}

\newpage

{\bf Pd-Y (Palladium - Yttrium).}
The phase diagram for the system Pd-Y is known with reasonable accuracy.
Several intermetallic compounds have been reported, but not 
all the structures have been determined experimentally 
\cite{BB,PP,PdY.73Loe,PdY.77Ler,PdY.80San,PdY.89Tak}.
We confirm the stability of the compounds Pd$_3$Y-L1$_2$ and PdY$_3$-D0$_{11}$.
The prototype of $\alpha$PdY is not known \cite{PdY.73Loe}. 
Our best guesses are PdY-B27 and PdY-B33 (CrB), 
which are degenerate in the calculation.
We also find PdY$_2$-C37, which occurs in a concentration between 
two known compounds Pd$_2$Y$_3$ and Pd$_2$Y$_5$ and is very close to
the tie-line of the two-phase field PdY$_3$\twophase PdY (5meV/atom)
Hence, if Pd$_2$Y$_3$ and Pd$_2$Y$_5$ were to be included, C37 
would likely not be stable.
We cannot check this prediction since we do not have Pd$_2$Y$_3$ 
and Pd$_2$Y$_5$ prototypes in the set of calculations.
{\it To address the degenerate structures,
we further investigate PdY with PAW-GGA potentials, 
\PawSection. 
With PAW,
PdY-B27 is the most stable compound and 
PdY-B33 is higher by 3.1meV/atom.} 

{\tablefont
  \begin{center}
    \begin{tabular}{||c||} \hline
      {\bf Pd-Y system} \\ \hline
      Low Temperature Phases comparison chart \\ \hline
      \begin{tabular}{c|c|c}
	Composition  & Experimental  & \tablelineone                      \\ 
	\% Pd        & (\massalski)  & \tablelinetwo                     \\ \hline
	25           & D0$_{11}$     & PdY$_3$-D0$_{11}$                 \\ \hline
        28.6     &Pd$_2$Y$_5$ unknown& unavailable                       \\ \hline 
	33.3       &two-phase region & PdY$_2$-C37 (uncertain)           \\ \hline
        40        & Pd$_2$Y$_3$-hR15 & unavailable                       \\ 
        \            &unknown        &                                   \\ \hline
	50     & $\alpha$PdY unknown & PdY-B27/B33 (\us)                  \\
        \            &               & {\it B27 stable (\paw)     }       \\ 
        \            &               & {\it B33$\sim$3.1meV/at.  }       \\
        \            &               & {\it above B27 (\paw)      }       \\ \hline 
	57.1      & Pd$_4$Y$_3$-hR14 & unavailable                       \\ 
        \            &unknown        &                                   \\ \hline
        60 &$\alpha$Pd$_3$Y$_2$ unknown& unavailable                     \\ \hline
	66.6         & unknown       & two phase region                           \\ \hline
        79.5 to 75   & L1$_2$        & Pd$_3$Y-L1$_2$                    \\ \hline
	87.5         & unknown       & unavailable                       \\ 
      \end{tabular} \\ \hline
    \end{tabular}
  \end{center}
}

\begin{center}
  \vspace{-3mm}
  \begin{figure}
    \epsfig{file=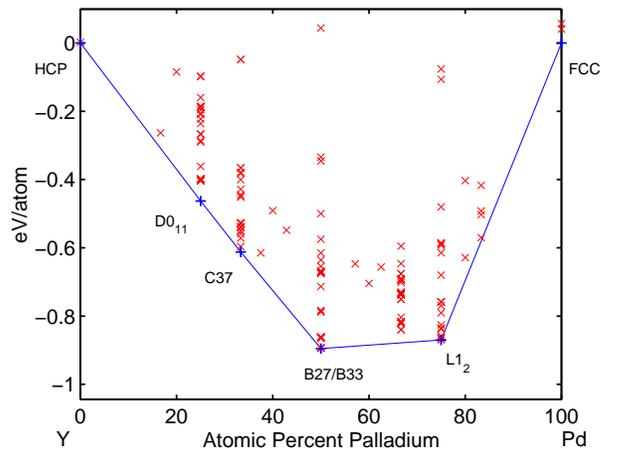,width=\picdim,clip=}
    \vspace{1mm}
    \caption{PdY (Palladium - Yttrium) ground state convex hull.}
    \label{label.fig.PdY}
  \end{figure}
  \vspace{-1mm}
\end{center}

\newpage

{\bf Pd-Zr (Palladium - Zirconium).}
The experimental phase diagram for the system Pd-Zr is based on limited information
\cite{BB,PP,PdZr.59And,PdZr.62Sto,PdZr.69Har,PdTi.70Har,PdZr.71Sav,PdZr.90Cha,PdZr.90Kuz,PdZr.92Sto,PdZr.96Ben}. 
We are able to confirm stability of
Pd$_3$Zr-D0$_{24}$ and PdZr$_2$-C11$_b$. 
The stable phase of PdZr is reported to be CrB (B33) \cite{PP,PdZr.96Ben}.
At that composition, we find two degenerate structures: PdZr-B27 or PdZr-B33 (CrB).
In the Zr-rich side of the phase diagram, we find PdZr$_3$-FCC$_{AB3}^{[001]}$.
However, the energy difference between PdZr$_3$-FCC$_{AB3}^{[001]}$ and the two-phase region
Zr\twophase PdZr$_2$-C11$_b$, is very small ($\sim$6meV/atom).
Hence, the existence of compound PdZr$_3$-FCC$_{AB3}^{[001]}$ remains uncertain.
At low-temperature, instead of a stable phase with 
stoichiometry Pd$_2$Zr, we find the two-phase field PdZr\twophase Pd$_3$Zr-D0$_{24}$.
The least unstable phases are Pd$_2$Zr-C49, Pd$_2$Zr-MoPt$_2$, Pd$_2$Zr-C11$_b$,
with energies $\sim$18meV/atom, $\sim$24meV/atom, $\sim$26meV/atom above the tie-line of the 
two-phase region, respectively. 
{\it {To address the degenerate structures and the erroneous two phase region PdZr\twophase Pd$_3$Zr,
we further investigate PdZr and Pd$_2$Zr with PAW-GGA potentials, 
\PawSection. 
With PAW,
PdZr-B33 is the most stable compound and 
PdZr-B27 is higher by 3.2meV/atom. 
In addition,
Pd$_2$Zr-C11$_b$ is degenerate with the tie line PdZr\twophase Pd$_3$Zr,
and MoPt$_2$ and C49 are higher by $\sim$1.3meV/atom and $\sim$7.1meV/atom with respect to C11$_b$
(also the tie-line PdZr\twophase Pd$_3$Zr has been recalculated with PAW-GGA potentials).
}}

{\tablefont
  \begin{center}
    \begin{tabular}{||c||} \hline
      {\bf Pd-Zr system} \\ \hline
      Low Temperature Phases comparison chart \\ \hline
      \begin{tabular}{c|c|c}
	Composition  & Experimental  & \tablelineone                      \\ 
	\% Pd        & (\massalski)  & \tablelinetwo                     \\ \hline
	25           & two-phase     & PdZr$_3$-FCC$_{AB3}^{[001]}$      \\
	\            & region        & (uncertain)                       \\ \hline
	33.3         & C11$_b$       & PdZr$_2$-C11$_b$                  \\ \hline
	50   &disorder fcc \cite{BB} & PdZr- B27/B33 (\us)               \\
	\            & CrB(B33-TlI)  & {\it B33 stable (\paw)      }     \\
	\ & \cite{PP,PdZr.96Ben}&{\it B27$\sim$3.2meV/at.}               \\ 
	\            &               & {\it above B33 (\paw)       }     \\ \hline
	66.6         & C11$_b$       & two-phase region (\us).           \\
	\            &               & C49$\sim$18meV/at.                \\
        \            &               & MoPt$_2$$\sim$24meV/at.           \\ 
        \            &               & C11$_b$$\sim$26meV/at.            \\ 
	\            &               & above the tie-line (\us).         \\
	\            &               & {\it C11$_b$/tie-line (\paw).    }\\
	\            &               & {\it MoPt$_2$$\sim$1.3meV/at.,   }\\
	\            &               & {\it C49$\sim$7.1meV/at.         }\\
	\            &               & {\it above tie-line (\paw).      }\\ \hline
        75           & D0$_{24}$     & Pd$_3$Zr-D0$_{24}$                \\ 
      \end{tabular} \\ \hline
    \end{tabular}
  \end{center}
}

\vspace{-4mm}
\begin{center}
  \vspace{-3mm}
  \begin{figure}
    \epsfig{file=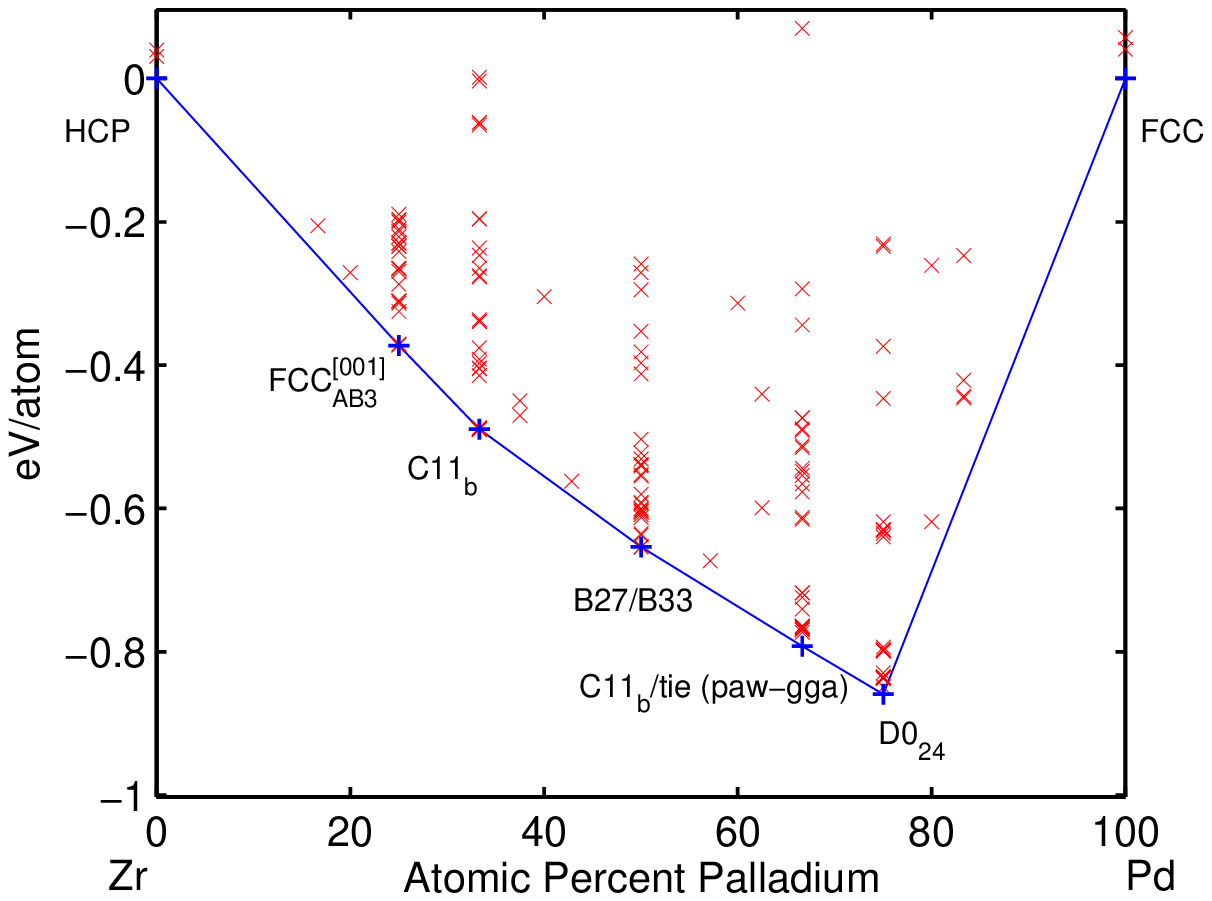,width=\picdim,clip=}
    \caption{PdZr (Palladium - Zirconium) ground state convex hull.}
    \label{label.fig.PdZr}
  \end{figure}
  \vspace{-3mm}
\end{center}

\newpage

{\bf Pt-Rh (Platinum - Rhodium).}
The experimental phase diagram of the system Pt-Rh is similar the system Pd-Pt.
The low-temperature part of the phase diagram is believed to have
a miscibility gap at a temperature of about 760$^\circ$C \cite{BB,PP,ref.Moffatt,PdRh.59Rau,PtRh.95Gag}. 
Instead of the gap, we find several stable phases,
all with fcc superstructure (Pt and Rh are both fcc),
similar to previous FLAPW-LDA calculations \cite{PdPt.91Zun}.
We find 
Pt$_4$Rh-D1$_{a}$, 
Pt$_3$Rh-D0$_{22}$, 
PtRh$_2$-C49, 
PtRh$_3$-D0$_{22}$, 
PtRh$_4$-D1$_{a}$, 
and, at 50\% concentration, PtRh-\CH40 (CH ``40'' in reference \cite{PdPt.91Zun}).
As shown in figure (\ref{label.fig.PtRh}), all the stable phases have very small 
formation energy ($<$ 30meV/atom) indicating that they may disorder at 
relatively low temperature. 

\ReferencePtRh

{\tablefont
  \begin{center}
    \begin{tabular}{||c||} \hline
      {\bf Pt-Rh system} \\ \hline
      Low Temperature Phases comparison chart \\ \hline
      \begin{tabular}{c|c|c}
	Composition  & Experimental  & \tablelineone                        \\ 
        \% Rh        & (\massalski)  & \tablelinetwo                       \\
	\            & Theoretical (\cite{PdPt.91Zun})  &                  \\ \hline
        20           &two-phase region \cite{BB}& Pt$_4$Rh-D1$_a$          \\
	\     & D1$_a$ \cite{PdPt.91Zun} &                                 \\ \hline
	25          &two-phase region \cite{BB}& Pt$_3$Rh-D0$_{22}$        \\ 
        \           &two-phase region \cite{PdPt.91Zun}&                   \\ \hline 
	50           &two-phase region \cite{BB} & PtRh-\CH40               \\ 
	\     & \CH40 \cite{PdPt.91Zun} &                                   \\ \hline
	66.6         &two-phase region \cite{BB}& PtRh$_2$-C49             \\ \hline
        71.4         &two-phase region \cite{BB}& unavailable              \\
	\     & X2 \cite{PdPt.91Zun} &                                     \\ \hline
        75           &two-phase region \cite{BB}& PtRh$_3$-D0$_{22}$       \\
	\     & D0$_{22}$ \cite{PdPt.91Zun} &                              \\ \hline
        80           &two-phase region \cite{BB}& PtRh$_4$-D1$_a$          \\
	\     & D1$_a$ \cite{PdPt.91Zun} &                                 \\ 
      \end{tabular} \\ \hline
    \end{tabular}
  \end{center}
}

\begin{center}
  \vspace{-3mm}
  \begin{figure}
    \epsfig{file=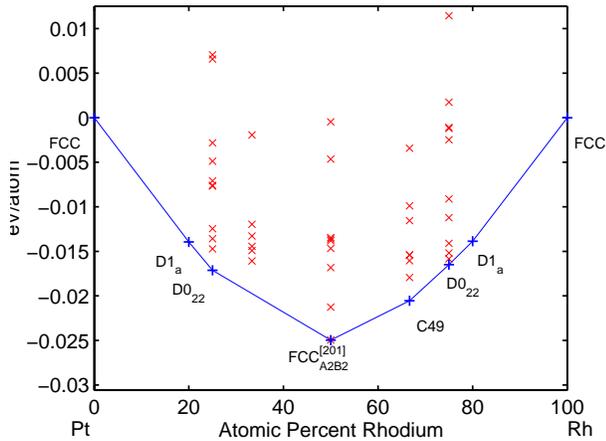,width=\picdim,clip=}
    \vspace{1mm}
    \caption{PtRh (Platinum - Rhodium) ground state convex hull.}
    \label{label.fig.PtRh}
  \end{figure}
  \vspace{-1mm}
\end{center}

\newpage

{\bf Pt-Ru (Platinum - Ruthenium).}
Only one compound has been found for the system Pt-Ru 
\cite{BB,PP,PtRu.Hut,PtRu.2002CamaraTicianelli}.
At low-temperature, the phase diagram reported in Massalski \cite{BB} 
has Platinum-rich and Ruthenium-rich solid solution with large solubilities of the other element, 
and a two-phase region for concentration between $\sim$70\% to $\sim$80\% of Ruthenium.
However, recent X-ray diffraction experimental work reported the existence
of a fcc phase at 50\% composition, with unknown prototype \cite{PtRu.2002CamaraTicianelli}.
For PtRu, our prediction is PtRu-FCC$_{A2B2}^{[001]}$.
At 25\% Ru composition, we find a stable phase Pt$_3$Ru-FCC$_{AB3}^{[001]}$, degenerate
with the two-phase field Pt\twophase PtRu.
Hence, the existence of compound Pt$_3$Ru-FCC$_{AB3}^{[001]}$ remains uncertain.
To our knowledge, PtRu is the first known system where the 
prototype structure FCC$_{A2B2}^{[001]}$ would be stable.

{\tablefont
  \begin{center}
    \begin{tabular}{||c||} \hline
      {\bf Pt-Ru system} \\ \hline
      Low Temperature Phases comparison chart \\ \hline
      \begin{tabular}{c|c|c}
	Composition  & Experimental  & \tablelineone                      \\ 
	\% Ru        & (\massalski)  & \tablelinetwo                     \\ \hline
	25& disorder Pt-A1 \cite{BB} & Pt$_3$Ru                          \\ 
	\            &               & FCC$_{AB3}^{[001]}$/tie-line      \\
	\            &               & (uncertain)                       \\ \hline
	50           &disorder Pt-A1 \cite{BB}&PtRu-FCC$_{A2B2}^{[001]}$ \\
        \            & PtRu-fcc \cite{PtRu.2002CamaraTicianelli} &   \\ 
      \end{tabular} \\ \hline
    \end{tabular}
  \end{center}
}

\begin{center}
  \vspace{-3mm}
  \begin{figure}
    \epsfig{file=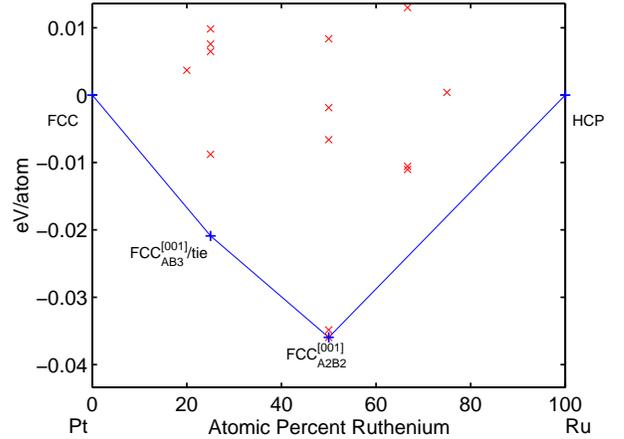,width=\picdim,clip=}
    \vspace{1mm}
    \caption{PtRu (Platinum - Ruthenium) ground state convex hull.}
    \label{label.fig.PtRu}
  \end{figure}
  \vspace{-1mm}
\end{center}

\newpage

{\bf Pt-Tc (Platinum - Technetium).}
The phase diagram for the system Pt-Tc has been determined
from experimental solid solubility data \cite{BB,PP,ref.Moffatt,Tc.62Dar,Tc.63Nie}.
No intermetallic compounds have been reported \cite{BB}.
However, we find two stable phases: Pt$_3$Tc-FCC$_{AB3}^{[001]}$ and PtTc$_3$-D0$_{19}$.
PtTc$_3$-D0$_{19}$ appears in the composition range of a two-phase 
region Pt-A1 and Tc-A3, that is present at temperatures 
higher than $\sim$ 1000\DEG.
\TcDO19

{\tablefont
  \begin{center}
    \begin{tabular}{||c||} \hline
      {\bf Pt-Tc system} \\ \hline
      Low Temperature Phases comparison chart \\ \hline
      \begin{tabular}{c|c|c}
	Composition  & Experimental  & \tablelineone                      \\ 
	\% Tc        & (\massalski)  & \tablelinetwo                     \\ \hline
        25           &disorder Pt-A1 & Pt$_3$Tc-FCC$_{AB3}^{[001]}$      \\ \hline
	75          &two-phase region& PtTc$_3$-D0$_{19}$                \\
        \           & above $\sim$1000\DEG&                              \\
        \           & Pt-A1\twophase Tc-A3&                                   \\ 
      \end{tabular} \\ \hline
    \end{tabular}
  \end{center}
}

\begin{center}
  \vspace{-3mm}
  \begin{figure}
    \epsfig{file=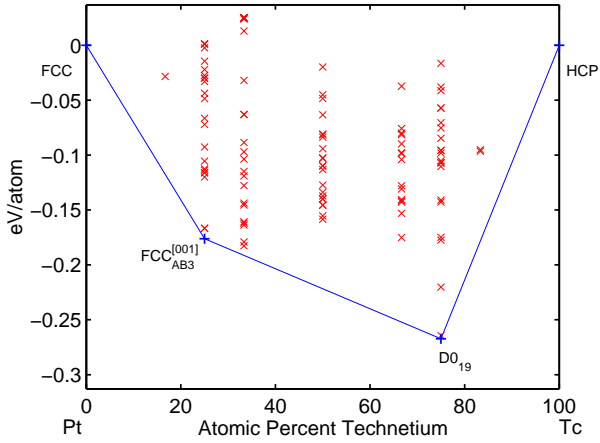,width=\picdim,clip=}
    \vspace{1mm}
    \caption{PtTc (Platinum - Technetium) ground state convex hull.}
    \label{label.fig.PtTc}
  \end{figure}
  \vspace{-1mm}
\end{center}

\newpage

{\bf Pt-Ti (Platinum - Titanium).}
Not much is known of Pt-Ti system to produce a precise phase diagram
\cite{BB,PP,ref.Titanium,PdTi.76Mie,PdTi.93Mik,PtTi.57Nis,PtTi.60Sat,PtTi.65Pie,PtTi.69Sin,PdTi.70Don,PtTi.76Mes}. 
Some intermetallic compounds are reported 
\cite{PtTi.57Nis,PtTi.60Sat,PtTi.65Pie,PtTi.69Sin,PdTi.70Don,PtTi.76Mes,PdTi.89Sel}.
We confirm the stability of phases PtTi$_3$-A15, 
$\alpha$PtTi-B19, Pt$_3$Ti-D0$_{24}$.
B19 is stable at 50\% composition, and L1$_0$ and B33 are higher by $\sim$20meV/atom and 
$\sim$30meV/atom, respectively.
At concentration 25\%Ti, we confirm the existence of $\gamma$-L1$_2$,
which has an energy that is $\sim$5meV/atom higher than Pt$_3$Ti-D0$_{24}$.
However, L1$_2$ has been reported to be stable away from stoichiometry $<25$\% Ti \cite{BB}. 
At composition 33\% Ti, Pt$_2$Ti, Massalski reports a 
two-phase region above 600\DEG$\,$ \cite{BB,PP}.
Instead of the two-phase field, we find a stable compound Pt$_2$Ti: 
two structures, C49 and C37, are degenerate.
We cannot say anything about Pt$_8$Ti-D1$_a$,
because our library does not contain off-stoichiometry D1$_a$.
{\it To address the degenerate structures, we further investigate 
Pt$_2$Ti with PAW-GGA potentials, 
\PawSection. 
With PAW,
Pt$_2$Ti-C49 is the most stable compound and 
Pt$_2$Ti-C37 is higher by 1.9meV/atom.}

\ReferencePtTi

{\tablefont
  \begin{center}
    \begin{tabular}{||c||} \hline
      {\bf Pt-Ti system} \\ \hline
      Low Temperature Phases comparison chart \\ \hline
      \begin{tabular}{c|c|c}
	Composition  & Experimental  & \tablelineone                      \\ 
	\% Ti        & (\massalski)  & \tablelinetwo                     \\ \hline
	1 to 12 &Pt$_8$Ti-D1$_a$& unavailable                       \\ \hline
	20 to 27 & D0$_{24}$ ($>$25\%Ti) & Pt$_3$Ti-D0$_{24}$            \\
	\            & $\gamma$-L1$_2$ ($<$25\%Ti) & L1$_2$$\sim$5meV/at.\\
        \            &               & above  D0$_{24}$                  \\ \hline
	33.3         &two-phase region& Pt$_2$Ti-C49/C37 (\us)            \\
	\            &above 600\DEG \cite{BB} & {\it C49 stable }        \\
	\	     &		     & {\it C37$\sim$1.9meV/at.}         \\
	\	     &		     & {\it above C49 (\paw) }            \\ \hline
	46 to 54   & $\alpha$PtTi-B19& PtTi-B19                          \\
	\            &               & L1$_0$$\sim$20meV/at.             \\
        \            &               & above B19                         \\ \hline
        71 to 78     & A15           & PtTi$_3$-A15                      \\ 
      \end{tabular} \\ \hline
    \end{tabular}
  \end{center}
}

\begin{center}
  \vspace{-3mm}
  \begin{figure}
    \epsfig{file=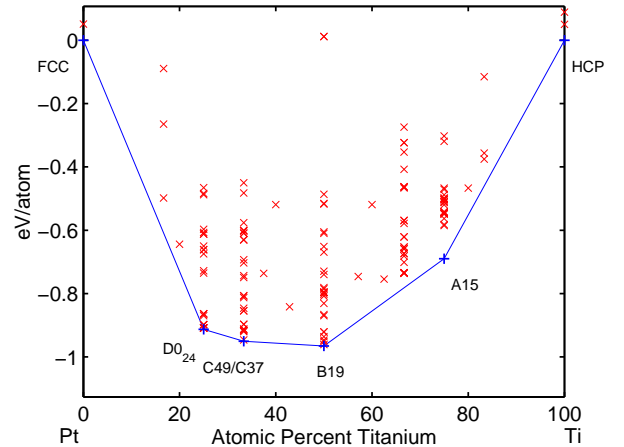,width=\picdim,clip=}
    \vspace{1mm}
    \caption{PtTi (Platinum - Titanium) ground state convex hull.}
    \label{label.fig.PtTi}
  \end{figure}
  \vspace{-1mm}
\end{center}

\newpage

{\bf Pt-Y (Platinum - Yttrium).}
The experimental phase diagram of the system Pt-Y has been sketched by analogy 
with other Rare Earth-Platinum diagrams \cite{BB}. 
Several intermetallic compounds have been reported 
\cite{BB,PP,ref.Moffatt,AuSc.76Mie,PdTi.65Dwi,PdTi.76Mie,PtY.59Com,PtY.65Geb,PtY.67Bro,PtY.71Kri,PtY.72Ram,PtY.73Erd,PtY.79Ler,PtY.88Yif,PtZr.75Mie}.
The stability of the compounds 
Pt$_3$Y-L1$_2$, Pt$_2$Y-C15, and PtY$_3$-D0$_{11}$ is confirmed.
At 50\% concentration, we do not find any stable PtY-B27.
Instead of B27, we find PtY-B33 (CrB prototype) to have lowest energy with 
B2 and B27 having energies $\sim$50meV/atom above B33.
At 66.6\% Y concentration, we find PtY$_2$-C37 which is 
the prototype of Co$_2$Si and Ni$_2$Si.
Hence our calculations confirm the correct experimental structure.
At concentration 62.5\% Y, 
we do not find Pt$_3$Y$_5$-D8$_8$, but the two-phase field
PtY\twophase PtY$_2$. 
D8$_8$ and W$_5$Si$_3$ are higher by $\sim$24meV/atom
and $\sim$80meV/atom with respect to the tie-line PtY\twophase PtY$_2$ 
(this problem is solved with PAW-GGA potentials).
As shown in figure (\ref{label.fig.PtY}), at concentration 16.6\% Y, 
we find a two-phase region instead of the reported stable compound 
Pt$_5$Y with unknown structure \cite{BB}.
Our best guess is Pt$_5$Y-D2$_d$ which is the 
least-metastable structure we have at such composition 
($\sim$18meV/atom above the tie-line Pt\twophase Pt$_3$Y).
We conclude that further experimental and theoretical investigations
are necessary to determine the behavior of PtY.
{\it {To address the disagreement with the experimental results for compounds
PtY and Pt$_3$Y$_5$, we further investigate the relevant structures with PAW-GGA potentials,
\PawSection.
With PAW,
PtY-B33 is stable and PtY-B27 is higher by 60meV/atom.
For compound Pt$_3$Y$_5$,
D8$_8$ is stable with an energy 6.9meV/atom lower than the
tie-line PtY\twophase PtY$_2$, which has also been recalculated with PAW-GGA potentials.
}}
{The disagreement at composition PtY is further discussed in Section (\ref{section.discussion}).}

{\tablefont
  \begin{center}
    \begin{tabular}{||c||} \hline
      {\bf Pt-Y system} \\ \hline
      Low Temperature Phases comparison chart \\ \hline
      \begin{tabular}{c|c|c}
	Composition  & Experimental  & \tablelineone                      \\ 
	\% Y         & (\massalski)  & \tablelinetwo                     \\ \hline
	16.6         & unknown       & Pt$_5$Y-D2$_d$                    \\
        \            &               & metastable.                       \\
        \            &               & D2$_d$ $\sim$18meV/at.            \\
        \            &               & above the tie-line                \\ \hline	
	25           & L1$_2$        & Pt$_3$Y-L1$_2$                    \\ \hline
	33.3         & C15           & Pt$_2$Y-C15                       \\ \hline
	42.9&Pt$_4$Y$_3$-Pd$_4$Pu$_3$&unavailable                        \\ \hline
	50           & B27           & PtY-B33                           \\
        \            &               & B2/B27 $\sim$50meV/at.            \\
        \            &               & above B33                         \\
	\            &               & {\it B33 stable (\paw)}           \\
        \            &               & {\it B27 $\sim$60meV/at.}         \\
        \            &               & {\it above B33.  (\paw)}          \\
	\            &               & See Section (\ref{section.discussion}). \\ \hline
	55.6&Pt$_4$Y$_5$-Pu$_5$Rh$_4$& unavailable                       \\ \hline
	62.5         & Pt$_3$Y$_5$-D8$_8$ & two-phase region             \\
        \            &               & PtY\twophase PtY$_2$. (\us)       \\ 
        \            &               & D8$_8$ $\sim$24meV/at.            \\
        \            &               & above the tie-line (\us)          \\
	\            &               & {\it D8$_8$ stable (\paw). }      \\
	\            &               & {\it $\sim$6.9meV/at. below    }  \\
	\            &               & {\it PtY\twophase PtY$_2$ (\paw).}\\ \hline
	66.6       & PtY$_2$-Ni$_2$Si \cite{BB}& PtY$_2$-C37             \\
        \            & (Co$_2$Si)    & (Co$_2$Si)                        \\ \hline        
        70       &Pt$_3$Y$_7$-D10$_2$& unavailable                       \\ \hline
        75           & D0$_{11}$     & PtY$_3$-D0$_{11}$                 \\ 
      \end{tabular} \\ \hline
    \end{tabular}
  \end{center}
}

\begin{center}
  \vspace{-3mm}
  \begin{figure}
    \epsfig{file=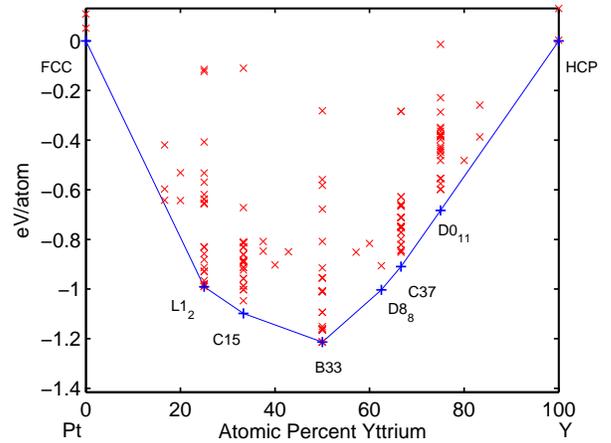,width=\picdim,clip=}
    \vspace{1mm}
    \caption{PtY (Platinum - Yttrium) ground state convex hull.}
    \label{label.fig.PtY}
  \end{figure}
  \vspace{-1mm}
\end{center}

\newpage

{\bf Pt-Zr (Platinum - Zirconium).}
The system Pt-Zr is quite interesting. Our {\it ab initio} method confirms the stability of 
$\alpha$PtZr-B33 (CrB prototype). Two crystal structures have been
reported for Pt$_3$Zr: D0$_{24}$ and L1$_2$ 
\cite{BB,PP,PtZr.43Wal,PtZr.68Sch,PtZr.64Ram,PtZr.75Mie,PdTi.76Mie,PdZr.90Kuz}.
We confirm the stability of Pt$_3$Zr-D0$_{24}$ and we find L1$_2$ 
to be higher by 10meV/atom with respect to D0$_{24}$. 
In the Zr-rich part of the phase diagram, we find two stable phases 
PtZr$_2$-C16 and PtZr$_3$-A15. At concentration 62.5\% Zr, 
we do not find Pt$_3$Zr$_5$-D8$_8$, but the two-phase field
PtZr\twophase PtZr$_2$. W$_5$Si$_3$ and D8$_8$ are higher by $\sim$26meV/atom
and $\sim$36meV/atom with respect to the tie-line PtZr\twophase PtZr$_2$.
{\it {To address the phase instability at composition Pt$_3$Zr$_5$,
we further investigate the relevant compounds with PAW-GGA potentials, 
\PawSection. 
With PAW,
at concentration Pt$_3$Zr$_5$,
there is a two-phase region PtZr\twophase PtZr$_2$.
In addition, 
Pt$_3$Zr$_5$-W$_5$Si$_3$ and Pt$_3$Zr$_5$-D8$_8$ 
have energies higher by 
23meV/atom and 26meV/atoms with respect 
to the tie-line PtZr\twophase PtZr$_2$, respectively.}}
{The disagreement at composition Pt$_3$Zr$_5$ is further discussed in Section (\ref{section.discussion}).}

{\tablefont
  \begin{center}
    \begin{tabular}{||c||} \hline
      {\bf Pt-Zr system} \\ \hline
      Low Temperature Phases comparison chart \\ \hline
      \begin{tabular}{c|c|c}
	Composition  & Experimental  & \tablelineone                      \\ 
	\% Zr        & (\massalski)  & \tablelinetwo                     \\ \hline
	25           & D0$_{24}$ in  & Pt$_3$Zr-D0$_{24}$                \\
	\ & \cite{PtZr.64Ram,PdZr.90Kuz,PtZr.75Mie,PdTi.76Mie} & L1$_2$$\sim$10meV/at. \\
        \ & L1$_2$ in \cite{PtZr.68Sch} & above D0$_{24}$                \\ \hline
	50  &$\alpha$PtZr-B33 (CrB)  & PtZr-B33                          \\ \hline
        62.5        & Pt$_3$Zr$_5$-D8$_8$ & two-phase region (\us)       \\
        \           &                & W$_5$Si$_3$$\sim$26meV/at.        \\
        \           &                & D8$_8$$\sim$36meV/at.             \\ 
	\           &                & above tie-line (\us).             \\
        \           &                & {\it two-phase region (\paw)    } \\
	\           &                & {\it W$_5$Si$_3$$\sim$23meV/at. } \\
        \           &                & {\it D8$_8$$\sim$26meV/at.      } \\ 
	\           &                & {\it above tie-line (\paw).     } \\	
	\            &               & See Section (\ref{section.discussion}).\\ \hline
	66.6        &two-phase region& PtZr$_2$-C16                      \\
        \            &above 600\DEG  & (uncertain)                       \\ \hline
        75          &two-phase region& PtZr$_3$-A15                      \\
        \            &above 600\DEG  & (uncertain)                       \\  
      \end{tabular} \\ \hline
    \end{tabular}
  \end{center}
}

\begin{center}
  \vspace{-3mm}
  \begin{figure}
    \epsfig{file=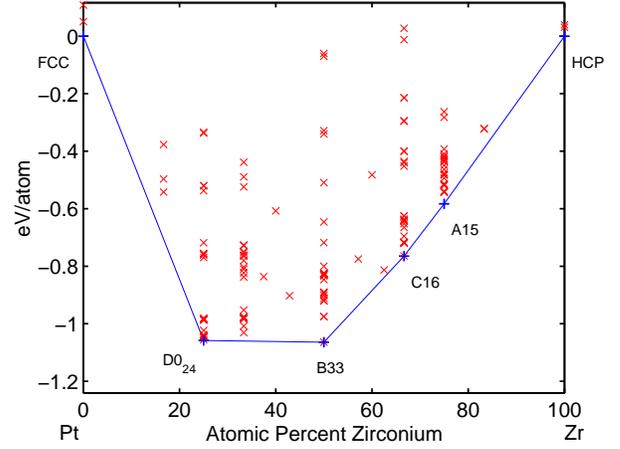,width=\picdim,clip=}
    \vspace{1mm}
    \caption{PtZr (Platinum - Zirconium) ground state convex hull.}
    \label{label.fig.PtZr}
  \end{figure}
  \vspace{-1mm}
\end{center}

\newpage

{\bf Rh-Ru (Rhodium - Ruthenium).}
The phase diagram for the system Rh-Ru is based on the 
solid solubility data \cite{BB,PP,ref.Moffatt,RhRu.84Pas}.
Experimental results report two solid solutions 
(fcc Rh-rich) and (hcp Ru-rich) with a two-phase region in between (from 34.5\% to 40\% atomic percent Ruthenium).
No intermetallic compounds have been reported \cite{RhRu.54Hel,RhRu.84Pas,RhRu.93Gur}
and the system is considered to be no compound forming \cite{PP}.
However, we find two stable phases: 
an orthorhombic oC12 RhRu$_2$$^{proto}$, with hcp superstructure and Cmcm \#63 space group,
and a trigonal hP4 RhRu$^{proto}$, with hcp superstructure and P$\bar{3}$m1 \#164 space group.
Both prototypes are described in Appendix (\ref{proto.AB2.RhRu142}).
As shown in figure (\ref{label.fig.RhRu}), all the stable phases have 
small formation energy ($<$ 10meV/atom) making them difficult to determine experimentally. 
{\it {To better describe the stability of this system,
we further investigate all the structures 
with negative formation energies with PAW-GGA potentials,
\PawSection.
Also with PAW, RhRu$_2$$^{proto}$ and RhRu$^{proto}$ are the most stable structures, 
with formation energies of -8.3meV/atom and -8.8meV/atom, respectively.}}

{\tablefont
  \begin{center}
    \begin{tabular}{||c||} \hline
      {\bf Rh-Ru system} \\ \hline
      Low Temperature Phases comparison chart \\ \hline
      \begin{tabular}{c|c|c}
	Composition  & Experimental  & \tablelineone                      \\ 
	\% Rh        & (\massalski)  & \tablelinetwo                     \\ \hline
	33           &solid solution & RhRu$_2$$^{proto}$ stable               \\
	\            & Ru-A3         & both \us \,and {\it \paw},              \\
	\            &               & {\it $E_f\sim-8.3$meV/atom }       \\
	\            &               & {\it (\paw).}                     \\ \hline	
	50           &solid solution & RhRu$^{proto}$ stable                   \\
	\            & Ru-A3         & both \us \,and {\it \paw},              \\
	\            &               & {\it $E_f\sim-8.8$meV/atom }       \\
	\            &               & {\it (\paw).}                     \\ 
      \end{tabular} \\ \hline
    \end{tabular}
  \end{center}
}

\begin{center}
  \vspace{-3mm}
  \begin{figure}
    \epsfig{file=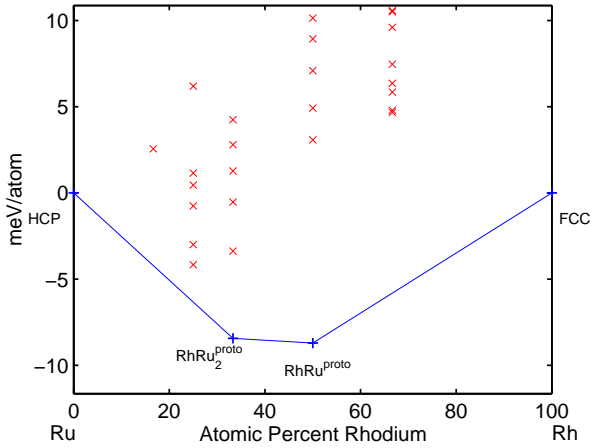,width=\picdim,clip=}
    \vspace{1mm}
    \caption{RhRu (Rhodium - Ruthenium) ground state convex hull.}
    \label{label.fig.RhRu}
  \end{figure}
  \vspace{-1mm}
\end{center}

\newpage

{\bf Rh-Tc (Rhodium - Technetium).}
The phase diagram for the system Rh-Tc is based on the 
solid solubility data \cite{BB,PP,ref.Moffatt,Tc.63Dar,Tc.63Nie}.
Experimental results report two solid solutions 
(fcc Rh-rich) and (hcp Tc-rich) with a two-phase region in between.
No intermetallic compounds have been reported \cite{Tc.63Dar,Tc.63Nie}. 
However, we find three stable phases: 
RhTc$_3$-D0$_{19}$, RhTc-B19, and Ru$_2$Tc-ZrSi$_2$.
\TcDO19

{\tablefont
  \begin{center}
    \begin{tabular}{||c||} \hline
      {\bf Rh-Tc system} \\ \hline
      Low Temperature Phases comparison chart \\ \hline
      \begin{tabular}{c|c|c}
	Composition  & Experimental  & \tablelineone                      \\ 
	\% Rh        & (\massalski)  & \tablelinetwo                     \\ \hline
	25           &solid solution Tc-A3 & RhTc$_3$-D0$_{19}$          \\ \hline
	50           &solid solution Tc-A3 & RhTc-B19                    \\
	\            &               & B27$\sim$29meV/at.                \\
        \            &               & above B19 (\us).                  \\ \hline
	66.6        &two-phase region& Rh$_2$Tc-ZrSi$_2$                 \\
        \      & above $\sim$1000\DEG&                                   \\ 
        \            & Pt-A1\twophase Tc-A3&                             \\ 
      \end{tabular} \\ \hline
    \end{tabular}
  \end{center}
}

\begin{center}
  \vspace{-3mm}
  \begin{figure}
    \epsfig{file=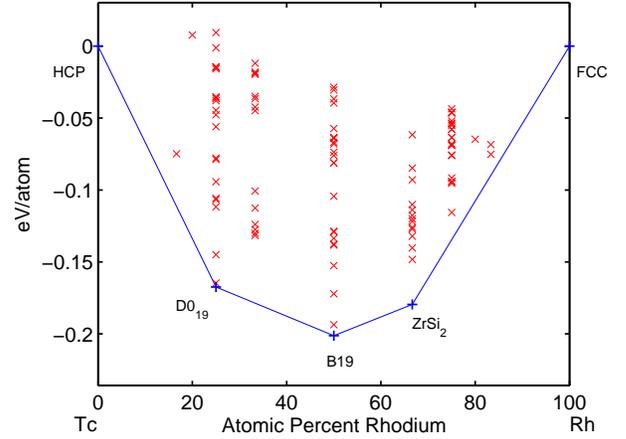,width=\picdim,clip=}
    \vspace{1mm}
    \caption{RhTc (Rhodium - Technetium) ground state convex hull.}
    \label{label.fig.RhTc}
  \end{figure}
  \vspace{-1mm}
\end{center}

\newpage

{\bf Rh-Ti (Rhodium - Titanium).}
There are qualitative disagreements about the phase diagram of the Rh-Ti system 
\cite{BB,PP,ref.Titanium,PtZr.64Ram,RhTi.66Rau,RhTi.66Ere,RhTi.72Ere,RhTi.75Sht,RhTi.88Kue}.
We confirm the stable phases that where found 
by all investigators \cite{RhTi.66Rau,RhTi.66Ere,RhTi.72Ere,RhTi.75Sht}: 
RhTi$_2$, $\alpha$RhTi-L1$_0$, and Rh$_3$Ti-L1$_2$.
References \cite{PP,PtZr.64Ram,RhTi.88Kue}
report RhTi$_2$-CuZr$_2$ instead of RhTi$_2$-C11$_b$. The prototype CuZr$_2$
is a distortion of C11$_b$ and has the same lattice type 
(tetragonal, tI6) and space group (I4/mmm \#139) \cite{PP}.
We find RhTi$_2$-C11$_b$ and RhTi$_2$-CuZr$_2$ to have degenerate energy.
In the Rh-rich part of the phase diagram we find a stable phase 
Rh$_2$Ti-C37. However, we do not have Rh$_5$Ti$_3$ in our library, 
so C37 might likely be unstable with respect to the two-phase field
Rh$_5$Ti$_3$\twophase L1$_2$.
At concentration $\approx$ 84\% Rh, we do not find
any stable compound, in agreement with \cite{RhTi.72Ere} 
and in contrast with \cite{RhTi.66Rau}.
{\it To address the degenerate structures RhTi$_2$-C11$_b$/CuZr$_2$, 
we further investigate RhTi$_2$ with PAW-GGA potentials, 
\PawSection. 
Also with PAW, 
C11$_b$ and CuZr$_2$ remain degenerate.  }

\ReferenceRhTi

{\tablefont
  \begin{center}
    \begin{tabular}{||c||} \hline
      {\bf Rh-Ti system} \\ \hline
      Low Temperature Phases comparison chart \\ \hline
      \begin{tabular}{c|c|c}
	Composition  & Experimental  & \tablelineone                      \\ 
	\% Rh        & (\massalski)  & \tablelinetwo                     \\ \hline
	33.3     & C11$_b$\cite{BB}/CuZr$_2$\cite{PP} & RhTi$_2$         \\
	\	     &              & C11$_b$/CuZr$_2$ (\us,)            \\
        \            &              & C11$_b$/CuZr$_2$ ({\it \paw})      \\ \hline
	$\sim$38 to 58 &$\alpha$RhTi-L1$_0$ & RhTi-L1$_0$                \\ \hline
	62.5 & Rh$_5$Ti$_3$-Ge$_3$Rh$_5$& unavailable                    \\ \hline
	66.6        &two-phase region& Rh$_2$Ti-C37                      \\ 
        \            &above 600\DEG  &                                   \\ \hline
        73 to 78     & L1$_2$        & Rh$_3$Ti-L1$_2$                   \\ \hline
	$\sim$83.8   & Rh$_5$Ti (unknown) & nothing stable               \\ 
      \end{tabular} \\ \hline
    \end{tabular}
  \end{center}
}

\begin{center}
  \vspace{-3mm}
  \begin{figure}
    \epsfig{file=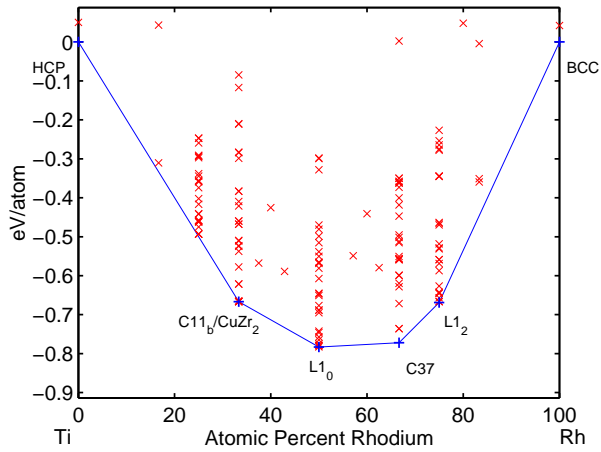,width=\picdim,clip=}
    \vspace{1mm}
    \caption{RhTi (Rhodium - Titanium) ground state convex hull.}
    \label{label.fig.RhTi}
  \end{figure}
  \vspace{-1mm}
\end{center}

\newpage

{\bf Rh-Y (Rhodium - Yttrium).}
Several compounds have been reported for the system Rh-Y at
low-temperature 
\cite{BB,PP,ref.Moffatt,RhY.59Com,PdTi.65Dwi,RhY.65Geb,RhY.72Ram,RhY.76Mor}.
The stability of Rh$_2$Y-C15, RhY-B2, RhY$_3$-D0$_{11}$ is confirmed.
At composition Rh$_5$Y, where a D2$_d$ structure has been seen 
at high-temperature (but not stable at low-temperature), we
find the D2$_d$ to be the lowest energy structure (of all the structures
at that composition), even though it is metastable with respect
to the phase separation into Rh$_2$Y-C15\twophase Rh.
At composition Rh$_3$Y one compound has been reported to be stable 
with prototype CeNi$_3$ and space group P6$_3$/mmc \#194 \cite{PP,RhY.73Gha,RhY.76Loe}.
We do not have such prototype in our library and we do not find any stable phase:
Rh$_3$Y-D0$_{19}$ is the least metastable prototype we obtain.
Rh$_3$Y-D0$_{19}$ is higher by 130 meV/atom with respect the tie-line, 
which is at least one order of magnitude bigger than the accuracy of the 
calculations. 
We also find RhY$_2$-C37, which appears in a concentration between 
two known compounds, Rh$_3$Y$_5$ and Rh$_3$Y$_7$, that are not present in 
our set of calculations. Therefore C37 might be unstable with 
respect to the two-phase region Rh$_3$Y$_5$\twophase Rh$_3$Y$_7$.

{\tablefont
  \begin{center}
    \begin{tabular}{||c||} \hline
      {\bf Rh-Y system} \\ \hline
      Low Temperature Phases comparison chart \\ \hline
      \begin{tabular}{c|c|c}
	Composition  & Experimental  & \tablelineone                      \\ 
	\% Rh        & (\massalski)  & \tablelinetwo                     \\ \hline
	25           & D0$_{11}$     & RhY$_3$-D0$_{11}$                 \\ \hline
	30     & Rh$_3$Y$_7$-D10$_2$ & unavailable                       \\ \hline
	33.3      &two-phase region& RhY$_2$-C37                         \\ 
	\            &above 0\DEG    & (uncertain)                       \\ \hline
	37.5  &Rh$_3$Y$_5$ (unknown) & unavailable                       \\ \hline
	40    &Rh$_2$Y$_3$ (unknown) & unavailable                       \\
        \            &tI140-I4/mcm   &                                   \\ \hline
	50           & B2            & RhY-B2                            \\ \hline
	66.6         & C15           & Rh$_2$Y-C15                       \\ \hline
        75       & Rh$_3$Y (unknown) & D0$_{19}$$\sim$130meV/at.         \\
        \            & hP24-P6$_3$/mmc & above tie-line                  \\ \hline
	83.5         & Rh$_5$Y-D2$_d$& D2$_d$$\sim$110meV/at.            \\ 
        \         &(high-temperature)& above tie-line                    \\ 
      \end{tabular} \\ \hline
    \end{tabular}
  \end{center}
}

\begin{center}
  \vspace{-3mm}
  \vspace{-3mm}
  \begin{figure}
    \epsfig{file=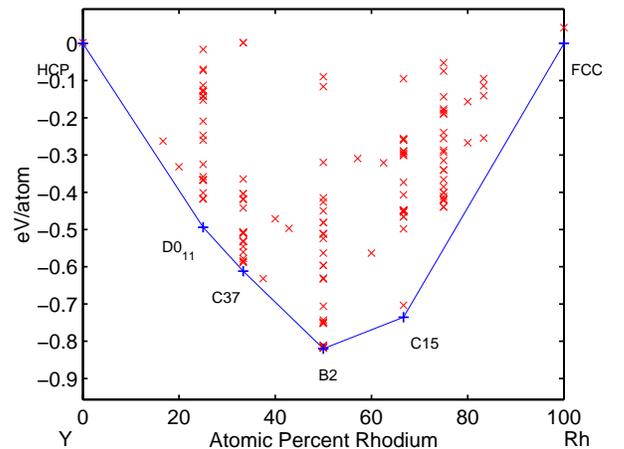,width=\picdim,clip=}
    \caption{RhY (Rhodium - Yttrium) ground state convex hull.}
    \label{label.fig.RhY}
  \end{figure}
  \vspace{-1mm}
\end{center}

\newpage

{\bf Rh-Zr (Rhodium - Zirconium).}
Although the system Zr-Rh is well known for its superconducting phases 
\cite{BB,PP,RhZr.65Zeg,RhZr.88Jor}, further investigations are needed to clarify
the stability and presence of intermediate phases \cite{RhZr.78Ere,RhZr.80Ere}.
Our {\it ab initio} method confirms the stability of RhZr$_2$-C16 and Rh$_3$Zr-L1$_2$.
Massalski does not report the prototype of the low-temperature phase 
$\alpha$RhZr \cite{BB}. References \cite{RhZr.88Jor,RhZr.88Kue,RhZr.94Sem} 
report $\alpha$RhZr-B27. 
We confirm the stability of $\alpha$RhZr-B27 (BFe prototype) with no other 
metastable compounds with similar energy.
In addition, we find three new phases Rh$_2$Zr-C37, RhZr$_4$-D1$_a$ and RhZr$_3$-FCC$_{AB3}^{[001]}$, 
which are degenerate with the two-phase fields RhZr\twophase Rh$_3$Zr, Zr \twophase RhZr, and Zr \twophase RhZr, respectively.

\ReferenceRhZr

{\tablefont
  \begin{center}
    \begin{tabular}{||c||} \hline
      {\bf Rh-Zr system} \\ \hline
      Low Temperature Phases comparison chart \\ \hline
      \begin{tabular}{c|c|c}
	Composition  & Experimental  & \tablelineone                      \\ 
	\% Rh        & (\massalski)  & \tablelinetwo                     \\ \hline
	20           &two-phase region& RhZr$_4$-D1$_a$/tie-line         \\
        \            & above 0\DEG   &                                   \\ \hline
	25           &two-phase region& RhZr$_3$-FCC$_{AB3}^{[001]}$/tie-line \\ 
        \            & above 0\DEG   &                                   \\ \hline
	33.3         & C16           & RhZr$_2$-C16                      \\ \hline
	50 to ? & $\alpha$RhZr &RhZr-B27                                 \\
	\            & unknown \cite{BB} &                               \\
	\  & B27 \cite{RhZr.88Jor,RhZr.88Kue,RhZr.94Sem}   &          \\ \hline
	57.1  &Rh$_4$Zr$_3$ (unknown)& unavailable                       \\ \hline
	62.5  &Rh$_5$Zr$_3$-Pd$_3$Pu$_3$ & unavailable                   \\ \hline
	66.6         &two-phase region& Rh$_2$Zr-C37/tie-line            \\
        \            &above 0\DEG    &                                   \\ \hline
        72 to 82     & L1$_2$        & Rh$_3$Zr-L1$_2$                   \\ 
      \end{tabular} \\ \hline
    \end{tabular}
  \end{center}
}

\begin{center}
  \vspace{-3mm}
  \begin{figure}
    \epsfig{file=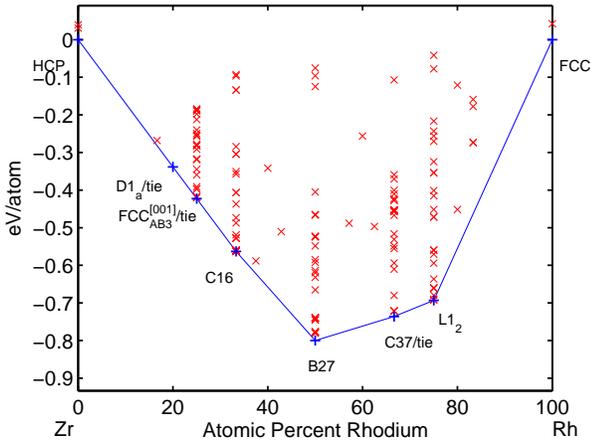,width=\picdim,clip=}
    \vspace{1mm}
    \caption{RhZr (Rhodium - Zirconium) ground state convex hull.}
    \label{label.fig.RhZr}
  \end{figure}
  \vspace{-1mm}
\end{center}

\newpage

{\bf Ru-Tc (Ruthenium - Technetium).}
The phase diagram for the system Ru-Tc is considered to have 
a continuous disordered hcp solid solution at low-temperature 
\cite{BB,PP,Tc.62Dar,Tc.63Nie,RuTc.75Ale}. We find three stable 
ordered phases Ru$_3$Tc-D0$_{19}$, RuTc-B19, and  RuTc$_3$-D0$_{19}$,
all of which are hcp superstructures. Hence, it is possible 
that they are low-temperature phases of the system Ru-Tc.
\TcDO19

{\tablefont
  \begin{center}
    \begin{tabular}{||c||} \hline
      {\bf Ru-Tc system} \\ \hline
      Low Temperature Phases comparison chart \\ \hline
      \begin{tabular}{c|c|c}
	Composition  & Experimental  & \tablelineone                      \\ 
	\% Ru        & (\massalski)  & \tablelinetwo                     \\ \hline
	25       & disorder solution & RuTc$_3$-D0$_{19}$                \\
	\            & (Ru,Tc)-A3    &                                   \\ \hline
	50           & same as above & RuTc-B19                          \\ \hline
        75           & same as above & Ru$_3$Tc-D0$_{19}$                \\ 
      \end{tabular} \\ \hline
    \end{tabular}
  \end{center}
}

\begin{center}
  \vspace{-3mm}
  \begin{figure}
    \epsfig{file=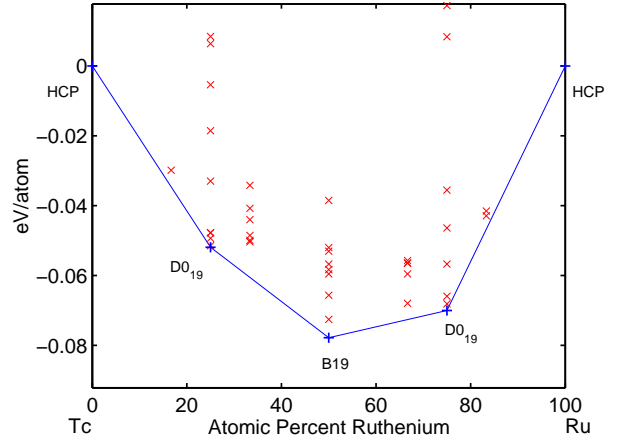,width=\picdim,clip=}
    \vspace{1mm}
    \caption{RuTc (Ruthenium - Technetium) ground state convex hull.}
    \label{label.fig.RuTc}
  \end{figure}
  \vspace{-1mm}
\end{center}

\newpage

{\bf Ru-Ti (Ruthenium - Titanium).}
The phase diagram of Ru-Ti is well determinated 
by several investigators.
A single low-temperature compound RuTi-B2 has been reported 
\cite{BB,PP,ref.Titanium,PdTi.76Mie,RuTi.55Jor,RuTi.59Dwi,RuTi.63Raub,RuTi.72Gul,RuTi.73Tam,RuTi.73Ere,RuTi.76Bor,RuTi.95Zha}, and
our {\it ab initio} method confirms its stability. 
In addition, we find two stable phases: 
RuTi$_2$-C49, and orthorhombic RuTi$_3$ 
with space group Immm \#71, bcc superstructure, and prototype 
RuTi$_3^{proto}$ described in Appendix (\ref{proto.AB3.RuTi81}).
Such compounds are close to the tie-line Ti\twophase RuTi.
In fact, for C49 and RuTi$_3^{proto}$, the formation energies 
are lower by $\sim$34meV/atom and $\sim$37meV/atom
with respect to the two-phase field  Ti\twophase RuTi.

{\tablefont
  \begin{center}
    \begin{tabular}{||c||} \hline
      {\bf Ru-Ti system} \\ \hline
      Low Temperature Phases comparison chart \\ \hline
      \begin{tabular}{c|c|c}
	Composition  & Experimental  & \tablelineone                      \\ 
	\% Ru        & (\massalski)  & \tablelinetwo                     \\ \hline
	25           &two-phase region& RuTi$_3^{proto}$                 \\
        \            &above 600\DEG  &Appendix (\ref{proto.AB3.RuTi81})  \\ 
        \            &               & $\sim$37meV/at.                   \\
	\            &               & below Ti\twophase RuTi            \\ \hline
	33.3         &same as above& RuTi$_2$-C49                        \\
        \            &               & $\sim$34meV/at.                   \\
	\            &               & below Ti\twophase RuTi            \\ \hline
	45 to 52 $\pm$1 & B2         & RuTi-B2                           \\ 
      \end{tabular} \\ \hline
    \end{tabular}
  \end{center}
}

\begin{center}
  \vspace{-3mm}
  \begin{figure}
    \epsfig{file=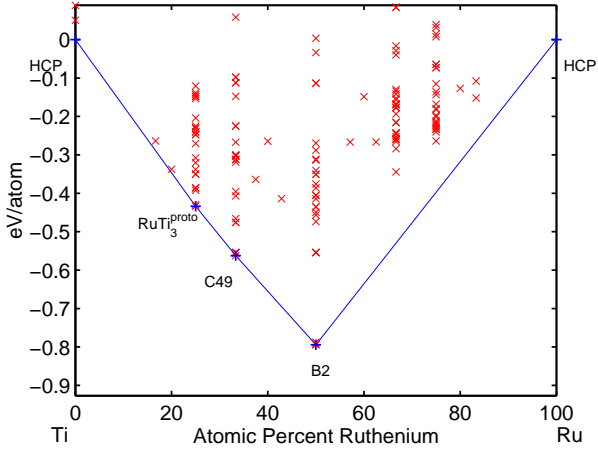,width=\picdim,clip=}
    \vspace{1mm}
    \caption{RuTi (Ruthenium - Titanium) ground state convex hull.}
    \label{label.fig.RuTi}
  \end{figure}
  \vspace{-1mm}
\end{center}

\newpage

{\bf Ru-Y (Ruthenium - Yttrium).}
Several compounds have been reported for the system Ru-Y 
\cite{BB,PP,ref.Moffatt,RhY.59Com,RuY.76Loe,RuY.80Cen,RuY.80San,RuY.84Sha,RuY.89For}.
We confirm the stability of RuY$_3$-D0$_{11}$ and Ru$_2$Y-C14.
Also, we find RuY$_2$-C16, which appears in a concentration between 
two known compounds Ru$_{25}$Y$_{44}$ and Ru$_2$Y$_5$ that are not in
our library of calculations. Hence, the existence of RuY$_2$-C16 is uncertain.
At concentration 50\%, we do not find any stable RuY compound, 
in agreement with \cite{RuY.84Sha}.

{\tablefont
  \begin{center}
    \begin{tabular}{||c||} \hline
      {\bf Ru-Y system} \\ \hline
      Low Temperature Phases comparison chart \\ \hline
      \begin{tabular}{c|c|c}
	Composition  & Experimental  & \tablelineone                      \\ 
	\% Ru        & (\massalski)  & \tablelinetwo                     \\ \hline
	25           & D0$_{11}$     & RuY$_3$-D0$_{11}$                 \\ \hline
	28.6 &Ru$_{2}$Y$_{5}$-C$_2$Mn$_5$& unavailable                   \\
        \            &  mS28 C12/c1  &                                   \\ 
        \            & \cite{PP,RuY.76Loe,RuY.80Cen,RuY.84Sha} &         \\ \hline
	33.3         &two-phase region& RuY$_2$-C16                      \\ 
        \            &above 0\DEG    & (uncertain)                       \\ \hline
	36.2 &Ru$_{25}$Y$_{44}$ (unknown)& unavailable                   \\
        \            &oP276 Pnna     &                                   \\
        \            & \cite{PP,RuY.76Loe,RuY.89For} &                   \\ \hline
 	40           & Ru$_2$Y$_3$-Er$_3$Ru$_2$ & unavailable            \\
	\            & hP10 P6$_3$/m \cite{RuY.90Pal} &                  \\ \hline
 	66.6         &Ru$_2$Y-C14    & Ru$_2$Y-C14                       \\ 
      \end{tabular} \\ \hline
    \end{tabular}
  \end{center}
}

\begin{center}
  \vspace{-3mm}
  \begin{figure}
    \epsfig{file=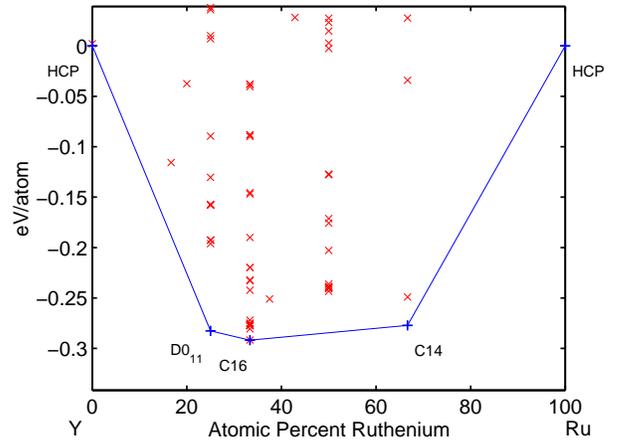,width=\picdim,clip=}
    \vspace{1mm}
    \caption{RuY (Ruthenium - Yttrium) ground state convex hull.}
    \label{label.fig.RuY}
  \end{figure}
  \vspace{-1mm}
\end{center}

\newpage

{\bf Ru-Zr (Ruthenium - Zirconium).}
The phase diagram of RuZr is known accurately \cite{BB,PP,RuZr.88Ere,RuZr.80Ere}.
Our {\it ab initio} method confirms the stability of the low-temperature phase RuZr-B2.
In agreement with experiments, we find no ground state 
at composition Ru$_2$Zr, though the lowest energy structure at that 
composition in our calculations is C14 which appears in the phase 
diagram at high-temperature. Ru$_2$Zr-C14 is higher by $\sim$60meV/atom
with respect to the two-phase field Ru\twophase RuZr. 
In the Zr-rich region we find one stable compound, RuZr$_4$-D1$_a$, previously unknown.
In addition, we find three metastable phases:
RuZr$_5$, RuZr$_3$, and RuZr$_2$-C49, with energies higher by
21meV/atom, 16.6meV/atom, and 14meV/atom, with respect to 
the two-phase fields RuZr$_4$\twophase Zr, RuZr\twophase RuZr$_4$, and RuZr\twophase RuZr$_4$, respectively.
The structure of RuZr$_5$ is similar to MoZr$_5^{proto}$ 
(Appendix (\ref{proto.AB5.MoZr132relaxed})), while 
the structure of RuZr$_3$ is similar to RuTi$_3^{proto}$ 
(Appendix (\ref{proto.AB3.RuTi81})).

\ReferenceRuZr

{\tablefont
  \begin{center}
    \begin{tabular}{||c||} \hline
      {\bf Ru-Zr system} \\ \hline
      Low Temperature Phases comparison chart \\ \hline
      \begin{tabular}{c|c|c}
	Composition  & Experimental  & \tablelineone                      \\ 
	\% Ru        & (\massalski)  & \tablelinetwo                     \\ \hline
	16.6        &two-phase region& RuZr$_5$$\approx$MoZr$_5^{proto}$ \\
        \           &RuZr\twophase Zr& $\sim$21meV/at.                   \\ 
        \            &above 400\DEG  & above RuZr$_4$\twophase Zr        \\ \hline
	20           &same as above  & RuZr$_4$-D1$_a$                   \\ \hline
	25           &same as above  & RuZr$_3$$\approx$RuTi$_3^{proto}$ \\
        \            &               & $\sim$16.6meV/at.                 \\
        \            &               & above RuZr\twophase RuZr$_4$      \\ \hline
	33.3         &same as above  & RuZr$_2$-C49                      \\
        \            &               & $\sim$14meV/at.                   \\ 
        \            &               & above RuZr\twophase RuZr$_4$      \\ \hline
	48 to 52     & B2            & RuZr-B2                           \\ \hline
	66 to 68     & C14 (high T)  & Ru$_2$Zr-C14                      \\ 
        \            &               & $\sim$60meV/at.                   \\
        \            &               & above Ru\twophase RuZr            \\ 
      \end{tabular} \\ \hline
    \end{tabular}
  \end{center}
}

\begin{center}
  \vspace{-3mm}
  \begin{figure}
    \epsfig{file=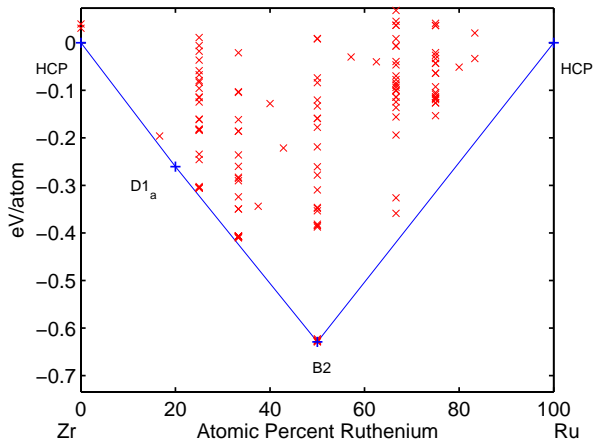,width=\picdim,clip=}
    \vspace{1mm}
    \caption{RuZr (Ruthenium - Zirconium) ground state convex hull.}
    \label{label.fig.RuZr}
  \end{figure}
  \vspace{-1mm}
\end{center}

\newpage

{\bf Tc-Ti (Technetium - Titanium).}
The phase diagram of the system TcTi has been constructed by analogy
with chemically related systems \cite{BB,PP,ref.Titanium,Tc.76Koc,Tc.62Dar}.
Two intermetallic compound, TcTi-B2 and $\chi$ are reported \cite{Tc.76Koc,Tc.62Dar}.
We confirm the stability of TcTi-B2, but we can not say anything about 
$\chi$ since we do not have prototypes at composition 85\% Ru.
In addition, we find Tc$_2$Ti-C11$_b$,
TcTi$_2$-C49, and an orthorhombic phase TcTi$_3$
with space group Immm \#71 and prototype TcTi$_3^{proto}$
described in Appendix (\ref{proto.AB3.TcTi81}).
These intermetallics have large negative formation energy, 
therefore they are expected to be very stable.
\TcDO19

{\tablefont
  \begin{center}
    \begin{tabular}{||c||} \hline
      {\bf Tc-Ti system} \\ \hline
      Low Temperature Phases comparison chart \\ \hline
      \begin{tabular}{c|c|c}
	Composition  & Experimental  & \tablelineone                      \\ 
	\% Tc        & (\massalski)  & \tablelinetwo                     \\ \hline
	25           & disorder      & TcTi$_3^{proto}$                  \\
        \            & $\beta$Ti-A2  &Appendix (\ref{proto.AB3.TcTi81})  \\ \hline
	33.3         & disorder      & TcTi$_2$-C49                      \\
        \            & $\beta$Ti-A2  &                                   \\ \hline
	$\sim$50     & B2            & TcTi-B2                           \\ \hline
	66.6         &two-phase region& Tc$_2$Ti-C11$_b$                 \\ 
        \      & TcTi\twophase $\chi$&                                   \\ \hline
        $\sim$85     & $\chi$-A12    & unavailable                       \\ 
      \end{tabular} \\ \hline
    \end{tabular}
  \end{center}
}

\begin{center}
  \vspace{-3mm}
  \begin{figure}
    \epsfig{file=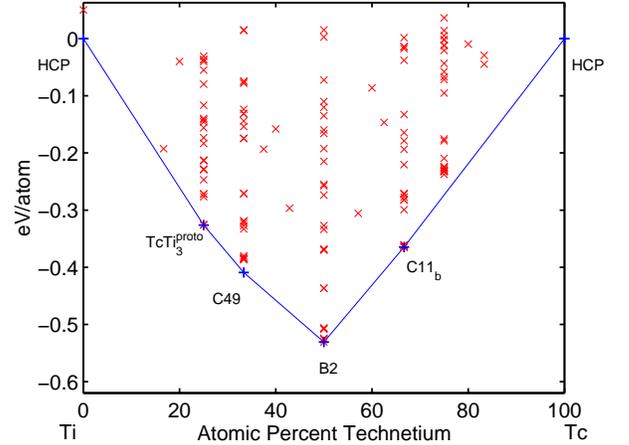,width=\picdim,clip=}
    \vspace{1mm}
    \caption{TcTi (Technetium - Titanium) ground state convex hull.}
    \label{label.fig.Ti_pvTc}
  \end{figure}
  \vspace{-1mm}
\end{center}

\newpage

{\bf Tc-Y (Technetium - Yttrium).}
Not enough information exists in order to construct 
a phase diagram for the system Tc-Y \cite{BB,PP}.
Only one intermetallic compound has been reported: Tc$_2$Y-C14 
(Friauf-Laves/Frank-Kasper phase) \cite{Tc.64Dar,Tc.81Szk}.
We confirm the stability of Tc$_2$Y-C14.
In addition, we find another stable phase TcY$_3$-D0$_{11}$.
\TcDO19

{\tablefont
  \begin{center}
    \begin{tabular}{||c||} \hline
      {\bf Tc-Y system} \\ \hline
      Low Temperature Phases comparison chart \\ \hline
      \begin{tabular}{c|c|c}
	Composition  & Experimental  & \tablelineone                      \\ 
	\% Tc        & (\massalski)  & \tablelinetwo                     \\ \hline
	25           & no information& TcY$_3$-D0$_{11}$                 \\ \hline
	66.6         & C14           & Tc$_2$Y-C14                       \\ 
        \            & \cite{Tc.64Dar,Tc.81Szk} & (Laves phase)         \\ 
      \end{tabular} \\ \hline
    \end{tabular}
  \end{center}
}

\begin{center}
  \vspace{-3mm}
  \begin{figure}
    \epsfig{file=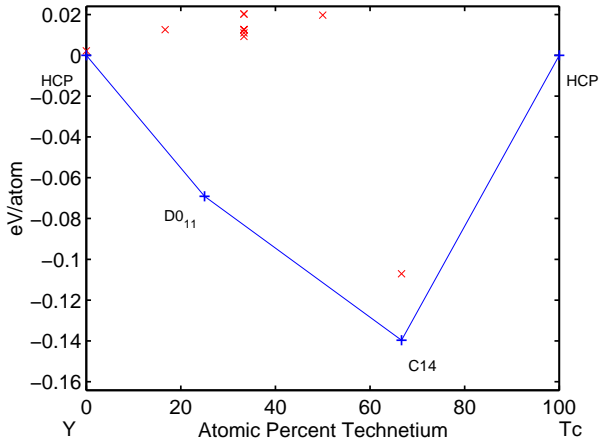,width=\picdim,clip=}
    \vspace{1mm}
    \caption{TcY (Technetium - Yttrium) ground state convex hull.}
    \label{label.fig.Y_pvTc}
  \end{figure}
  \vspace{-1mm}
\end{center}

\newpage

{\bf Tc-Zr (Technetium  - Zirconium).}
Not enough information exists in order to construct 
a phase diagram for the system Tc-Zr \cite{BB,PP}. 
Only two intermetallic compounds have been reported: 
Tc$_2$Zr-C14 (Friauf-Laves/Frank-Kasper phase) and Tc$_6$Zr-A12 \cite{BB,PP,Tc.62Dar,Tc.70Gio}.
We confirm the stability of Tc$_2$Zr-C14, but we cannot
determine A12 since we do not have the proper prototype in our library.
At 50\% composition, Miedema {\it et-al.} reported the existence 
of a TcZr compound with unknown structure \cite{PP,PdTi.76Mie}.
At such composition we find TcZr-B2.
In addition, we find other stable phases: 
TcZr$_4$-D1$_a$ and TcZr$_2$-C49.
These intermetallics have large negative formation energy, 
therefore they are expected to be very stable.
\TcDO19 

{\tablefont
  \begin{center}
    \begin{tabular}{||c||} \hline
      {\bf Tc-Zr system} \\ \hline
      Low Temperature Phases comparison chart \\ \hline
      \begin{tabular}{c|c|c}
	Composition  & Experimental  & \tablelineone                      \\ 
	\% Tc        & (\massalski)  & \tablelinetwo                     \\ \hline
	20           &no information & TcZr$_4$-D1$_a$                   \\ \hline
	33.3         &no information & TcZr$_2$-C49                      \\ \hline
	50           &TcZr (unknown) & TcZr-B2                           \\
        \          &\cite{PdTi.76Mie}&                                   \\ \hline
	66.6         & C14           & Tc$_2$Zr-C14                      \\
        \  &\cite{Tc.62Dar,Tc.70Gio} & (Laves phase)                     \\ \hline
        85.7         & Tc$_6$Zr-A12  & unavailable                       \\ 
      \end{tabular} \\ \hline
    \end{tabular}
  \end{center}
}

\begin{center}
  \vspace{-3mm}
  \begin{figure}
    \epsfig{file=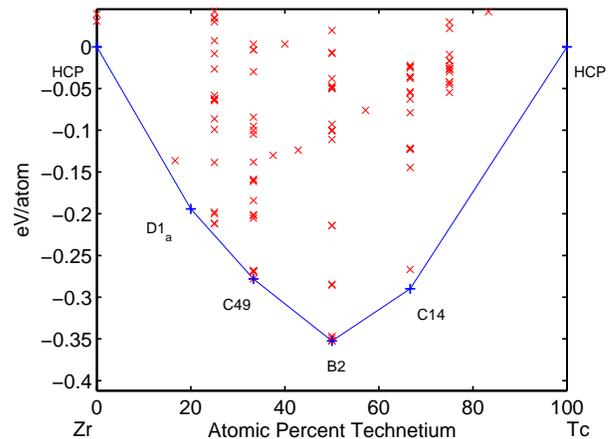,width=\picdim,clip=}
    \vspace{1mm}
    \caption{TcZr (Technetium - Zirconium) ground state convex hull.}
    \label{label.fig.Zr_pvTc}
  \end{figure}
  \vspace{-1mm}
\end{center}

\section{Trend for Technetium alloys}
\label{section.Tc.trend}
In our set of calculations, we have noticed that the phase
D0$_{19}$ appears in systems MTc$_3$ where M is a transition metal 
in the columns on the right of Tc (Tc is in column 7B) while
D0$_{19}$ is not present if M is in the columns on the left of Tc:
D0$_{19}$ is stable in PdTc$_3$, PtTc$_3$, RhTc$_3$, RuTc$_3$,
and unstable in NbTc, TcTi, TcY, and TcZr.

\section{Acknowledgments}
\label{section.acknowledgments}

The research was supported by the 
Department of Energy, Office of Basic Energy Science under Contract No. DE-FG02-96ER45571 and
National Science Foundation Information Technology Research (NSF-ITR) Grant No. DMR-0312537.

It has benefited from discussion with 
John Rodgers, Kristin Persson, Frank Hadley Cocks, 
Chris Fischer, Aleksey Kolmogorov, Vincent Crespi and Zi-Kui Liu.

\onecolumn

\section{Tables of Results}

\subsection{Experimental compounds in agreement with {ab initio} solutions }
{\ftnsz
\begin{center}
    \begin{tabular}{||c||} \hline
      {\bf Experimental compound $\Leftrightarrow$ {\it ab initio} compound}     \\ \hline
      \begin{tabular}{l|l|l|||l|l|l}
        System & Experimental result                         & \tablelineone\,\,\tablelinetwo   & System & Experimental result                         & \tablelineone\,\,\tablelinetwo  \\ \hline 
        Ag-Cd  & AgCd-B2 \cite{PP,AgCd.74Ton,AgCd.78Zar,AgCd.78Mat,AgCd.81Mat} & AgCd-B2/B19/B27 & Nb-Pd  & NbPd$_3$-D0$_{22}$                          & NbPd$_3$-D0$_{22}$               \\ \hline 
        Ag-Mg  & Ag$_3$Mg-D0$_{23}$ \cite{AgMg.87Kul}        & AgMg-D0$_{23}$/D0$_{24}$          & Nb-Pt  & Nb$_3$Pt-A15                                & Nb$_3$Pt-A15                     \\ \hline 
        Ag-Mg  & AgMg-B2                                     & AgMg-B2                           & Nb-Pt  & NbPt$_2$-MoPt$_2$                           & NbPt$_2$-MoPt$_2$                \\ \hline 
        Ag-Na  & Ag$_2$Na-C15                                & Ag$_2$Na-C15                      & Nb-Pt  & NbPt$_3$-D0$_\alpha$                        & NbPt$_3$-D0$_\alpha$             \\ \hline 
        Ag-Ti  & AgTi$_2$-C11$_b$                            & AgTi$_2$-C11$_b$                  & Nb-Rh  & Nb$_3$Rh-$\alpha'$-A15                      & Nb$_3$Rh-A15                     \\ \hline 
        Ag-Ti  & AgTi-B11                                    & AgTi-B11 ($\gamma$CuTi)           & Nb-Rh  & NbRh-L1$_0$                                 & NbRh-L1$_0$                      \\ \hline 
        Ag-Y   & AgY-B2                                      & AgY-B2                            & Nb-Rh  & $\epsilon$(Nb$_{2}$Rh$_{3}$)-B19 (high T)   & NbRh-B19                         \\ \hline 
        Ag-Y   & Ag$_2$Y-C11$_b$                             & Ag$_2$Y-C11$_b$                   & Nb-Rh  & $\eta$(Nb$_{13}$Rh$_{27}$)-Al$_3$Pu         & NbRh$_{3}$-Al$_3$Pu              \\ \hline 
        Ag-Zr  & AgZr$_2$-C11$_b$                            & AgZr$_2$-C11$_b$                  & Pd-Ti  & PdTi$_3$-A15 at 20\%                        & PdTi$_3$-A15 at 25\%             \\ \hline 
        Ag-Zr  & AgZr-B11                                    & AgZr-B11 ($\gamma$CuTi)           & Pd-Ti  & PdTi$_2$-C11$_b$                            & PdTi$_2$-C11$_b$                 \\ \hline 
        Al-Sc  & AlSc$_2$-B8$_2$                             & AlSc$_2$-B8$_2$                   & Pd-Ti  & $\alpha$(TiPd)-B19                          & PdTi-B19                         \\ \hline 
        Al-Sc  & AlSc-B2                                     & AlSc-B2                           & Pd-Ti  & Pd$_3$Ti-D0$_{24}$                          & Pd$_3$Ti-D0$_{24}$               \\ \hline 
        Al-Sc  & Al$_2$Sc-C15                                & Al$_2$Sc-C15                      & Pd-Y   & PdY$_3$-D0$_{11}$                           & PdY$_3$-D0$_{11}$                \\ \hline 
        Al-Sc  & Al$_3$Sc-L1$_2$                             & Al$_3$Sc-L1$_2$                   & Pd-Y   & Pd$_3$Y-L1$_2$                              & Pd$_3$Y-L1$_2$                   \\ \hline 
        Au-Cd  & $\beta'$AuCd-B19                            & AuCd-B19                          & Pd-Zr  & PdZr$_2$-C11$_b$                            & PdZr$_2$-C11$_b$                 \\ \hline 
        Au-Cd  & $\beta$AuCd-B2 (high T)                     & AuCd-B2                           & Pd-Zr  & PdZr-B33                                    & PdZr-B33                         \\ \hline 
        Au-Nb  & Au$_2$Nb-C32                                & Au$_2$Nb-C32 (\paw)               & Pd-Zr  & Pd$_2$Zr-C11$_b$                            & Pd$_2$Zr-C11$_b$/tie-line (\paw) \\ \hline 
        Au-Sc  & Au$_4$Sc-D1$_a$                             & Au$_4$Sc-D1$_a$                   & Pd-Zr  & Pd$_3$Zr-D0$_{24}$                          & Pd$_3$Zr-D0$_{24}$               \\ \hline 
        Au-Sc  & Au$_2$Sc-C11$_b$                            & Au$_2$Sc-C11$_b$/MoPt$_2$         & Pt-Ti  & Pt$_3$Ti-D0$_{24}$                          & Pt$_3$Ti-D0$_{24}$               \\ \hline 
        Au-Sc  & AuSc-B2                                     & AuSc-B2/B19                       & Pt-Ti  & $\alpha$PtTi-B19                            & PtTi-B19                         \\ \hline 
        Au-Ti  & Au$_4$Ti-D1$_a$                             & Au$_4$Ti-D1$_a$                   & Pt-Ti  & PtTi$_3$-A15                                & PtTi$_3$-A15                     \\ \hline 
        Au-Ti  & Au$_2$Ti-C11$_b$                            & Au$_2$Ti-C11$_b$/MoPt$_2$         & Pt-Y   & Pt$_3$Y$_5$-D8$_8$                          & Pt$_3$Y$_5$-D8$_8$ (\paw)        \\ \hline 
        Au-Ti  & $\alpha$AuTi-B11                            & AuTi-B11                          & Pt-Y   & Pt$_3$Y-L1$_2$                              & Pt$_3$Y-L1$_2$                   \\ \hline 
        Au-Ti  & AuTi$_3$-A15                                & AuTi$_3$-A15                      & Pt-Y   & Pt$_2$Y-C15                                 & Pt$_2$Y-C15                      \\ \hline 
        Au-Y   & Au$_3$Y-D0$_a$                              & Au$_3$Y-D0$_a$                    & Pt-Y   & PtY$_2$-C37                                 & PtY$_2$-C37                      \\ \hline 
        Au-Y   & Au$_2$Y-C11$_b$                             & Au$_2$Y-C11$_b$                   & Pt-Y   & PtY$_3$-D0$_{11}$                           & PtY$_3$-D0$_{11}$                \\ \hline 
        Au-Zr  & Au$_3$Zr-D0$_a$                             & Au$_3$Zr-D0$_a$                   & Pt-Zr  & Pt$_3$Zr-D0$_{24}$                          & Pt$_3$Zr-D0$_{24}$               \\ \hline 
        Au-Zr  & Au$_2$Zr-C11$_b$                            & Au$_2$Zr-C11$_b$                  & Pt-Zr  & PtZr-B33                                    & PtZr-B33                         \\ \hline 
        Au-Zr  & AuZr$_2$-CuZr$_2$/C11$_b$                   & AuZr$_2$-CuZr$_2$ (\paw)          & Rh-Ti  & RhTi$_2$-C11$_b$\cite{BB}/CuZr$_2$\cite{PP} & RhTi$_2$-C11$_b$/CuZr$_2$        \\ \hline 
        Au-Zr  & AuZr$_3$-A15                                & AuZr$_3$-A15                      & Rh-Ti  & $\alpha$RhTi-L1$_0$                         & RhTi-L1$_0$                      \\ \hline 
        Cd-Pd  & CdPd-$\beta_1$-L1$_0$                       & CdPd-L1$_0$                       & Rh-Ti  & Rh$_3$Ti-L1$_2$                             & Rh$_3$Ti-L1$_2$                  \\ \hline 
        Cd-Pt  & CdPt-$\alpha'_1$-L1$_0$                     & CdPt-L1$_0$                       & Rh-Y   & RhY$_3$-D0$_{11}$                           & RhY$_3$-D0$_{11}$                \\ \hline 
        Cd-Ti  & CdTi$_2$-C11$_b$                            & CdTi$_2$-C11$_b$                  & Rh-Y   & RhY-B2                                      & RhY-B2                           \\ \hline 
        Cd-Ti  & CdTi-B11                                    & CdTi-B11 ($\gamma$CuTi)           & Rh-Y   & Rh$_2$Y-C15                                 & Rh$_2$Y-C15                      \\ \hline 
        Cd-Y   & CdY-B2                                      & CdY-B2                            & Rh-Zr  & RhZr$_2$-C16                                & RhZr$_2$-C16                     \\ \hline 
        Cd-Y   & Cd$_2$Y-C6                                  & Cd$_2$Y-C6                        & Rh-Zr  & $\alpha$RhZr-B27                            & RhZr-B27                         \\ \hline 
        Cd-Y   & Cd$_3$Y-Cd$_3$Er                            & Cd$_3$Y-Cd$_3$Er                  & Rh-Zr  & Rh$_3$Zr-L1$_2$                             & Rh$_3$Zr-L1$_2$                  \\ \hline 
        Cd-Zr  & CdZr$_2$-C11$_b$                            & CdZr$_2$-C11$_b$                  & Ru-Ti  & RuTi-B2                                     & RuTi-B2                          \\ \hline 
        Cd-Zr  & CdZr-B11                                    & CdZr-B11 (\paw)                   & Ru-Y   & RuY$_3$-D0$_{11}$                           & RuY$_3$-D0$_{11}$                \\ \hline 
        Cd-Zr  & Cd$_3$Zr-L1$_2$                             & Cd$_3$Zr-L1$_2$                   & Ru-Y   & Ru$_2$Y-C14                                 & Ru$_2$Y-C14                      \\ \hline 
        Mo-Pt  & MoPt-B19                                    & MoPt-B19                          & Ru-Zr  & RuZr-B2                                     & RuZr-B2                          \\ \hline 
        Mo-Pt  & MoPt$_2$                                    & MoPt$_2$                          & Tc-Ti  & TcTi-B2                                     & TcTi-B2                          \\ \hline 
        Mo-Rh  & MoRh-B19                                    & MoRh-B19                          & Tc-Y   & Tc$_2$Y-C14                                 & Tc$_2$Y-C14                      \\ \hline 
        Mo-Zr  & Mo$_2$Zr-C15                                & Mo$_2$Zr-C15                      & Tc-Zr  & Tc$_2$Zr-C14                                & Tc$_2$Zr-C14                     \\ \hline 
        Nb-Pd  & NbPd$_2$-MoPt$_2$                           & NbPd$_2$-MoPt$_2$                 &        &                                             &                                  \\ 
      \end{tabular} \\ \hline
    \end{tabular}
  \end{center}
}

  {TABLE 5: Experimental compounds in agreement with {\it ab initio} solutions (\NNagreementsGGA \,entries). }  
\clearpage


\subsection{Experimentally unknown, non-identified or speculated compounds and {ab initio} predictions}
{\ftnsz
\begin{center}
    \begin{tabular}{||c||} \hline
      {\bf Experimental unknown/speculated compound $\Leftrightarrow$ {\it ab initio} compound}     \\ \hline
      \begin{tabular}{l|l|l}
        System     & Experimental result               & \tablelineone\,\,\tablelinetwo                 \\ \hline
        Ag-Cd      & $\beta'$-bcc ordered              & AgCd-B2/B19/B27                                 \\ \hline 
        Ag-Mg      & cF* (unknown)                     & AgMg$_3$-D0$_{19}$ (hP8) / D0$_{a}$ (oP8)       \\ \hline 
	Au-Cd      & Au$_3$Cd-$\alpha_2$-hP? (unknown) & Au$_3$Cd-D0$_{24}$/D0$_{19}$/Al$_3$Pu           \\ \hline 
	Au-Cd      & $\alpha''$ $\sim$ AuCd (unknown)  & AuCd-L1$_0$/\CH40                               \\ \hline 
	Au-Cd      &$\epsilon'$$\sim$AuCd$_3$ (unknown)& AuCd$_3$-L6$_0$                                 \\ \hline 
        Au-Pd      & Au$_3$Pd-L1$_2$ (speculated)      & Au$_3$Pd-D0$_{23}$/D0$_{22}$/L1$_2$             \\
	           & (wide concentration range)        & Au$_4$Pd-D1$_a$/tie-line                        \\ \hline 
        Au-Pd      & AuPd (speculated and unknown)     & AuPd-\CH40                                      \\ \hline 
        Au-Pd      & AuPd$_3$-L1$_2$ (speculated)      & AuPd$_3$-D0$_{23}$/D0$_{22}$/L1$_2$/tie-line    \\ \hline 
        Au-Sc      & AuSc$_2$-C37 (speculated)         & AuSc$_2$-C37                                    \\ \hline 
	Cd-Pd      & $\gamma'$-(unknown)               & Cd$_3$Pd-D0$_{19}$/D0$_{24}$/NbPd$_3$/Al$_3$Pu  \\ \hline 
        Cd-Pt      & $\gamma_1$-(unknown)              & Cd$_3$Pt-D0$_{11}$/D0$_{a}$/D0$_{22}$           \\ \hline 
        Cd-Pt      & Cd$_2$Pt-(unknown)                & Cd$_2$Pt-C37/C16/tie-line                       \\ \hline 
        Cd-Zr      & cubic (unknown)                   & Cd$_2$Zr-C11$_b$                                \\ \hline 
        Mo-Pd      & MoPd$_2\sim$MoPt$_2$              & MoPd$_2$-ZrSi$_2$                               \\ \hline 
        Mo-Rh      & MoRh$_3$ (unknown)                & MoRh$_3$-CdMg$_3$                               \\ \hline 
        Nb-Pt      & Nb two-phase region$_{1-x}$Pt$_{1+x}$-B19 & NbPt-L1$_0$                            \\ \hline 
        Pd-Ti &orthorhombic distortion of Pd$_2$Ti-C11$_b$ & Pd$_2$Ti-MoPt$_2$ (distortion of C11$_b$)   \\ \hline 
        Pd-Y       & $\alpha$PdY (unknown)             & PdY-B27                                         \\ \hline 
        Pt-Ru      & PtRu-FCC (unknown)                & PtRu-FCC$_{A2B2}^{[001]}$                       \\ \hline 
        Pt-Y       & Pt$_5$Y (unknown)                 & Pt$_5$Y-D2$_d$ metastable                       \\ \hline 
        Tc-Zr      & TcZr (unknown)                    & TcZr-B2                                         \\ 
      \end{tabular} \\ \hline
    \end{tabular}
  \end{center}
}
{TABLE 6: Experimentally unknown, non-identified or speculated compounds and {\it ab initio} predictions (\NNgoodpredictions \,entries).} 


\subsection{Experimental solid solutions, two-phases and ``not studied'' regions and possible {ab initio} predictions}
{\ftnsz
\begin{center}
    \begin{tabular}{||c||} \hline
      {\bf  Experimental non-compound $\Leftrightarrow$ {\it ab initio} compound}     \\ \hline
      \begin{tabular}{l|l|l}
        System     & Experimental result               & \tablelineone\,\,\tablelinetwo     \\ \hline
	Ag-Au      & short-range order $\gtrsim$950\DEG& Ag$_4$Au-D1$_a$/tie-line                                          \\ \hline
	Ag-Au      & short-range order $\gtrsim$950\DEG& Ag$_3$Au-L1$_2$/D0$_{23}$/Al$_3$Pu/NbPd$_3$/D0$_{22}$/D0$_{24}$   \\ \hline 
	Ag-Au      & short-range order $\gtrsim$950\DEG& Ag$_2$Au-C37/MoPt$_2$                                             \\ \hline 
	Ag-Au      & short-range order $\gtrsim$950\DEG& AgAu-L1$_0$                                                       \\ \hline 
	Ag-Au      & short-range order $\gtrsim$950\DEG& AgAu$_2$-C37/MoPt$_2$                                             \\ \hline 
	Ag-Au      & short-range order $\gtrsim$950\DEG& AgAu$_3$-L1$_2$/D0$_{22}$/D0$_{23}$                               \\ \hline 
        Ag-Cd      & fcc solid solution                & Ag$_3$Cd-D0$_{22}$/D0$_{24}$                                      \\ \hline 
        Ag-Cd      & fcc solid solution                & Ag$_2$Cd-C37                                                      \\ \hline 
        Ag-Cd      & none                              & AgCd$_2$-ZrSi$_2$                                                 \\ \hline 
        Ag-Cd      & hcp solid solution                & AgCd$_3$-D0$_{19}$                                                \\ \hline 
        Ag-Pd      & solid solution $\gtrsim$900\DEG   & AgPd-L1$_1$                                                       \\ \hline 
        Ag-Pd      & solid solution $\gtrsim$900\DEG   & Ag$_2$Pd-C37                                                      \\ \hline 
        Ag-Pd      & solid solution $\gtrsim$900\DEG   & Ag$_3$Pd- L1$_2$/D0$_{22}$                                        \\ \hline 
        Ag-Y       & two-phase region above 200\DEG    & AgY$_2$-C37/tie-line (uncertain)                                  \\ \hline 
        Ag-Y       & two-phase region above 200\DEG    & Ag$_3$Y-D0$_a$ (uncertain)                                        \\ \hline 
        Ag-Zr      & two-phase region above 700\DEG    & AgZr$_3$-FCC$_{AB3}^{[001]}$/tie-line (uncertain)                 \\ \hline 
        Ag-Zr      & two-phase region above 700\DEG    & Ag$_2$Zr-C32                                                      \\ \hline 
        Al-Sc      & two-phase region above 0\DEG      & AlSc$_3$-D0$_{19}$                                                \\ \hline 
	Au-Nb      & two-phase region (calculated)     & AuNb$_2$-\str64                                                   \\ \hline 
        Au-Pd      & solid solution                    & Au$_2$Pd-C49/C37                                                  \\ \hline 
	Au-Ti      & two-phase region above 500\DEG    & Au$_4$Ti$_3$-Cu$_4$Ti$_3$/tie                                     \\ \hline 
	Au-Y       & not studied                       & AuY$_2$-C37                                                       \\ \hline 
	Au-Zr      & two-phase region                  & Au$_4$Zr$_3$-Cu$_4$Ti$_3$ (unreliable)                            \\ \hline 
	Au-Zr      & two-phase region                  & AuZr-B11/FCC$_{A2B2}^{[001]}$ (unreliable)                        \\ \hline 
        Cd-Pd      & Pd phase above 100\DEG            & CdPd$_3$-D0$_{22}$/NbPd$_3$                                       \\ \hline 
        Cd-Pd      & two-phase region above 100\DEG    & CdPd$_2$-C37                                                      \\ \hline 
        Cd-Rh      & not studied                       & Cd$_2$Rh-C37                                                      \\ \hline 
        Cd-Rh      & not studied                       & Cd$_3$Rh-Al$_3$Pu                                                 \\ \hline 
	Cd-Ti      & two-phase region above 200\DEG    & Cu$_4$Ti$_3$/tie-line                                             \\ \hline 
        Cd-Y       & not studied/two-phase region      & CdY$_2$-C37                                                       \\ \hline 
        Cd-Zr      & not studied/two-phase region      & CdZr$_3$-A15                                                      \\ \hline 
        Mo-Nb      & not studied                       & MoNb$_2$-C11$_b$                                                  \\ \hline 
        Mo-Nb      & not studied                       & MoNb-B2                                                           \\ \hline 
        Mo-Nb      & not studied                       & Mo$_2$Nb-C11$_b$                                                  \\ \hline 
        Mo-Nb      & not studied                       & Mo$_3$Nb -D0$_3$                                                  \\ \hline 
        Mo-Pd      & disorder fcc Pd-A1                & MoPd$_4$-D1$_a$                                                   \\ \hline 
        Mo-Pt      & two-phase region MoPt$_2$\twophase Pt &MoPt$_4$-D1$_a$                                                \\ \hline 
        Mo-Rh      & two-phase region above 900\DEG    & Mo$_2$Rh-C37                                                      \\ \hline 
        Mo-Ru      & disorder hcp Ru-A3 above 800\DEG  & MoRu$_3$-D0$_{19}$                                                \\ \hline 
        Mo-Ti      & not studied/two-phase region above 400\DEG&MoTi$_3^{proto}$/tie-line (uncertain), Appendix (\ref{proto.AB3.MoTi81}) \\ \hline 
        Mo-Ti      & not studied/two-phase region above 400\DEG&MoTi$_2$-BCC$_{AB2}^{[211]}$                               \\ \hline 
        Mo-Ti      & not studied/two-phase region above 400\DEG&MoTi$^{proto}$, Appendix (\ref{proto.AB.MoTi80})           \\ \hline 
        Mo-Ti      & not studied/two-phase region above 400\DEG&Mo$_2$Ti-C11$_b$                                           \\ \hline 
        Mo-Ti      & not studied/two-phase region above 400\DEG&Mo$_3$Ti$^{proto}$, Appendix (\ref{proto.AB3.MoTi81})      \\ \hline 
        Mo-Ti      & not studied/two-phase region above 400\DEG&Mo$_4$Ti-D1$_a$/tie-line                                   \\ \hline 
        Mo-Ti      & not studied/two-phase region above 400\DEG&Mo$_5$Ti$^{proto}$, Appendix (\ref{proto.AB5.MoTi128relaxed})\\ \hline 
        Mo-Zr      & two-phase region above 400\DEG& MoZr$_5^{proto}$/tie-line, Appendix (\ref{proto.AB5.MoZr132relaxed}) (uncertain)\\ \hline 
        Mo-Zr      & two-phase region above 400\DEG& MoZr$_3^{proto}$/tie-line, Appendix (\ref{proto.AB5.MoZr132relaxed}) (uncertain)\\ \hline 
        Mo-Zr      & two-phase region above 400\DEG& $\sim$ MoTi$^{proto}$ (uncertain)                                     \\
                   &                                   & ($\sim$2meV above two-phase reg.)                                 \\ 
      \end{tabular} \\ \hline
    \end{tabular}
  \end{center}}
  {TABLE 7a: Experimental solid solutions, two-phases and ``not studied'' regions and possible {\it ab initio} 
   predictions. The table continues in the next page.}

 \clearpage

{\ftnsz
\begin{center}
    \begin{tabular}{||c||} \hline
      {\bf  Experimental non-compound $\Leftrightarrow$ {\it ab initio} compound}     \\ \hline
      \begin{tabular}{l|l|l}
        System     & Experimental result               & \tablelineone\,\,\tablelinetwo                                   \\ \hline
        Nb-Pd      & two-phase region above 700\DEG    & Nb$_2$Pd-BCC$_{AB2}^{[011]}$                                      \\ \hline 
        Nb-Ru      & disorder Nb-A2                    & Nb$_5$Ru $^{proto}$, Appendix (\ref{proto.AB5.NbRu132relaxed})    \\ \hline 
        Nb-Ru      & disorder Nb-A2                    & Nb$_3$Ru-D0$_3$                                                   \\ \hline 
        Nb-Ru      & two-phase region above 700\DEG    & NbRu$_2$-C37                                                      \\ \hline 
        Nb-Tc      & not studied/unknown               & Nb$_3$Tc$^{proto}$, Appendix (\ref{proto.AB3.NbTc81})             \\ \hline 
        Nb-Tc      & not studied/unknown               & Nb$_2$Tc-C11$_b$                                                  \\ \hline 
        Nb-Tc      & not studied/unknown               & NbTc-B2                                                           \\ \hline 
        Pd-Pt      & two-phase region Pd\twophase Pt   & Pd$_3$Pt$^{proto}$/tie-line (uncertain), Appendix (\ref{proto.AB3.PdPt11relaxed})\\ \hline 
        Pd-Pt      & two-phase region Pd\twophase Pt   & PdPt-L1$_1$                                                       \\ \hline 
        Pd-Pt      & two-phase region Pd\twophase Pt   & PdPt$_3^{proto}$, Appendix (\ref{proto.AB3.PdPt11relaxed})        \\ \hline 
	Pd-Tc      & two-phase region above $\sim$1000\DEG& PdTc$_3$-D0$_{19}$                                             \\ \hline 
        Pd-Y       & two-phase region                  & PdY$_2$-C37 (uncertain)                                           \\ \hline 
        Pd-Zr      & two-phase region                  & PdZr$_3$-FCC$_{AB3}^{[001]}$ (uncertain)                          \\ \hline 
        Pt-Rh      & two-phase region Pt\twophase Rh \cite{BB}, D1$_a$ \cite{PdPt.91Zun} & Pt$_4$Rh-D1$_a$                 \\ \hline 
        Pt-Rh      & two-phase region Pt\twophase Rh   & Pt$_3$Rh-D0$_{22}$                                                \\ \hline 
        Pt-Rh      & two-phase region Pt\twophase Rh   & PtRh-NbP                                                          \\ \hline 
        Pt-Rh      & two-phase region Pt\twophase Rh   & PtRh$_2$-C49                                                      \\ \hline 
        Pt-Rh      & two-phase region Pt\twophase Rh   & PtRh$_3$-D0$_{22}$                                                \\ \hline 
        Pt-Rh      & two-phase region Pt\twophase Rh \cite{BB}, D1$_a$ \cite{PdPt.91Zun} & PtRh$_4$-D1$_a$                 \\ \hline 
        Pt-Ru      & disorder Pt-A1                    & Pt$_3$Ru-FCC$_{AB3}^{[001]}$/tie-line (uncertain)                 \\ \hline 
        Pt-Tc      & disorder Pt-A1                    & Pt$_3$Tc-FCC$_{AB3}^{[001]}$                                      \\ \hline 
        Pt-Tc      &two-phase region above $\sim$1000\DEG& PtTc$_3$-D0$_{19}$                                              \\ \hline 
        Pt-Zr      & two-phase region above 600\DEG    & PtZr$_2$-C16 (uncertain)                                          \\ \hline 
        Pt-Zr      & two-phase region above 600\DEG    & PtZr$_3$-A15 (uncertain)                                          \\ \hline 
	Rh-Ru      & solid solution Ru-A3              & RhRu$_2$$^{proto}$                                                \\ \hline 
	Rh-Ru      & solid solution Ru-A3              & RhRu$^{proto}$                                                    \\ \hline 
	Rh-Tc      & solid solution Tc-A3              & RhTc$_3$-D0$_{19}$                                                \\ \hline 
	Rh-Tc      & solid solution Tc-A3              & RhTc-B19                                                          \\ \hline 
	Rh-Tc      & two-phase region above 1000\DEG   & Rh$_2$Tc$_2$-ZrSi$_2$                                             \\ \hline 
        Rh-Ti      & two-phase region above 600\DEG    & Rh$_2$Ti-C37                                                      \\ \hline 
        Rh-Y       & two-phase region above 0\DEG      & RhY$_2$-C37 (uncertain)                                           \\ \hline 
        Rh-Zr      & two-phase region above 0\DEG      & RhZr$_4$-D1$_a$/tie-line                                          \\ \hline 
        Rh-Zr      & two-phase region above 0\DEG      & RhZr$_3$-FCC$_{AB3}^{[001]}$/tie-line                             \\ \hline 
        Rh-Zr      & two-phase region above 0\DEG      & Rh$_2$Zr-C37/tie-line                                             \\ \hline 
        Ru-Tc      & disorder solution (Ru,Tc)-A3      & RuTc$_3$-D0$_{19}$                                                \\ \hline 
        Ru-Tc      & disorder solution (Ru,Tc)-A3      & RuTc-B19                                                          \\ \hline 
        Ru-Tc      & disorder solution (Ru,Tc)-A3      & Ru$_3$Tc-D0$_{19}$                                                \\ \hline 
        Ru-Ti      & two-phase region above 600\DEG    & RuTi$_3^{proto}$, Appendix (\ref{proto.AB3.RuTi81})               \\ \hline 
        Ru-Ti      & two-phase region above 600\DEG    & RuTi$_2$-C49                                                      \\ \hline 
        Ru-Y       & two-phase region above 0\DEG      & RuY$_2$-C16 (uncertain)                                           \\ \hline 
	Ru-Zr      & two-phase region above 400\DEG    & RuZr$_4$-D1$_a$                                                   \\ \hline 
        Tc-Ti      & disorder $\beta$Ti-A2             & TcTi$_3^{proto}$, Appendix (\ref{proto.AB3.TcTi81})               \\ \hline 
        Tc-Ti      & disorder $\beta$Ti-A2             & TcTi$_2$-C49                                                      \\ \hline 
        Tc-Ti      & two-phase region TcTi\twophase $\chi$ & Tc$_2$Ti-C11$_b$                                              \\ \hline 
        Tc-Y       & no information                    & TcY$_3$-D0$_{11}$                                                 \\ \hline 
        Tc-Zr      & no information                    & TcZr$_4$-D1$_a$                                                   \\ \hline 
        Tc-Zr      & no information                    & TcZr$_2$-C49                                                      \\ 
      \end{tabular} \\ \hline
    \end{tabular}
  \end{center}
}
  {TABLE 7b: Experimental solid solutions, two-phases and ``not studied'' regions and possible {\it ab initio} 
         predictions (Table 7a + Table 7b = \NNpossiblepredictions \,entries). 
         The table starts in the previous page.}


\subsection{Experimental compounds that could not be checked by our calculations}
{\ftnsz
\begin{center}
    \begin{tabular}{||c||} \hline
      {\bf  Experimental compound $\Leftrightarrow$ unavailable compound}     \\ \hline
      \begin{tabular}{l|l|l}
        System     & Experimental result                         & \tablelineone\,\,\tablelinetwo                     \\ \hline
        Ag-Mg      & AgMg$_4$-hP* (unknown)                      & \unkexp                           \\ \hline 
        Ag-Y       & Ag$_{51}$Y$_{14}$-Ag$_{51}$Gd$_{14}$        & \unvproto                         \\ \hline 
	Au-Cd      & $\eta'$ (unknown)                           & \unkexp                           \\ \hline 
	Au-Cd      & $\alpha_1$Au$_3$Cd-Ag$_3$Mg                 & \unvproto                         \\ \hline 
	Au-Cd      & Au$_3$Cd$_5$-D8$_m$                         & \unkexp                           \\ \hline 
	Au-Nb      & Au$_2$Nb$_3$ (unknown)                      & \unkexp                           \\ \hline 
	Au-Ti      & $\beta$AuTi-B19                             & \unvproto \,off-stoichiometry B19 \\ \hline 
	Au-Y       & AuY$_3$ (unknown)                           & \unkexp                           \\ \hline 
	Au-Zr      & Au$_4$Zr (oP20 - Pnma)                      & \unvproto                         \\ \hline 
	Au-Zr      & Au$_{10}$Zr$_7$ (tI34)                      & \unvproto                         \\ \hline 
	Au-Zr      & Au$_4$Zr$_5$ (unknown prototype)            & \unkexp                           \\ \hline 
        Cd-Pd      & $\gamma_1$-(unknown)                        & \unkexp                           \\ \hline 
        Cd-Pd      & $\gamma$-D8$_3$                             & \unvproto                         \\ \hline 
        Cd-Pt      & $\gamma$-(unknown)                          & \unkexp                           \\ \hline 
        Cd-Pt      & $\gamma_2$-(unknown)                        & \unkexp                           \\ \hline 
        Cd-Y       & Cd$_{45}$Y$_{11}$-Cd$_{45}$Sm$_{11}$        & \unvproto                         \\ \hline 
        Cd-Y       & Cd$_{58}$Y$_{13}$-Pu$_{13}$Zn$_{58}$        & \unvproto                         \\ \hline 
        Cd-Y       & Cd$_{6}$Y                                   & \unvproto                         \\ \hline 
        Mo-Pt      & $\epsilon'$-D0$_{19}$ above 1000\DEG        & \unvproto \,off-stoichiometry D0$_{19}$ \\ \hline 
        Nb-Pt      & D8$_b$                                      & \unvproto \,$\sigma$ phase        \\ \hline 
        Nb-Rh      & $\sigma$(Nb$_{13}$Rh$_{7}$)-D8$_b$          & \unvproto                         \\ \hline 
        Nb-Rh      & $\xi$(Nb$_{2}$Rh$_{3}$)-Nb$_{2}$Rh$_{3}$    & \unvproto                         \\ \hline 
        Nb-Tc      & NbTc$_3$,Nb$_{0.15}$Tc$_{0.85}$-$\alpha$Mn  & \unvproto                         \\ \hline 
        Pd-Ti      & Pd$_3$Ti$_2$ $\sim$ Au$_2$V                 & \unvproto                         \\ \hline 
        Pd-Ti      & Pd$_5$Ti$_3$ $\sim$ C11$_b$                 & \unvproto                         \\ \hline 
        Pd-Y       & unknown at 66.6\% Pd                        & \unkexp                           \\ \hline 
        Pd-Y       & unknown at 87.5\% Pd                        & \unkexp                           \\ \hline 
        Pd-Y       & Pd$_2$Y$_5$ (unknown)                       & \unkexp                           \\ \hline 
        Pd-Y       & Pd$_2$Y$_3$-hR15 (unknown)                  & \unkexp                           \\ \hline 
        Pd-Y       & Pd$_4$Y$_3$-hR14 (unknown)                  & \unkexp                           \\ \hline 
        Pd-Y       & $\alpha$Pd$_3$Y$_2$ (unknown)               & \unkexp                           \\ \hline 
        Pt-Ti      & Pt$_8$Ti-D1$_\alpha$                        & \unvproto \,off-stoichiometry D1$_\alpha$ \\ \hline 
        Pt-Ti      & Pt$_2$Ti-ReSi$_2$                           & \unvproto                         \\ \hline 
        Pt-Y       & Pt$_4$Y$_3$-Pd$_4$Pu$_3$                    & \unvproto                         \\ \hline 
        Pt-Y       & Pt$_4$Y$_5$-Pu$_5$Rh$_4$                    & \unvproto                         \\ \hline 
        Pt-Y       & Pt$_3$Y$_7$-D10$_2$                         & \unvproto                         \\ \hline 
        Rh-Ti      & Rh$_5$Ti (unknown)                          & \unkexp                           \\ \hline 
        Rh-Ti      & Rh$_5$Ti$_3$-Ge$_3$Rh$_5$                   & \unvproto                         \\ \hline 
        Rh-Y       & Rh$_3$Y (unknown), hP24-P6$_3$/mmc          & \unkexp                           \\ \hline 
        Rh-Y       & Rh$_3$Y$_7$-D10$_2$                         & \unvproto                         \\ \hline 
        Rh-Y       & Rh$_3$Y$_5$ (unknown)                       & \unkexp                           \\ \hline 
        Rh-Y       & Rh$_2$Y$_3$ (unknown), tI140-I4/mcm         & \unkexp                           \\ \hline 
        Rh-Zr      & Rh$_4$Zr$_3$ (unknown)                      & \unkexp                           \\ \hline 
        Rh-Zr      & Rh$_5$Zr$_3$-Pd$_3$Pu$_3$                   & \unvproto                         \\ \hline 
        Ru-Y       & Ru$_{2}$Y$_{5}$-C$_2$Mn$_5$                 & \unvproto                         \\ \hline 
        Ru-Y       & Ru$_{25}$Y$_{44}$ (unknown)                 & \unkexp                           \\ \hline 
        Tc-Ti      & $\chi$-A12                                  & \unvproto                         \\ \hline 
        Tc-Zr      & Tc$_6$Zr-A12                                & \unvproto                         \\ 
      \end{tabular}  \\ \hline
    \end{tabular}
  \end{center}
}
  {TABLE 8: Experimental compounds that could not be checked by our calculations, because the proper 
    structure prototype or concentration is not known or not in our library. 
    The table contains \NNimpossibleunkexp {\it ``\unkexp s''} and \NNimpossibleunvproto {\it ``\unvproto s''} (\NNimpossible \,entries total)}. 

\clearpage

\def\sss{\ }

\subsection{Experimental compounds in disagreement with other {ab initio} compounds or two-phase regions}
{\ftnsz
\begin{center}
    \begin{tabular}{||c||} \hline
      {\bf  Experimental compound  $\Leftrightarrow$ {\it ab initio} wrong compound or two-phases}   \\ \hline
      \begin{tabular}{l|l|l|r|r|c}
        System     & Experimental result               & \tablelineone\,\,\tablelinetwo & $\Delta$E$_f$ (\us) & {\it $\Delta$E$_f$ (\paw)}   & Note \\
                   &                                   &                                 & {\ftnsz(meV/atom )} & {\ftnsz \it (meV/atom) }     &      \\ \hline
	Au-Nb      & AuNb$_3$-A15                      & two-phase region                          & \sss 7  & 6.5                            &  (a) \\ \hline 
	Au-Y       & AuY-B2 \cite{AuY.63Cha}           & AuY-B33                                   & \sss 26 & 25                             &  (b) \\ \hline 
        Cd-Nb      & Cd$_3$Nb-L1$_2$                   & immiscible system                         & \sss 70 &  $>$100                        &  (c) \\ \hline 
        Cd-Pt&CdPt$_3-\alpha'$-L1$_2$ \cite{CdPt.52Now}& CdPt$_3^{proto}$, Appendix (\ref{proto.AB3.CdPt11relaxed})& \sss 25 & 10.4           &  (b) \\ \hline 
        Nb-Rh      & $\kappa$(NbRh$_{3}$)-L1$_2$       & NbRh$_{3}$-Al$_3$Pu                       & \sss 8  &  5.3                           &  (a) \\ \hline 
        Nb-Ru      & NbRu$_3$-L1$_2$                   & NbRu$_3$-D0$_{24}$                        & \sss 8  &  2.5                           &  (a) \\ \hline 
        Nb-Ru      & NbRu$'$-L1$_0$                    & two-phase region                          & \sss 20 &  4                             &  (a) \\ \hline 
        Pt-Y       & PtY-B27                           & PtY-B33                                   & \sss 50 & 60                             &  (c) \\ \hline 
        Pt-Zr      & Pt$_3$Zr$_5$-D8$_8$               & two-phase region                          & \sss 36 &  23                            &  (c) \\ 
      \end{tabular} \\ \hline
    \end{tabular}
  \end{center}
}
  {TABLE 9: Experimental compounds (which we have in the library as prototypes) in disagreement with
other {\it ab initio} compounds or two-phase regions. 
We include the differences between the formation energies of the phases in disagreement, both for US-LDA and PAW-GGA potentials.
(\NNdisagreementsGGA \,entries). 
Note {\bf (a)}: the {\it ab initio} ground state is within less than 10meV/atom of the experimental ground state (4).
Note {\bf (b)}: the assigned experimental structures are poorly justified (2) \cite{AuY.63Cha,CdPt.52Now}.
Note {\bf (c)}: unambiguous significant disagreement (3).
}





\newpage
\section{Appendix: new structure prototypes which are superstructures of FCC, BCC, HCP}
\label{section.structures}

\label{proto.AB3.CdPt11relaxed} 
\label{proto.AB3.PdPt12relaxed} 
\label{proto.AB3.PdPt11relaxed} 
\label{proto.AB.MoTi80}
\label{proto.AB3.MoTi81}
\label{proto.AB3.NbTc81}
\label{proto.AB3.RuTi81}
\label{proto.AB3.TcTi81}
\label{proto.AB2.RhRu142}
\label{proto.AB.RhRu126}

\def\pro{{}}


{\begin{center}
{\footnotesize
\hspace{-8mm}\begin{tabular}{||c||c||c|c||c|c||}\hline  
System           & CdPt$_3$, PdPt$_3$,            &     MoTi,                 & MoTi$_3$, Mo$_3$Ti, Nb$_3$Tc, &  RhRu                    &  RhRu$_2$                \\
                 &  Pd$_3$Pt                      &                           & RuTi$_3$, TcTi$_3$            &                          &                          \\ \hline 
Superlattice     &  FCC AB$_3$                    &  BCC A$_2$B$_2$           &  BCC AB$_3$                   & HCP A$_2$B$_2$           & HCP A$_2$B$_4$           \\ \hline 
Lattice          &   Orthorhombic                 &  Orthorhombic             &  Orthorhombic                 & Trigonal                 & Orthorhombic             \\ \hline 
Space Group      &    Cmmm \#65                   &  Imma \#74                &  Immm \#71                    & {\ftnsz P$\bar{3}$m1 \#164}       & Cmcm \#63                \\ \hline 
Pearson symbol   &        oC8                     &        oI8                &          oI8                  &  hP4                     & oC12                     \\ \hline
Primitive        &                                &                           &                               &                          &                          \\ 
vectors (cart.)  &                                &                           &                               &                          &                          \\
  {\bf a$_1$}/a  & $(  1, -1/2,  1/2)$            & $( 3/2,  1/2,-1/2)$       & $( 3/2,   1/2,  -1/2 )$       & $(1/2,-\sqrt{3}/2,0)$    & $(-1/2,3\sqrt{3}/2,0)$   \\
  {\bf a$_2$}/a  & $( -1, -1/2,  1/2)$            & $( 1/2,  3/2, 1/2)$       & $( 1/2,   3/2,   1/2 )$       & $(1/2,\sqrt{3}/2,0)$     & $(-1/2,-3\sqrt{3}/2,0)$  \\
  {\bf a$_3$}/a  & $(  0  -1/2, -1/2)$            & $(-1/2, -3/2, 1/2)$       & $(-1/2,  -3/2,   1/2 )$       & $(0,0,2\sqrt{8/3})$      & $(0,0,\sqrt{8/3})$       \\  \hline 
Atomic           &                                &                           &                               &                          &                          \\ 
positions (fract.)       &                                &                           &                               &                          &                          \\ 
{\bf A$1$}       & $(  0,   0,   0)$              & $(   0,   0,   0)$        &  $(   0,   0,   0 )$          & $(   0,   0,   0 )$      & $(   0,   0,   0 )$      \\
{\bf A$2$}       &        $-$                     & $( 1/4, 3/4, 1/2)$        &           $-$                 & $( 1/3, 2/3, 1/4 )$      & $(5/9,4/9,1/2)$          \\ 
{\bf B$1$}       & $(  0, 1/2, 1/2)$              & $( 1/2, 1/2,   0)$        &  $(1/4, 3/4, 1/2)$            & $(   0,   0, 1/2 )$      & $(2/9,7/9,1/2)$          \\
{\bf B$2$}       & $(1/2,   0, 1/2)$              & $( 3/4, 1/4, 1/2)$        &  $(1/2, 1/2,   0)$            & $( 1/3, 2/3, 3/4 )$      & $(1/3,2/3,0)$            \\
{\bf B$3$}       & $(1/2, 1/2,   0)$              &        $-$                &  $(3/4, 1/4, 1/2)$            &          $-$             & $(2/3,1/3,0)$            \\ 
{\bf B$4$}       &        $-$                     &        $-$                &           $-$                 &          $-$             & $(8/9,1/9,1/2)$          \\ \hline 
\end{tabular}}   
\end{center}
}
{
TABLE 10. Geometry of FCC, BCC, HCP superstructures which appear as new prototypes in our study. 
Positions are given as unrelaxed positions in the parent lattice.

\section{Appendix: relaxed structure prototypes}
\label{section.structures2}

\label{proto.AB3.MoZr124relaxed} 
\label{proto.AB5.MoTi128relaxed} 
\label{proto.AB5.MoZr132relaxed} 
\label{proto.AB5.NbRu132relaxed} 
\def\abg{\alpha,\beta,\gamma}

{\begin{center}
{\footnotesize
\hspace{-9mm}\begin{tabular}{||c||c||c|c|c||}\hline   
System           & MoZr$_3$ (AB$_3$)      & Mo$_5$Ti$^*$ (AB$_5$)      &  MoZr$_5$$^*$ (AB$_5$)      & Nb$_5$Ru$^*$ (AB$_5$)        \\ \hline  
Lattice          &     Orthorhombic       &    Monoclinic          &    Monoclinic           &    Monoclinic            \\ \hline 
Space Group      &     Imma \#74          &     C2/m \#12          &     C2/m \#12           &     C2/m \#12            \\ \hline 
Prototype        &     MoZr$_3$           & Mo$_5$Ti see note$^*$  & Mo$_5$Ti see note$^*$   & Mo$_5$Ti see note$^*$    \\ \hline 
Primitive vectors&                        &                        &                         &                          \\ 
$(a,b,c)$ (\AA)  &$ (5.678, 5.678, 5.678)$&$ (5.192, 5.192, 9.879)$&$ (5.695, 5.695, 10.881)$&$ (5.339, 5.339, 10.178)$ \\ 
$(\abg)$ degrees &$ (145.5, 128.9, 63.1) $&$(139.8, 139.8, 35.0)  $&$ (138.5, 138.5, 34.1)  $&$ (139.8, 139.8, 35.1)  $ \\ \hline
Atomic positions (fract.) &               &                        &                         &                          \\ 
{\bf A$1$}       &$(0,0,0)$               &$(0,0,0)$               &$(0,0,0)$                &$(0,0,0)$                 \\
{\bf B$1$}       &$(0.259,0.722,0.537)$   &$(0.166,0.166,0.833)$   &$(0.205,0.205,0.864)$    &$(0.170,0.170,0.836)$     \\
{\bf B$2$}       &$(0.524,0.500,0.024)$   &$(0.334,0.334,0.665)$   &$(0.349,0.349,0.692)$    &$(0.335,0.335,0.672)$     \\
{\bf B$3$}       &$(0.815,0.278,0.537)$   &$(0.500,0.500,0.499)$   &$(0.495,0.495,0.513)$    &$(0.501,0.501,0.500)$     \\
{\bf B$4$}       &        $-$             &$(0.667,0.667,0.334)$   &$(0.642,0.642,0.313)$    &$(0.666,0.666,0.328)$     \\
{\bf B$5$}       &        $-$             &$(0.834,0.834,0.166)$   &$(0.800,0.800,0.160)$    &$(0.831,0.831,0.164)$     \\  \hline 
\end{tabular}} 
\end{center}
}
{
TABLE 11. Geometry of relaxed structures which appear as new prototypes in our study. 
Positions are given as fractional positions of the primitive vectors. 
Note $^*$: Mo$_5$Ti, MoZr$_5$, and Nb$_5$Ru are slight distorsion of the prototype 
Mo$_5$Ti$^{proto}$, monoclinic lattice, space group C2/m \#12,
with atoms in the following Wyckoff positions: Ti in 2a, Mo(1) in 2c, Mo(2) in 4i ($x=1/3,z=1/6$), Mo(3) in 4i ($x=2/3,z=1/3$), and primitive vectors
{\bf a$_1$}/a=$(3.171,1,0)$, 
{\bf a$_2$}/a=$(-0.315,1,0)$,
{\bf a$_3$}/a=$(-0.638,-0.201,1.987)$
\cite{ref.cristallography}.
}


\twocolumn


\end{document}